\documentclass[11pt, a4paper, oneside]{Thesis} 

\usepackage{caption}
\usepackage{subcaption}
\usepackage{amssymb}
\usepackage{amsmath,array}
\usepackage{algorithm,algorithmicx,algpseudocode}
\usepackage{tikz}
\def\checkmark{\tikz\fill[scale=0.4](0,.35) -- (.25,0) -- (1,.7) -- (.25,.15) -- cycle;}
\tikzset{
  treenode/.style = {shape=rectangle, rounded corners,
                     draw, align=center,
                     top color=white, bottom color=blue!20},
  root/.style     = {treenode, font=\Large, bottom color=red!30},
  env/.style      = {treenode, font=\ttfamily\normalsize},
  dummy/.style    = {circle,draw}
}

\usepackage{xcolor,colortbl}
\definecolor{green}{rgb}{0.1,0.1,0.1}

\usepackage{array}
\usepackage{framed}
\usepackage{multirow}

\usepackage{pgfplots}
\pgfplotsset{width=15cm,compat=1.9}
\usepackage[squa   re, numbers, comma, sort&compress]{natbib} 
\hypersetup{urlcolor=black, colorlinks=true} 
\title{\ttitle} 

\begin{document}

\frontmatter 

\setstretch{1.3} 

\fancyhead{} 
\rhead{\thepage} 
\lhead{} 

\pagestyle{fancy} 

\newcommand{\HRule}{\rule{\linewidth}{0.5mm}} 

\hypersetup{pdftitle={\ttitle}}
\hypersetup{pdfsubject=\subjectname}
\hypersetup{pdfauthor=\authornames}
\hypersetup{pdfkeywords=\keywordnames}


\maketitle 
\makeapprovalpage 
 
%

\Declaration{

\addtocontents{toc}{\vspace{0em}} 

I, \authornames, declare that this thesis titled, '\ttitle' and the work presented in it are my own. I confirm that:

\begin{itemize} 
\item[\tiny{$\blacksquare$}] This work was done wholly or mainly while in candidature for a research degree at this University.
\item[\tiny{$\blacksquare$}] Where any part of this thesis has previously been submitted for a degree or any other qualification at this University or any other institution, this has been clearly stated.
\item[\tiny{$\blacksquare$}] Where I have consulted the published work of others, this is always clearly attributed.
\item[\tiny{$\blacksquare$}] Where I have quoted from the work of others, the source is always given. With the exception of such quotations, this thesis is entirely my own work.
\item[\tiny{$\blacksquare$}] I have acknowledged all main sources of help.
\item[\tiny{$\blacksquare$}] Where the thesis is based on work done by myself jointly with others, I have made clear exactly what was done by others and what I have contributed myself.\\
\end{itemize}
 
Signed:\\
\rule[1em]{25em}{0.5pt} 
 
Date:\\
\rule[1em]{25em}{0.5pt} 
}

\clearpage 


\pagestyle{empty} 

\null\vfill 

\textit{``The scientist is not a person who gives the right answers, he is one who asks the right questions.'' } 

\begin{flushright}
Claude Levi-Strauss
\end{flushright}

\vfill\vfill\vfill\vfill\vfill\vfill\null 

\clearpage 


\addtotoc{Abstract} 

\abstract{\addtocontents{toc}{\vspace{0em}} 

The advance of cloud computing and big data technologies brings out major changes in the ways that people make use of information systems. While those technologies extremely ease our lives, they impose the danger of compromising privacy and security of data due to performing the computation on an untrusted remote server. Moreover, there are also many other real-world scenarios requiring two or more (possibly distrustful) parties to securely compute a function without leaking their respective inputs to each other. In this respect, various secure computation mechanisms have been proposed in order to protect users' data privacy. Yao's garbled circuit protocol is one of the most powerful solutions for this problem.  In this thesis, we first describe the Yao's protocol in detail, and include the complete list of optimizations over the Yao's protocol. We also compare their advantages in terms of communication and computation complexities, and analyse their compatibility with each other. We also look into generic Yao implementations (including garbled RAM) to demonstrate the use of this powerful tool in practice. We compare those generic implementations in terms of their use of garbled circuit optimizations. We also cover the specific real-world applications for further illustration. Moreover, in some scenarios, the functionality itself may also need to be kept private which leads to an ideal solution of secure computation problem. In this direction, we finally cover the problem of Private Function Evaluation, in particular for the 2-party case where garbled circuits have an important role. We finally analyse the generic mechanism of Mohassel \textit{et al.} and contribute to it by proposing a new technique for the computation of the number of possible circuit mappings. 

}

\clearpage 

\addtotoc{\"Oz} 

\oz{\addtocontents{toc}{\vspace{0em}} 

Bulut bili\c{s}im ve b\"uy\"uk veri teknolojilerinin ilerlemesi insanlar{\i}n bili\c{s}im sistemlerini kullanma yollar{\i}nda b\"uy\"uk de\u{g}i\c{s}imler getirmi\c{s}tir. Bu teknolojiler hayat{\i}m{\i}z{\i} b\"uy\"uk \"olç\"ude kolayla\c{s}t{\i}r{\i}rken, ayn{\i} zamanda hesaplamalar{\i}n uzak bir sunucuda yap{\i}lmas{\i} nedeniyle bilgilerin mahremiyetini ve g\"uvenli\u{g}ini tehlikeye atmaktad{\i}rlar. Birbirine yeterince g\"uvenemeyen iki veya daha fazla taraf{\i}n bir fonksiyonu g\"uvenli olarak hesaplamas{\i}n{\i} gerektiren gerçek hayatta kar\c{s}{\i}la\c{s}{\i}labilecek birçok durum vard{\i}r. Bu sebeple, kullan{\i}c{\i}lar{\i}n veri mahremiyetini koruyan çe\c{s}itli g\"uvenli hesaplama y\"ontemleri \"onerilmi\c{s}tir. Yao'nun kar{\i}\c{s}t{\i}r{\i}lm{\i}\c{s} devresi protokol\"u bu g\"uvenli hesaplama problemine kar\c{s}{\i} \"onerilmi\c{s} en g\"uçl\"u ç\"oz\"umlerden biridir. Bu tezde, \"oncelikle Yao protokol\"un\"u ve bu protokol\"un optimizasyonu için \"onerilmi\c{s} geli\c{s}melerin tam listesini anlatmaktay{\i}z. Ayn{\i} zamanda, bu geli\c{s}meleri ileti\c{s}im ve hesaplama zorlu\u{g}u olarak k{\i}yasl{\i}yoruz ve birbirleriyle uyumluluklar{\i}n{\i} analiz ediyoruz. Bu g\"uçl\"u protokol\"un pratikteki kullan{\i}m{\i}n{\i} g\"ostermek amac{\i}yla çe\c{s}itli genel Yao uygulamalar{\i}n{\i} (kar{\i}\c{s}t{\i}r{\i}lm{\i}\c{s} RAM dahil) inceliyoruz. Bu uygulamalar{\i} kulland{\i}klar{\i} kar{\i}\c{s}{\i}k devre optimizasyonlar{\i}na g\"ore k{\i}yasl{\i}yoruz. \"ozel olarak baz{\i} gerçek-hayat uygulamalar{\i}yla Yao protokol\"un\"u daha da \"orneklendiriyoruz. Hesaplanacak fonksiyonun da gizli bir bilgi olmas{\i} durumunda, onun da gizlenmesinin tam bi mahremiyet için gerekli oldu\u{g}u unutulmamal{\i}d{\i}r. Bu do\u{g}rultuda geli\c{s}tirilmi\c{s} olan gizli fonksiyon hesaplama y\"ontemlerini, \"ozellikle kar{\i}\c{s}{\i}k devrelerin \"onemli bir rol\"un\"un oldu\u{g}u iki tarafl{\i} durum i\c{c}in tezimizde anlat{\i}yoruz. Son olarak Mohassel ve Sadeghian'{\i}n geli\c{s}tirmi\c{s} oldu\u{g}u mekanizmay{\i} ele al{\i}yoruz ve olas{\i} devre haritalar{\i}n{\i}n say{\i}s{\i}n{\i} hesaplamak i\c{c}in kullan{\i}lacak yeni bir teknik \"onererek buna katk{\i}da bulunuyoruz.

{\noindent \bf Anahtar S\"ozc\"ukler:} \turkishkeywordnames
}

\clearpage 


\setstretch{1.3} 

\pagestyle{empty} 

\dedicatory{To my mom and dad for their patience and love} 

\addtocontents{toc}{\vspace{0em}} 

\setstretch{1.3} 

\acknowledgements{\addtocontents{toc}{\vspace{0em}} 

First of all, I would like to thank to my advisor Prof. Dr. Ensar G\"{u}l for his contributions and support. I would also thank to my co-advisor Dr. Mehmet Sab{\i}r Kiraz to whom I cannot express my gratitude  for introducing the area of cryptographic protocols to me, for encouraging me studying Yao's garbled circuits, for his contributions during the preparation of my thesis, his perfect guidance, suggestions and feedbacks. Without his input, this thesis could not have been completed as good as it is right now. He is quite enthusiastic in teaching and guiding which makes him the best mentor that a master student wish to have during this hard process.

I would also thank to Muhammed Ali Bing\"{o}l for interesting discussions, insightful suggestions, and successful observations on my thesis. His ideas and feedbacks have helped me a lot  to improve the thesis significantly. I am also grateful to Dr. Osmanbey Uzunkol and Dr. \.{I}sa Sertkaya for reviewing my thesis, making comments and giving quite helpful feedbacks.

I would also thank to Mike Rosulek although we have not met personally. His talk in Simons Institute, University of California, Berkeley, namely A Brief History of Practical Garbled Circuit Optimizations gave me the basic understanding of the Yao's protocol and the starting point for my research of garbled circuit optimizations.

The last but not the least, I cannot express my gratitude for my mom and dad for supporting me in my all decisions, including going after the area of secure computation. They have always helped me, guided me and been the perfect parents whom any child would ever hope for.
}
\clearpage 


\pagestyle{fancy} 

\lhead{\emph{Contents}} 
\tableofcontents 

\lhead{\emph{ List of Figures}} 
\listoffigures 

\lhead{\emph{List of Tables}} 
\listoftables 


\clearpage 

\setstretch{1.5} 

\lhead{\emph{Abbreviations}} 
\listofsymbols{ll} 
{
\textbf{�} & described in [chapter/section/subsection] \\
\textbf{MPC} & Secure \textbf{M}ulti-\textbf{P}arty \textbf{C}omputation \\
\textbf{SFE} &  \textbf{S}ecure \textbf{F}unction \textbf{E}valuation \\
\textbf{GMW} &  \textbf{G}oldreich-\textbf{M}icali-\textbf{W}idgerson \\
\textbf{2PC} &  Secure \textbf{2}-\textbf{P}arty \textbf{C}omputation \\
\textbf{RAM} &  \textbf{R}andom \textbf{A}ccess \textbf{M}emory \\
\textbf{{PRF}} &  \textbf{P}seudo-\textbf{R}andom \textbf{F}unction \\
\textbf{MAC} &  \textbf{M}essage \textbf{A}uthentication \textbf{C}ode \\
\textbf{DKC} &  \textbf{D}ual \textbf{K}ey \textbf{C}ipher \\
\textbf{OT} &  \textbf{O}blivious \textbf{T}ransfer \\
\textbf{HE} &  \textbf{H}omomorphic \textbf{E}ncryption \\
\textbf{CPU} &  \textbf{C}entral \textbf{P}rocessing \textbf{U}nit \\
\textbf{ct} &  \textbf{c}ipher\textbf{t}ext \\
\textbf{edt} &  total \textbf{e}ncryption and/or \textbf{d}ecryption \textbf{t}ime \\
\textbf{P\&P} &  \textbf{P}oint \& \textbf{P}ermute \\
\textbf{GRR3} &  \textbf{G}arbled \textbf{R}ow \textbf{R}eduction \textbf{3} ciphertexts \\
\textbf{GRR2} &  \textbf{G}arbled \textbf{R}ow \textbf{R}eduction \textbf{2} ciphertexts \\
\textbf{lsb} &  \textbf{l}east \textbf{s}ignificant \textbf{b}it \\
\textbf{cpg} &  \textbf{c}ycles \textbf{p}er \textbf{g}ate \\
\textbf{bpg} &  \textbf{b}ytes \textbf{p}er \textbf{g}ate \\
\textbf{PFE} &  \textbf{P}rivate \textbf{F}unction \textbf{E}valuation \\
\textbf{IBE} &  \textbf{I}dentity \textbf{B}ased \textbf{E}ncryption \\
\textbf{ORAM} &  \textbf{O}blivious \textbf{R}andom \textbf{A}ccess \textbf{M}emory \\
\textbf{erf} &  \textbf{e}rror \textbf{f}unction \\
\textbf{CTH} &  \textbf{C}ircuit \textbf{T}opology \textbf{H}iding \\
\textbf{PGE} &  \textbf{P}rivate \textbf{G}ate \textbf{E}valuation \\
\textbf{ow} &  \textbf{o}utgoing \textbf{w}ire \\
\textbf{iw} &  \textbf{i}ncoming \textbf{w}ire \\
\textbf{EP} &  \textbf{E}xtended \textbf{P}ermutation \\
\textbf{OMAP} &  \textbf{O}blivious \textbf{M}apping \\
\textbf{OEP} &  \textbf{O}blivious Evaluation of \textbf{E}xtended \textbf{P}ermutation \\
\textbf{SN} &  \textbf{S}witching \textbf{N}etwork \\
\textbf{PN} &  \textbf{P}ermutation \textbf{N}etwork \\
\textbf{OSN} &  \textbf{O}blivious Evaluation of \textbf{S}witching \textbf{N}etworks \\
}


\mainmatter 

\pagestyle{fancy} 


%


\chapter{Introduction} 

\label{Chapter1} 

\lhead{Chapter 1. \emph{Introduction}} 


Two rich people want to determine which one of them is richer so that he would pay the bill for the dinner. However, none of them is willing to permit the other learn more information about his personal wealth than what the mere knowledge of who is richer does. They start discussing how they could achieve this just by talking to each other. They are quite sure that both will always tell the truth since they are honourable businessmen who cannot take the risk of being caught while lying. On the other hand, both suspect that the other may try to deduce information about his wealth from the conversation. After some time of discussion, they come to the conclusion that it is impossible to decide who is richer under these conditions since they do not know much about \textit{secure computation} techniques.

This  famous problem is known as ``millionaires' problem''  proposed by Andrew Yao \cite{Yao82}. He has also proposed a cryptographic solution for this problem, and generalized it to the secure computation of any function  \cite{Yao82, Yao86}.  His later work has showed that any function that can be computed by a polynomial-size circuit can be computed securely \cite{Yao86}. The problem has further widened and solved for the case of more than two parties \cite{Yao82, GMW87}. Yao's research is followed by many others' in constituting an active subfield of cryptography known as \textit{secure multi-party computation} (MPC) or \textit{secure function evaluation} (SFE), which aims solving the problem of two or more parties computing a function jointly without revealing their secret inputs to each other.

There are many real-life examples where MPC techniques can be applied, including financial systems \cite{BTW12}, cooperation of intelligence agencies, companies and governments \cite{LP08, HLOW16}, electronic elections \cite{CGS97}, electronic auctions \cite{BCD+09, NPS99}, secure biometric identification \cite{KN13, BCP13, KGK15}, secure e-mail filtering \cite{LADM14}, \textit{etc.} In fact, there is no bound for the areas where MPC may be used, and it can be adopted in any case some parties are required to compute a function on their private data.

Various methods have been proposed for MPC, including generic methods and function specific methods. Although function specific methods usually run more efficiently, they are limited in use due to the fact that each of them works for only one function. It is quite inefficient to design a method and to prove its security for each different function unless the function will be used many times. An example of frequently used functions is the Hamming distance calculation which is used in many scenarios, including biometric checks \cite{KN13} $etc.$ Hence, designing a specific protocol for it while proving its security makes sense \cite{BCP13, KGK15}. However, general research approach is  towards the generic methods which can be applied to arbitrary functions.

Generic methods have been developed for usage in an unlimited set of functions. Usually one method is better than the other for different computational settings. For instance, \textit{homomorphic encryption} will be a very good fit for arithmetic circuits if an efficient fully homomorphic encryption scheme become available in the future \cite{Sch11}. However, currently the proposed fully homomorphic encryption schemes  are inefficient for practical secure computation. 

The most efficient methods for secure computation of functions represented as boolean circuits include \textit{GMW protocol} \cite{GMW87} and \textit{Yao's garbled circuit protocol} (Yao's protocol).  The former usually gives better results in the presence of at least three parties, while the latter is usually better for two-party case.  

Yao's protocol remains one of the most important paradigms for MPC, especially in the case of \textit{secure two-party computation} (2PC) \cite{ZRE15}. In particular, it is valuable for its constant round complexity. Since the time it was proposed by Andrew Yao in \cite{Yao86}, it has become one of the major fields in modern cryptographic research. It is constantly being  optimized in terms of communication complexity and computation complexity.

While the research for optimizing Yao's protocol scheme continues, various practical applications using Yao's protocol have also been developed. These applications demonstrate that it is a promising cryptographic primitive for a wide range of applications, including privacy preserving data mining, efficient secure two-party computation, private function evaluation \textit{etc}.

In this thesis, we first describe the Yao's protocol in detail, and include the complete list of optimizations over the Yao's protocol. We also {compare their advantages in terms of communication and computation complexities, and analyse their compatibility with each other}. We also look into generic Yao implementations (including garbled RAM) to demonstrate the use of this powerful tool in practice. We compare those generic implementations in terms of their use of garbled circuit optimizations. We also cover the specific real-world applications for further illustration. Moreover, in some scenarios, the functionality itself may also need to be kept private which leads to an ideal solution of secure computation problem. In this direction, we finally cover the problem of {Private Function Evaluation}, in particular for the 2-party case where garbled circuits have an important role. We finally analyse the generic mechanism of Mohassel \textit{et al.} and contribute to it by proposing {a new technique for the computation of the number of possible circuit mappings}.

\section{Overview of the Thesis} \label{sec:overviewofthesis}{}
\begin{large}
\textbf{Research goal:}
\end{large}

\begin{framed}
Our goal in this thesis is to compare the advantages of currently known Yao's protocol optimizations in terms of communication and computation complexities, to analyse their compatibility with each other, to demonstrate their role with a view towards its practical and real-world applications and in private function evaluation. We intend to describe  the current state of the art for Yao's protocol, since it is hard to find many comprehensive works about it. We believe that this work will be quite useful to cryptography community as a study material as well.
\end{framed}

\begin{large}
\textbf{Organization of the thesis:}
\end{large}

\textbf{Chapter \ref{Chapter1}: Introduction} \\
Chapter \ref{Chapter1} is dedicated to introduction and overview of the thesis.

\textbf{Chapter \ref{Chapter2}: Preliminaries} \\
Chapter \ref{Chapter2} is dedicated to generic MPC methods, and to cryptographic basis. We also included a section for circuit concepts which is assumed to be helpful for the people with potentially different backgrounds.

\textbf{Chapter \ref{Chapter3}: Yao's Garbled Circuit Protocol} \\
Chapter \ref{Chapter3} includes general description and formal definition of Yao's protocol, as well as the generic Yao's protocol template together with its security properties.

\textbf{Chapter \ref{Chapter4}: Garbled Circuit Optimizations} \\
Chapter \ref{Chapter4} presents known garbled circuit optimizations in a chronological order (\textit{i.e.}, P\&P (\ref{sec:PP}), GRR3 (\ref{sec:GRR3}{}), free \texttt{XOR} (\ref{sec:freeXOR}), GRR2 (\ref{sec:GRR2}{}), fle\texttt{XOR} (\ref{sec:fleXOR}{}), half gates (\ref{sec:halfgates}{})). We analyze these optimizations in terms of their relations and contradictions as well as their compatibility with each other. One of our aims is to give a clear overview, therefore, we did not get involved with proofs and other related complex formulas. 

\textbf{Chapter \ref{chap:implementations}: Practical Implementations of Yao's Protocol} \\
Chapter \ref{chap:implementations} composes of generic Yao's protocol applications and some real-world examples, including pipelining method, garbled RAM, MPC for satellite collusion probability, and privacy preserving data mining.

\textbf{Chapter \ref{Chapter6}: Private Function Evaluation} \\
Chapter \ref{Chapter6} is dedicated to private function evaluation. We intend to describe Mohassel \textit{et al.}'s generic PFE scheme, which is the most efficient to date, and its application to Yao's protocol. We contribute to it by proposing {a new technique for the computation of the number of possible circuit mappings}.

\textbf{Chapter \ref{chap:conclusion}: Conclusion and Discussions} \\
Chapter \ref{chap:conclusion} concludes with general discussions of garbled circuit optimization techniques, Yao's protocol applications and private function evaluation.


\chapter{Preliminaries} 

\label{Chapter2} 

\lhead{Chapter 2. \emph{Preliminaries}} 


In this chapter, we will present the basic concepts of secure computation techniques. First, we will show the required properties for a secure computation scheme. We will continue with general adversary models in cryptographic protocols. This will be followed by circuit concepts useful for MPC techniques which, we suppose, will be quite helpful for people new to the area. Then, we will present general cryptographic primitives. We will  also give the summary of oblivious transfer protocol, homomorphic encryption, and GMW protocol. 

\section{Requirements of Secure Multi-Party Computation} \label{sec:mpc}{}  

To formally claim and prove the security of an MPC protocol, some general security properties are required \cite{LP08}. The most central of these properties are described in \cite{LP08} by Lindell \textit{et al.} as follows:

\begin{enumerate}
\item{\textit{Correctness:} The output that is delivered to each party (\textit{i.e.} each participant of the MPC protocol) is guaranteed to be correct.}
\item{\textit{Privacy}: None of the participants is allowed to learn anything more about other participants' inputs than what he can learn from the output itself.}
\item{\textit{Independence of inputs:} The protocol may not allow any of the parties to choose his input based on other parties' inputs. This property is different from privacy since choosing an input dependent on another party's unknown input is possible .}
\item{\textit{Guaranteed output delivery:} In the end of the protocol, honest parties should receive their outputs no matter how hard corrupt parties try to prevent it.}
\item{\textit{Fairness:} A party whether he is corrupt or not can receive his output if all of the parties receive their outputs. For detailed information about how to achieve efficient fair MPC, we  refer the reader to \cite{KiS08, Kir08}.}
\end{enumerate}

Lindell \textit{et al.} stress that this list does not define security, but rather compose of the requirements that any secure protocol must conform \cite{LP08}.

 \section{Adversary Models} \label{sec:adversarymodels}{}

Security of cryptographic protocols are formalized and proved against adversaries with different capabilities \cite{Sch11}.

 \subsection{Semi-Honest Adversaries} \label{sub:semi-honest}{}
 
The \textit{semi-honest} (also known as passive, or  honest-but-curious) threat model is the standard adversary model for MPC. Here parties typically follow the protocol as they are supposed to but may try to deduce information about another party's input from the protocol transcript \citep{HEKM11}. If a protocol is secure against semi-honest adversaries, it does not allow them to learn any extra information from the protocol.

\subsection{Covert Adversaries} \label{sub:covert}{}
 
\textit{Covert adversaries} constitute the type of adversaries that are allowed to deviate from the protocol with a restriction that they must evade being caught while they are doing so \cite{Sch11}. It can be safely assumed that in many political, social and business scenarios, the gain from cheating is overweighted by the results of being caught. If those deviations are detected with a certain frequency (\textit{e.g.}, 1 out of 10 times), such a protocol can be considered secure enough.  If a protocol is secure against covert adversaries, it allows catching those adversaries with a certain probability if they deviate from the protocol.

\subsection{Malicious Adversaries} \label{sub:malicious}{}

The strongest type of adversaries is the \textit{malicious adversaries} (also known as active adversaries), which may deviate from the protocol arbitrarily so that they can extract the other parties private inputs or alter the computation outcome \cite{Sch11}. If a protocol is secure against malicious adversaries, a corrupt party will be caught whenever he deviates from the protocol.

Throughout this thesis, we focus on the security against semi-honest adversaries due to the following reasons  \citep{HEKM11}:

\begin{enumerate}
\item{There are many real-world situations where modelling the parties as semi-honest adversaries is appropriate: 
\begin{enumerate}
\item{where parties are legitimately trusted but there is a legal need for preventing them from divulging information, or for protection against break-ins in the future.}
\item{where the software used for MPC can hardly be changed by participants without being detected, either due to software attestation use or the fact that internal controls are in place (\textit{e.g.}, when parties are government agencies, or large corporations).}
\end{enumerate}}
\item{Securing protocols against semi-honest adversaries is an important step toward construction of secure protocols against stronger adversaries. There are generic ways of altering them to achieve security against covert or malicious adversaries \cite{Kir08, KiS06}.}
\end{enumerate}

 \section{Corruption Models} \label{sec:corruptionmodels}{}
 
Apart from the above adversary models, there also exist static and adaptive corruption models.

\textit{Static corruption model:} This model implies that if a party is honest in the beginning, he always remains honest; whereas if a party is corrupted in the beginning, he always remains corrupted \cite{LP08}.

\textit{Adaptive corruption model:} Instead of including a fixed number of corrupted parties, adaptive corruption model suggests that the number of corrupted parties may increase during the computation. However, if a party gets corrupted, it remains that way from then on \cite{LP08}. Therefore, there may never be a decrease in the number of corrupted parties.

\section{Circuit Concepts}\label{sec:circuitconsepts}{}

For a generic MPC protocol to take place, first a function must be written  as a combination of common building blocks, \textit{i.e.}, they must be represented as \textit{circuits}. The number of types of building blocks is limited. Therefore, by showing how to compute each building block, a generic MPC scheme permits calculation of unlimited functions. Standard circuit representations generally used in MPC protocols are boolean circuits and arithmetic circuits \cite{Sch11}.

\subsection{Boolean circuits} 

\begin{figure}
\includegraphics[height=6.2cm, angle=0]{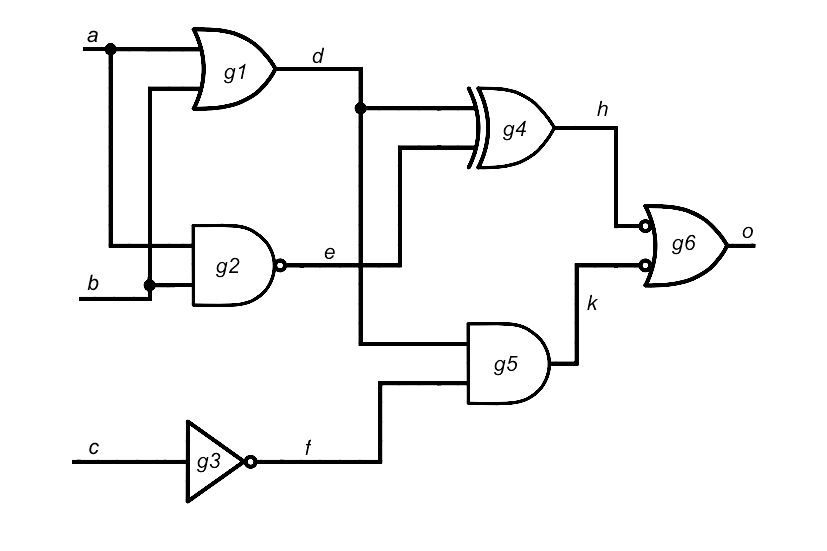}
\caption{ An example boolean circuit.}
\label{fig:booleancirc}
\end{figure}

In engineering and computer science, functions are classically represented as \textit{Boolean circuits} \cite{Sch11}. A boolean circuit basically composes of \textit{logic gates} and \textit{wires} connecting them \cite{Ana14}. Figure \ref{fig:booleancirc} shows an example boolean circuit whose wires are $a$, $b$, $c$, $d$, $e$, $f$, $h$, $k$, and $o$, and gates are $g1$, $g2$, $g3$, $g4$, and $g5$.

$a$, $b$, and $c$ are the \textit{input}s of the circuit in Figure \ref{fig:booleancirc}, $d$, $e$, $f$, $h$, and $k$ are the \textit{intermediate wires}, and $o$ is the \textit{output} wire. A boolean circuit may have more than one output as well. A wire is exactly 1 bit that may have one of the two truth values, \textit{i.e.}, either \texttt{TRUE} (also denoted as 1 or \texttt{High}) or \texttt{FALSE} (also denoted as 0 or \texttt{Low}). When 2 wires cross each other, they are connected if there is a big dot in the connection point, otherwise they are not connected. For example $a$ and $b$ cross each other but not connected (the same applies to $d$ and $e$ in Figure \ref{fig:booleancirc}).

A logic or boolean gate generally takes 1 or 2 wires as input (although there is no certain limitation) and outputs exactly 1 wire. Formally a $d$-input gate $G_d$ is a boolean function mapping $d>0$ bits input to 1-bit output, \textit{i.e.} \cite{Sch11}:

\begin{equation}
 G_d: (in_1,\ldots,in_d) \in \{0,1\}^d \rightarrow (out)\in \{0,1\}
\end{equation}

For the gates of the circuit in Figure \ref{fig:booleancirc}, the left sides are used for the input, the right side are used for the output. For example $g1$ in Figure \ref{fig:booleancirc} takes $a$ and $b$ as inputs and outputs $d$. However, gates may be rotated in a different circuit. In this case, one needs to look at the two asymmetric sides of a gate. Generally, the larger assymetric side of the gate is the side of inputs and the narrower assymetric side is for the output. \textit{A wire can only be an output of exactly 1 gate}, although it can be input to multiple gates  \cite{Ana14}.

In Figure \ref{fig:booleancirc}, $g1$ is an \texttt{OR} gate ($d\gets a\vee b$), $g4$ is an \texttt{XOR} gate ($h\gets d\oplus e$), and $g5$ is an \texttt{AND} gate ($k\gets d\wedge f$). If there is a bubble on the wire, its truth value is inverted after the bubble. For example, $g2$ would have been an \texttt{AND} gate without the bubble on its output. But the bubble means the output is inverted. Actually, there is a special name for the type of $g2$, it is a \texttt{NAND} gate ($e\gets (a\wedge b)'$). $g3$ would have been a \texttt{buffer} gate without the bubble on its output. A \texttt{buffer} gate outputs the input as it is. However, with the bubble $g2$ is a \texttt{NOT} gate ($f=c'$). $g6$ would have been an \texttt{OR} gate without the bubbles on its inputs. Now, it takes the inputs inverted, and \texttt{OR}s them afterwards ($o\gets h'\vee k'$). Actually $g6$ is another representation of a \texttt{NAND} gate due to the logic identity $h'\vee k'= (h\wedge k)'$. There also exist \texttt{NOR} gates represented as an \texttt{OR} gate with a bubble on its output.

The truth table of a gate shows the relation between its possible inputs and its possible outputs. The truth table of a gate has $2^k$ rows where $k$ is the number of its input wires. The truth table of the \texttt{AND} gate $g5$ in Figure \ref{fig:booleancirc} can be seen in Table \ref{tab:truthtable}.

\begin{table}[]
\centering
\caption{Truth table of an \texttt{AND} gate ($g5$ in Figure \ref{fig:booleancirc}).}
\label{tab:truthtable}
\begin{tabular}{|cc|c|}
\hline
$d$ & $f$ & $k=d\wedge f$ \\ \hline
0 & 0 & 0 \\
0 & 1 & 0 \\
1 & 0 & 0 \\
1 & 1 & 1 \\ \hline
\end{tabular}
\end{table}

In fact, there are basically $2^4$ different 2-input gates in total. However, some of them are trivial (\textit{i.e.}, the ones whose output depends only one of the inputs and the ones whose output depends none of the inputs). Those gates can be replaced by more efficient representations, \textit{e.g.}, wires, \texttt{NOT} gates, \textit{etc.} The remaining non-trivial gates fall into the category of either even gates or odd gates \cite{PSSW09}.

\begin{definition}{Even gates}
are the 2-input gates whose truth table has 2 \texttt{FALSE} outputs and 2 \texttt{TRUE} outputs. 
\end{definition}

\begin{definition}{Odd gates}
are the 2-input gates whose truth table has either 3 \texttt{FALSE} outputs and 1 \texttt{TRUE} output or 1 \texttt{FALSE} output and 3 \texttt{TRUE} outputs. 
\end{definition}

There are only 2 non-trivial even gates which are \texttt{XOR} and \texttt{XNOR}, and 8 non-trivial odd gates, including \texttt{OR}, \texttt{AND}, \texttt{NOR}, \texttt{NAND}, \textit{etc} \cite{PSSW09}.

The \textit{size} of a boolean circuit means the number of its gates \cite{Vol99}. The \textit{depth} of a boolean circuit means the number of gates in the longest path that must be taken from any input to any output \cite{Vol99}. The \textit{topology} of a boolean circuit means the connections between its gates \cite{Vol99}. A boolean circuit can uniquely be defined by its topology and its gates.

The \textit{topological order} of a boolean circuit is that when its gates are indexed as $G_1,\ldots , G_n$, $i^{th}$, a gate $G_i$ does not get the output of a succeeding gate $G_{j>i}$ as its input \cite{Sch11}. Intuitively, in order to compute a gate, all of its input wires must be known, which can be ensured by computing the gates in topological order. By computing the gates one-by-one in topological order the whole boolean circuit can be computed. The topological order is not necessarily unique for a given boolean circuit \cite{Sch11}. 

A group of gate types ($G_1,\ldots,G_n$) is \textit{Turing-complete}, if and only if any {probabilistic polynomial time} algorithm can be represented by a combination of those gates \cite{Sip96}. Examples are (\texttt{AND},\texttt{XOR}) and (\texttt{NAND}). Building a \texttt{NAND} gate from a group of gates is an easy way to see whether that group of gates is Turing-complete or not.

A decrease in the number of gates in a circuit also means a decrease in overall cost of an MPC protocol in terms of computation complexity, and communication complexity. There are various techniques for circuit optimizations. Some circuit optimization techniques intend to reduce the number of odd gates at the cost of increasing the even gates. They could  also be useful in some MPC techniques \cite{KK10,BPP00}.

\subsection{Arithmetic circuits} \label{sub:arth}

\begin{figure}
\includegraphics[height=3.5cm, angle=0]{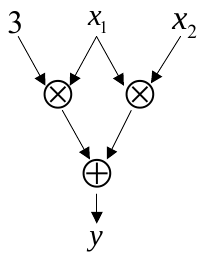}
\caption{ An example arithmetic circuit \cite{Sch11}.}
\label{fig:arithmeticcirc}
\end{figure}

A more compact representation for functions is \textit{arithmetic circuits} \cite{Sch11}. Unlike boolean circuits where wires are chosen from $ \mathbb{Z}_2$, here wires have values chosen from $\mathbb{Z}_{m\geq2}$. The gates operations are either modular addition $+$ or modular multiplication $\times$. Figure \ref{fig:arithmeticcirc} shows an example arithmetic circuit.

One can express any boolean circuit as an arithmetic circuit over $ \mathbb{Z}_2$. However, if $\mathbb{Z}_m$ has a modulus $m$ which is sufficiently large, then the resulting arithmetic circuit representation of a function will probably have much lower size than its boolean circuit representation, since a single operation will be enough for each integer addition or multiplication \cite{Sch11}.

Computations on both positive and negative integers $x$ can be simulated by arithmetic circuits, since one can map them into elements of $\mathbb{Z}_m: \mathbb{Z}\rightarrow\mathbb{Z}_m, x\rightarrow x \texttt{ mod } m$ \cite{Sch11}.

\section{Cryptographic Basis} \label{sec:cryptographicalbasis}{}

As the cryptographic basis for this thesis, we  present symmetric and public key encryptions, cryptographic hash functions, pseudo-random functions and message authentication codes. We will include only a brief summaries of them due to the fact that the details of them are not necessary for understanding protocols and that vast majority of our readers will probably have an acquaintance with them. However, at the end of this section we present dual-key ciphers in more detail because of their intensive use in Yao's protocol and supposed unfamiliarity of some readers with them.

\subsection{Symmetric Encryption} \label{sub:SymetricEncryptions}{} 
A \textit{symmetric encryption} scheme uses the same cryptographic key $k$ for both encryption of plaintext and decryption of ciphertext \cite{AES-FIPS}. A well-known example is \texttt{AES} encryption \cite{AES-FIPS}. The notation $c\gets E_k(m)$ means that a plaintext message $m$ is encrypted with a key $k$ resulting in a ciphertext $c$.

Decryption is generally denoted as either $m\gets D_k(c)$ or the inverse of $E$, namely $m\gets E_k^{-1}(c)$ .

\subsection{Public Key Encryption} \label{sub:publickeycryptography}{} 
A \textit{public key encryption} scheme uses different keys for encryption and decryption. \textit{Public keys} which are known publicly as their name implies are used for encryption, while private keys which are known only to their owners are used for decryption \cite{RSA78}. Any user can encrypt a message with the public key of the receiver, but the resulting ciphertext can be decrypted only with the receiver's private key. The notation $c\gets E_{pk_i}(m)$ means that a plaintext message $m$ encrypted with a public key $pk_i$ of $i^{th}$ person results in a ciphertext $c$. 

Decryption with the secret key $sk_i$ of the $i^{th}$ person is denoted as either $m\gets D_{sk_i}(c)$ or $m\gets E_{pk_i}^{-1}(c)$. The well-known public key cryptosystems  are \texttt{ElGamal} \cite{ElGamal85} and \texttt{RSA} \cite{RSA78}.

\subsection{Cryptographic Hash Function} \label{sub:cryptographichashfunction}{}  
 A \textit{cryptographic hash function} $H(m)$ maps an arbitrary size message $m$ to a fixed size $\ell$-bit string $c\gets H(m)$ \cite{RS04}. Throughout this thesis when we say \textit{hash function}, we refer to a cryptographic hash function. 

Hash functions are ideally modelled in the \textit{random oracle model} \cite{KM15}. A random oracle is a theoretical black-box responding to every unique query with a true random number picked from its output domain. It records its responses to unique queries so that it can respond a query the same way every time it is repeated. A well-known hash function scheme is $\texttt{SHA}256$ \cite{Han11}.

\subsection{Pseudo-Random Function} \label{sub:pseudorandomfunction}{}

A \textit{pseudo-random function} (\texttt{PRF}) is a function that can be used for pseudo-random generation, \textit{i.e.}, it can be modelled as random oracle. It is denoted as $\texttt{PRF}(x)$ on an input $x$. Its representation can be extended as $ \texttt{PRF}_k(x)$  to include the use of a private key $k$ \cite{Sch11}.

An instantiation of \texttt{PRF} can be achieved with a block cipher, \textit{e.g.}, \texttt{AES}, or a hash function, \textit{e.g.}, $\texttt{SHA}256$.  In case a \texttt{PRF} with the same key $k$ is  repeatedly  used, the \texttt{AES} instantiation would be more efficient since its key schedule needs to be run just once \cite{Sch11}.

\subsection{Message Authentication Code (\texttt{MAC})} \label{sub:messageauthenticationcode}{}

A \textit{message authentication code} (\texttt{MAC}) is a fixed-sized data that is used for authentication of a message. It is denoted as $\texttt{MAC}_k(m)$ on an input message $m$ that needs to be authenticated and a private key $k$ \cite{Sch11}.

The \texttt{MAC} value provides  protection for both data integrity and  authenticity of a message since it  allows the detection of any changes in the message content by the verifiers  possessing  the private key $k$.

 \subsection{Dual-Key Cipher} \label{part:dkcschemes}{}
A \textit{dual-key cipher} (\texttt{DKC}) is a cryptographic notion proposed by Bellare \textit{et al.} in \cite{BHR12}. A \texttt{DKC} formally represents a two-key lockbox where  both keys are required for openning the box. A \texttt{DKC} is a function $E$ associating a security parameter $k \in N$ where $N$ is the set of positive integers and keys $A, B \in \{0, 1\}^k$ with a $k$-bit pseudo-random number $E_{A,B} : \{0, 1\}^k \rightarrow \{0, 1\}^k$. Let $D_{A,B} : \{0, 1\}^k \rightarrow \{0, 1\}^k$  denote the inverse of this function \cite{BHR12}.

Decryption of \texttt{DKC} may also be denoted by the inverse function notation $E^{-1}_{A,B}: \{0, 1\}^k \rightarrow \{0, 1\}^k$ instead of $D_{A,B} : \{0, 1\}^k \rightarrow \{0, 1\}^k$.

Throughout this thesis an encryption with two keys mean a \texttt{DKC} unless it is stated otherwise.

So far, a variety of \texttt{DKC} schemes have been proposed. Among them, an earlier one is Equation~\eqref{eq:encscheme2hashes} proposed by Naor \textit{et al.} in \citep{NPS99}. For every encryption, \texttt{PRF} is called twice. \texttt{PRF} may be implemented as a keyed hash.

\begin{equation} 
\label{eq:encscheme2hashes}
 E_{A,B}(C)\rightarrow \texttt{PRF}(A, \texttt{gateID})\oplus \texttt{PRF}(B, \texttt{gateID}) \oplus C
\end{equation}

Lindell \textit{et al.} proposed a more efficient \texttt{DKC} scheme Equation~\eqref{eq:encscheme1hash} in \cite{LPS08}. It requires one hash per encryption, which reduces the computational cost significantly. 

\begin{equation} 
\label{eq:encscheme1hash}
E_{A,B}(C)\rightarrow H(A||B||\texttt{gateID})\oplus C
\end{equation}

Kreuter \textit{et al.} proposed the \texttt{DKC} scheme Equation~\eqref{eq:encschemeaes} in \citep{KsS12}. An $\texttt{AES}256$ encryption is used instead of a hash function. Kreuter \textit{et al.} shows that this improvement reduces the computational cost around 25\%.

\begin{equation} 
\label{eq:encschemeaes}
E_{A,B}(C)\rightarrow\texttt{AES}256(A||B||\texttt{gateID})\oplus C
\end{equation}

Bellare \textit{et al.} proposed the \textit{state-of-the-art} \texttt{DKC} scheme Equation~\eqref{eq:encschemeaesconstkey}\footnote{$K=2A\oplus4B\oplus gateID$} in \citep{BHKR13} which eliminates the need for key precessing in each AES encryption by using a constant key $k_c$ for all of them.

\begin{equation} 
\label{eq:encschemeaesconstkey}
 E_{A,B}(C)\rightarrow\texttt{AES}128_{k_c}(K)\oplus K \oplus C 
\end{equation}

\section{Secret Sharing} \label{sec:secretshare}{}

\textit{Secret sharing} refers to the methods where a secret value is distributed amongst a group of parties, each having a share from the secret \cite{Sha79}. To reconstruct the secret, parties need to combine a sufficient number of shares together; since individual share of a party is useless on its own. There have been various secret sharing schemes proposed so far. Here we will introduce only some of them which will be helpful throughout this thesis.

\subsection{\texttt{XOR} Sharing} \label{sub:xorshare}{}

\texttt{XOR} \textit{sharing} (also known as \textit{boolean sharing}) is a secret sharing type where for an $\ell$-bit value $x$ shared by $m$ parties, the share of a party $i$ is an $\ell$-bit value $x_i$, and when the shares of all $m$ parties \texttt{XOR}ed  bitwise together the result is $x$, \textit{i.e.},  $x=x_1\oplus\ldots\oplus x_m$ \cite{DSZ15}. There is no number limit for parties in \texttt{XOR} sharing. However, if any of the parties keeps his share, the rest of the parties cannot even get close to learning the shared value.

\subsection{Arithmetic Sharing} \label{sub:arithshare}{}

\textit{Arithmetic sharing} is similar to \texttt{XOR} \textit{sharing} in that there is no number limit for parties and that if any of the parties keeps his share, the rest of the parties cannot even get close to learning the shared value \cite{DSZ15}. It is a secret sharing type where for an $\ell$-bit value $x$ shared by $m$ parties, the share of a party $i$ is an $\ell$-bit value $x_i$, and when the shares of all $m$ parties added together in a modulus $n$ which conforms $2\leq n\leq 2^\ell$ the result is $x$, \textit{i.e.},  $x=x_1+\ldots+ x_m \texttt{ mod } n$.

\subsection{Yao Sharing} \label{sub:yaoshare}{}

\textit{Yao sharing} is a secret sharing type where 1 bit is shared by 2 parties \cite{DSZ15}. In order to share a bit $b$, the first party $P_1$ picks 2 random $\ell$-bit strings $B_0$ and $B_1$. The second party $P_2$, without knowing $b$, keeps only $B_b$. $P_1$ does not know which of the 2 strings kept by $P_2$, and $P_2$ does not know the other string picked by $P_1$. Only together, they can evaluate $b$. Although keeping costly strings for a bit does not look very efficient at first, Yao sharing has certain advantages for 2PC which will be obvious when we describe Yao's protocol in \ref{Chapter3}.

\subsection{Shamir's Secret Sharing} \label{sub:shamirshare}{}

\textit{Shamir's secret sharing} is an effective secret sharing scheme proposed by Adi Shamir \cite{Sha79, NS10} where a group of $n$ users share a secret data $D$. The scheme permits any predefined $(k + 1)\leq n$ or more users to reconstruct the secret. However, no information about $D$ can be recovered by $k$ or less users. This scheme can also be referred to as $(k + 1,n)$-threshold secret sharing scheme, where $(k + 1)$ is the threshold and $n$ is the number of users sharing the secret.

All users have a different point in two-dimensional plane, $(x_1,y_1),\ldots,(x_n,y_n)$. All of the points must be chosen such that they are on a $k$-degree polynomial. Therefore, any $k + 1$ of these shares  suffices for Lagrange's interpolation. The secret value is the evaluation of the polynomial on axis $x=0$.

\section{Oblivious Transfer} \label{sec:oblivioustrans}{}

An \textit{1-out-of-$m$ oblivious transfer} (1-out-of-$m$ OT) protocol is a two-party asymmetric\footnote{An asymmetric protocol means that parties play different roles during the protocol.} protocol where one of the parties is the sender, and the other one is the receiver \cite{Pil15}. The sender has the set of values $\{x_1,\ldots,x_m\}$ and the receiver has an index $i$. At the end of the protocol, the receiver should only learn one of the sender's inputs, which is $x_i$; whereas the sender should not learn anything about the index $i$. An efficient 1-out-of-$m$ OT technique can be found in \cite{CO15}.

The high computational complexity of OT is a major source of inefficiency. In order to reduce this cost, some optimizations (\textit{e.g.}, extended OT \cite{YKNP03}) have been proposed.

There also exist OT protocols for settings with more parties, known as \textit{multi-party oblivious transfer}. A multi-party OT is a protocol where one of the parties holds the values $x_1, \ldots, x_m$, but multiple parties secret share the choice index $i$. At the end of the protocol, the parties learn shares of $x_i$ instead of learning it as a whole. The party holding the initial values is called the sender, whereas the other ones are called the receivers.

\section{Homomorphic Encryption} \label{sec:homoenc}{} 

\textit{Homomorphic Encryption} (HE) schemes are used for secure evaluation of arithmetic circuits since they permit computation of multiplication and addition on ciphertexts \cite{Sch11}. An \textit{additively} HE scheme allows only unlimited addition on encrypted data; whereas a \textit{multiplicative} HE scheme allows only unlimited multiplication on it. An encryption scheme having both multiplicatively and additively HE property is called \textit{fully homomorphic encryption} (FHE).

There was a wide-spread belief that FHE does not exist until recently. Gentry has been the inventor of the first FHE scheme \cite{Gen09}. Unfortunately, huge sizes and computational costs of current FHE schemes make them too inefficient to be used in practical applications no matter how much effort has been given for improving their performances. The problem is that a FHE scheme must allow algebraic operations while providing strong security assumptions, which makes the costs grow substantially.

\section{Goldreich-Micali-Wigderson (GMW) Protocol for MPC} \label{sec:gmwprotocol}{} 

One of the commonly used MPC schemes is \textit{Goldreich-Micali-Wigderson} (GMW) protocol that uses \texttt{XOR} sharing (\ref{sec:secretshare}), and is proposed in \cite{GMW87}. It proposes MPC of boolean circuits with gates \texttt{AND} and \texttt{XOR} against semi-honest adversaries (\ref{sub:semi-honest}). 

\texttt{XOR} gates can be computed locally and are communication free \cite{Pil15}. To illustrate, to compute $c=a \oplus b$, each party $i$ only needs to use its shares $c_i = a_i \oplus b_i$ in order to receive his output share $c_i$. However, to compute an \texttt{AND} gate, parties are required to communicate for 1-out-of-4 OT (\ref{sec:oblivioustrans}). In the case of $2$ parties, to compute their output shares of $a \wedge b$, $P_1$ constructs the evaluation table for both input shares of $P_2$ and they engage in a 1-out-of-4 OT (\ref{sec:oblivioustrans}) where $P_2$'s inputs are used as the choice index. To extend the protocol for $m$ parties, ${m\choose 2}$  runs of the OT protocol is required. One can also see it as one run of a multi-party 1-out-of-4 OT protocol where the choice indices are $a$ and $b$ \cite{Pil15}.


\chapter{Yao's Garbled Circuit Protocol} 

\label{Chapter3} 

\lhead{Chapter 3. \emph{Yao's Garbled Circuit Protocol}} 

\begin{figure}[b]
\includegraphics[height=5cm, angle=0]{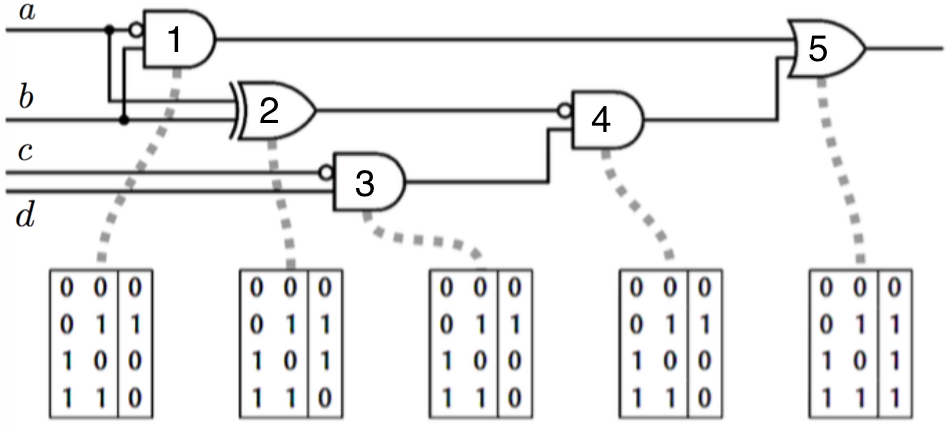}
\caption{ A boolean circuit of a function $f$ with the truth table of the gates included.}
\label{fig:abooleancircuit}
\end{figure}
\begin{figure}
\includegraphics[height=12cm, angle=-90]{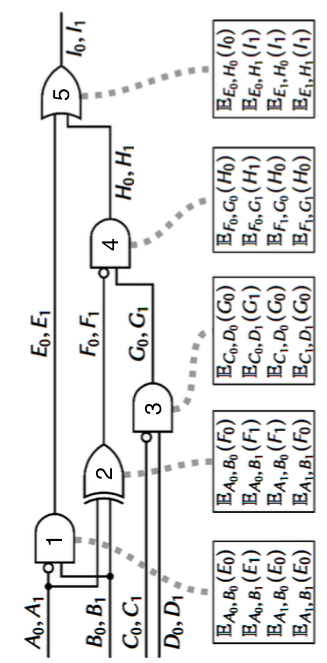}
\caption{Garbling the circuit in Figure \ref{fig:abooleancircuit}}
\label{fig:anencryptedcircuit}
\end{figure}
\begin{figure}[b]
\includegraphics[height=10cm, angle=-90]{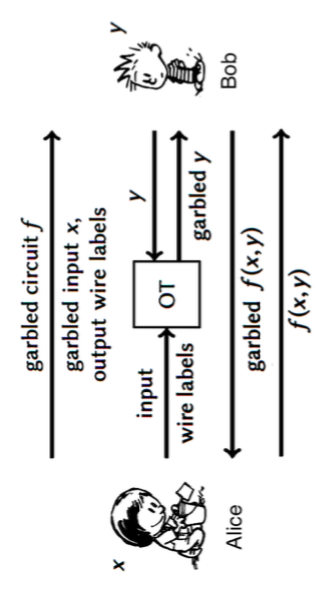}
\caption{ Communication flow in the semi-honest Yao's protocol.}
\label{fig:yaosgcpflow}
\end{figure}

Even though Yao's protocol has more than two-party applications, its use will be held limited to 2PC. It is an asymmetric protocol, which means that parties play different roles while the protocol is running. One of the parties has the role of the \textit{garbler}, whereas the other one becomes the \textit{evaluator}. The protocol is intended to be secure in the semi-honest model (\ref{sub:semi-honest}). It runs on boolean functions, so first a function must be converted to a boolean circuit. Figures \ref{fig:abooleancircuit}, \ref{fig:anencryptedcircuit} and \ref{fig:yaosgcpflow} have been taken from Mike Rosulek's presentation in Simons Institute, University of California, Berkeley, namely \textit{A Brief History of Practical Garbled Circuit Optimizations}.

\textbf{A gentle introduction}. Yao's garbled circuit protocol is briefly as follows (later we propose it in a more formal model):

Assume Alice and Bob are trying to compute a function $f$ whose boolean circuit is given in Figure \ref{fig:abooleancircuit}. Throughout this thesis, Alice will be the {garbler}, Bob will be the {evaluator}. Alice's input is $x$ including bits $a$ and $c$, and Bob's input is $y$ including bits $b$ and $d$.

\texttt{Garbling:}
\begin{enumerate}
\setcounter{enumi}{0}
\item{Alice picks random and computationally indistinguishable masking values for possible truth values \texttt{FALSE} and \texttt{TRUE} of each wire.}
\item{She encrypts the output masking values of each gate using their corresponding input masking values as the \texttt{DKC} key (\ref{part:dkcschemes}). This way she gets four ciphertexts for each gate in the circuit as in Figure \ref{fig:anencryptedcircuit}.}
\end{enumerate}

\texttt{Input Transfer:}
\begin{enumerate}
\setcounter{enumi}{2}
\item{She sends all ciphertexts for each gate, as well as her masked input values for $a$ and $c$ to Bob. He takes his own masked input values from Alice using 1-out-of-2 OT (\ref{sec:oblivioustrans}).}
\end{enumerate}

\texttt{Evaluating:}
\begin{enumerate}
\setcounter{enumi}{3}
\item{Bob decrypts the related ciphertext (we will come to this later) gate-by-gate in \textit{topological order}, reaching the output masking values of the circuit. Topological order means from the inputs to the output. The rule is that if the output of a gate $g_1$ is input to another gate $g_2$, $g_1$ must be evaluated before $g_2$. In this case the gate order might be chosen as 1, 2, 3, 4, 5.}
\end{enumerate}

\texttt{Output Reveal:}
\begin{enumerate}
\setcounter{enumi}{4}
\item{Bob tells Alice the output masking values, and Alice sends the output of the function $f(x,y)$ to Bob.}
\end{enumerate}

The flow of communication between the garbler and the evaluator is summed up in Figure \ref{fig:yaosgcpflow}.

\section{Formal Definiton of Yao's Protocol} \label{sec:FormalDefinitonandSecurityofYao'sProtocol}{}

The Yao's protocol scheme proposed by Bellare \textit{et al.} in \citep{BHR12} brought a significant jump by defining the procedures involved in a secure Yao's protocol. A conventional circuit can be defined as $f = (n, m, q, A, B, G)$ where the numbers of its inputs, its outputs, and its gates are $n \geq 2$, $m \geq 1$, and $q \geq 1$, respectively. The number of its wires is denoted as $r = n + q$. The sets of the circuit \texttt{Inputs}, \texttt{Wires}, \texttt{OutputWires} and \texttt{Gates} are defined as $\texttt{Inputs}=\{1,\ldots,n\}$, $\texttt{Wires}=\{1,\ldots,n+q\}$, $\texttt{OutputWires}= \{n+q-m+1,\ldots,n+q\}$, and $\texttt{Gates}= \{n + 1,\ldots, n + q\}$. Then the function identifying each gate's first incoming wire is $A : \texttt{Gates} \rightarrow \texttt{Wires}\backslash \texttt{OutputWires}$. The function identifying each gate's second incoming wire is $B : \texttt{Gates} \rightarrow \texttt{Wires}\backslash \texttt{OutputWires}$. The function determining the functionality of each gate is $G : \texttt{Gates} \times  \{0, 1\}^2 \rightarrow \{0, 1\}$. The requirement is that $A(g) < B(g) < g$ for all $g \in \texttt{Gates}$.


Bellare \textit{et al.} defines the generic garbling scheme  consisting of \texttt{\textsc{Gb}}, \texttt{\textsc{En}}, \texttt{\textsc{Ev}}, and \texttt{\textsc{De}} algorithms which are described as folows (see also Figure \ref{fig:gcpprocflow} and Algorithm \ref{alg:yaosgcpproc}) \citep{BHR12}:

\begin{figure}[t]
\includegraphics[height=2.8cm, angle=0]{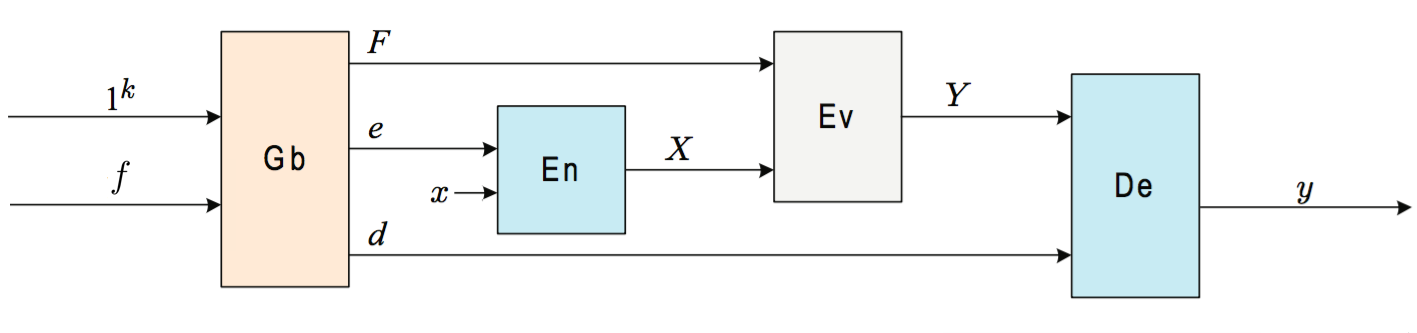}
\caption{The flow of procedures in Yao's protocol in \citep{BHR12}.}
\label{fig:gcpprocflow}
\end{figure}

\begin{enumerate}
\item{\textbf{Garble} (\texttt{\textsc{Gb}}): \texttt{\textsc{Gb}} procedure takes $1^k$ and a boolean circuit $f$ as input, and outputs $(F,e,d)$, where $F$ is a garbled circuit, $e$ is the encoding information, and $d$ is the decoding information. The \textbf{for}-loop on Line \ref{lst:assigningmaskingvalue} of Algorithm \ref{alg:yaosgcpproc} assigns masking values for every wire in the circuit for both \texttt{TRUE} and \texttt{FALSE}. It also assures that the last bits of the assigned masking values for a wire, which we call label bits, differ from each other. The \textbf{for}-loop on Line \ref{lst:encryptingoutputs} of Algorithm \ref{alg:yaosgcpproc}  encrypts the possible output masked with their corresponding input masking values for each gate. It also orders the ciphertexts with respect to the label bits (\texttt{lsb}) of input masking values so that the order does not leak information (we will call this technique point and permute).}
\item{\textbf{Encode} (\texttt{\textsc{En}}): \texttt{\textsc{En}} procedure takes $(e,x)$ as input, where $e$ is as we mentioned above and $x$ is a suitable input for $f$, and outputs a garbled input $X$. In this scheme, encoding is directly assigning the pre-known masking values for the inputs.}
\item{\textbf{Evaluate} (\texttt{\textsc{Ev}}): \texttt{\textsc{Ev}} procedure takes $(F, X)$ as input, and outputs a garbled output $Y$. The \textbf{for}-loop on Line \ref{lst:evaluategates} of Algorithm \ref{alg:yaosgcpproc} decrypts only one ciphertext related to a gate with its input masking values and with respect to their label bits.}
\item{\textbf{Decode} (\texttt{\textsc{De}}): \texttt{\textsc{De}} procedure takes $(d, Y )$ as input, and outputs a plain output $y$. In this scheme, decoding is directly assigning the pre-known outputs for the masking values obtained by the \texttt{\textsc{Ev}} procedure.}
\end{enumerate}

\begin{algorithm}
	\scriptsize
	\caption{Garbled circuit scheme \citep{BHR12}.}
	\label{alg:yaosgcpproc}
	\begin{algorithmic}[1] 
	\Procedure{\texttt{Gb}}{$1^k,f$} \Comment \texttt{Garbling phase}
		\State $(n,m,q,A',B',G)\gets f$
		\For{$i\in\{1,\ldots,n+q\}$}  \label{lst:assigningmaskingvalue}
			\State $t\gets \{0,1\}$, $X_i^0\gets \{0,1\}^{k-1}t$, $X_i^1\gets \{0,1\}^{k-1}\overline{t}$
		\EndFor
		\For{$(g,i,j)\in\{n+1,\ldots,n+q\}\times \{0,1\}\times \{0,1\}$} \label{lst:encryptingoutputs}
                		\State $a \gets A'(g)$, $b \gets B'(g)$,  $A \gets X_a^i$, $\text{a} \gets \texttt{lsb}(A)$, $B \gets X_b^i$, $\texttt{b} \gets \texttt{lsb}(B)$
			\State $T  \gets g\parallel\texttt{a}\parallel\text{b}$, $P[g,\texttt{a},\texttt{b}] \gets E_{A,B}(X_g^{G_g(i,j)})$
           	\EndFor
		\State $F\gets(n,m,q,A',B',P)$
		\State $e\gets(X_1^0,X_1^1,\ldots,X_n^0,X_n^1)$
		\State $d\gets(X_{n+q-m+1}^0,X_{n+q-m+1}^1,\ldots,X_{n+q}^0,X_{n+q}^1)$
            	\State \textbf{return} $(F,e,d)$ 
        \EndProcedure
        \Statex
        \Procedure{\texttt{En}}{$e,x$} \Comment \texttt{Encoding phase}
		\State $(X_1^0,X_1^1,\ldots,X_n^0,X_n^1)\gets e$
		\State $x_1\ldots x_n\gets x$, $X\gets (X_1^{x_1},\ldots,X_n^{x_n})$
		\State \textbf{return} $X$ 
        \EndProcedure
        \Statex
        \Procedure{\texttt{Ev}}{$F,X$} \Comment \texttt{Evaluating phase}
		\State $(n,m,q,A',B',P)\gets F$, $(X_1,\ldots,X_n)\gets X$
		\For{$g\gets n+1 \textbf{ to } n+q$}   \label{lst:evaluategates}
                		\State $a \gets A'(g)$, $b \gets B'(g)$,  $A \gets X_a^i$, $\text{a} \gets \texttt{lsb}(A)$, $B \gets X_b^i$, $\text{b} \gets \texttt{lsb}(B)$
			\State $T  \gets g\parallel\texttt{a}\parallel\texttt{b}$, $X_g \gets D_{A,B}(P[g,\texttt{a},\texttt{b}])$
           	\EndFor
		\State \textbf{return} $(X_{n+q-m+1},\ldots,X_{n+q})$ 
        \EndProcedure
         \Statex
        \Procedure{\texttt{De}}{$d,Y$} \Comment \texttt{Decoding phase}
		\State $(Y_1,\ldots,Y_m)\gets Y$, $(Y_1^0,Y_1^1,\ldots,Y_m^0,Y_m^1)\gets d$
		\For{$i\in\{1,\ldots,m\}$}
                		\If{$Y_i=Y_i^0$} $y_i\gets 0$
			\ElsIf{$Y_i=Y_i^1$} $y_i\gets 1$
			\Else \textbf{ return} $\bot$
			\EndIf
           	\EndFor
		\State \textbf{return} $y\gets y_1\ldots y_m$
        \EndProcedure
    \end{algorithmic}
\end{algorithm}

\textit{Correctness} property is that  Equation~\eqref{eq:correctnessofGCP} holds for all possible input $x$ where $(F,e,d)\gets\texttt{\textsc{Gb}} (1^k,f)$.

\begin{equation} 
\label{eq:correctnessofGCP}
\texttt{\textsc{De}}(d, \texttt{\textsc{Ev}}(F, \texttt{\textsc{En}}(e, x))) = f (x)
\end{equation}

\section{Security Properties of Yao's Protocol}\label{sec:yaosec}{}
We need some parameters in order to appreciate the security of a garbling scheme. The security parameters defined by Bellare \textit{et al.} are \textit{privacy}, \textit{obliviousness}, and \textit{authenticity} \citep{BHR12}.

\subsection{Privacy}\label{sub:Privacy}{}

\textit{Privacy} is achieved by a garbling scheme if no more information about the input $x$ must be revealed by the collection $(F,X,d)$ than that is revealed by $f (x)$ \citep{ZRE15, BHR12}.  Let $(f, x)$ be chosen by the adversary. Then either the circuit is garbled  to $(F,e,d)\gets\texttt{\textsc{Gb}}(1^k,f)$, the input is encoded as $X\gets\texttt{\textsc{En}}(e,x)$, the adversary getting $(F,X,d)$; or the simulator $S$ devises a \textit{fake} $(\bar{F}, \bar{X}, \bar{d})$ depending solely on the security parameter $k$, the side information\footnote{Side-information means any information about the circuit which the protocol does not intend to hide, like its size or its topology. $\Phi(f)$ is the side-information function which maps $f$ to $\phi$. \label{ftnt:si}} $\phi = \Phi(f )$, and the output $y = \texttt{\textsc{Ev}}(f, x)$. The $(\bar{F}, \bar{X}, \bar{d})$ produced by the simulator must be indistinguishable from the ones coming from the actual garbling scheme.

 \subsection{Obliviousness} \label{sub:Obliviousness}{}

\textit{Obliviousness} is achieved by a garbling scheme if $(F, X)$ reveals nothing more than the side information\textsuperscript{\ref{ftnt:si}} $\Phi(f )$ about $f$ or $x$ \citep{ZRE15, BHR12}. To compare obliviousness with privacy (\ref{sub:Privacy}), where the output is learned by the evaluator, here, he does not learn that since $d$ is kept hidden. The output can be revealed by a private scheme even without $d$, while $x$  can be revealed by an oblivious scheme once $d$ is exposed. Let $(f, x)$ be chosen by the adversary. Either the circuit is garbled to $(F,e,d)\gets\texttt{\textsc{Gb}}(1^k,f)$, the input is encoded as $X\gets\texttt{\textsc{En}}(e,x)$, and the adversary getting $(F,X)$; or the simulator $S$ to devises a \textit{fake} $(\bar{F}, \bar{X})$ depending solely on $k$, and $\phi = \Phi(f )$. The $(\bar{F}, \bar{X})$ produced by the simulator must be indistinguishable from from the ones coming from the actual garbling scheme.

\subsection{Authenticity} \label{sub:Authenticity}{} 

\textit{Authenticity} is achieved by a garbling scheme if from $(F, X)$, an adversary cannot construct a garbled output $\bar{Y}$ which is not authentic, \textit{i.e.} $\texttt{\textsc{De}}(d, \bar{Y} )\neq \bot$ only if $\bar{Y}= \texttt{\textsc{Ev}}(F, X)$, except for negligible probability \citep{ZRE15, BHR12}.


\chapter{Garbled Circuit Optimizations} 

\label{Chapter4} 

\lhead{Chapter 4. \emph{Garbled Circuit Optimizations}} 

Since we have introduced the generic garbled circuit framework, it is time to present the optimizations on it in detail. We start with describing the parameters of a garbled circuit scheme that can be optimized and their relevant importance. We then continue with optimization techniques, along with comparing them with each other and presenting the relations between them. At the end, we have included a useful table to show the compatibility of various garbling techniques. Figures \ref{fig:pointandpermute}, \ref{fig:GRR3}, \ref{fig:freexor}, \ref{fig:freexorand}, \ref{fig:grr2} and \ref{fig:flexor} have been taken from Mike Rosulek's presentation in Simons Institute, University of California, Berkeley, namely \textit{A Brief History of Practical Garbled Circuit Optimizations}.

Mainly, there are three parameters related to Yao's protocol that can be optimized: the \textit{size} of the garbled circuit which limits the communication complexity cost, the \textit{computation time} required both for encryption and decryption, and the \textit{security} of the protocol \cite{KMR14}. The size of the garbled circuit is important because it usually needs to be transmitted to the evaluator over a limited channel. Clearly, the computation time required is also an important parameter for both parties.

\section{General Focus}

\subsection{The Size Parameter} \label{sec:size}{}

The size of the garbled circuit  is  usually the primary parameter due to the limits of the communication channel. The most effort in the garbled circuit research has been dedicated to make it smaller. Reducing it even in the expense of worse computation times or weaker hardness assumptions is often preferable \cite{KMR14}.

A reduction in the size of a garbled circuit generally comes from a decrease in the number of ciphertexts needed per gate. Circuits can grow to contain billions of gates, meaning each garbled circuit can be gigabytes in size. Our primary goal in this chapter is to cover garbled gate size optimization techniques.

\subsection{The Computation Time Parameter} \label{sec:comptime}{}

Computation time is related to time consumptions of \texttt{\textsc{Gb}} and \texttt{\textsc{Ev}} procedures. Naturally the research aims to make them shorter. The computation time may be even more important when the \texttt{CPU} resource of a party is restricted, such as a mobile device. The improvements in \texttt{DKC} schemes (\ref{part:dkcschemes}) schemes proposed are also for this parameter. The gate garbling techniques may also improved for this parameter as well \cite{KMR14}.

\subsection{Security Parameter} \label{sec:hardassump}{}

A garbling scheme must conform the security properties (\ref{sec:yaosec}) although in some cases authenticity parameter may be omitted. If the hardness assumptions of the building blocks of a scheme (\textit{e.g.}, \texttt{DKC} scheme (\ref{part:dkcschemes}), gate garbling technique) is stronger, the protocol will also be more secure \cite{KMR14}.

The rest of this chapter is especially dedicated to the techniques related to the optimizations in the size parameter. However, the techniques will also be compared for the other parameters whenever it is necessary. After the description of each technique, there will be a size and computation time scoreboard for comparing that technique with the previous ones (see Tables \ref{tab:scrbrdpp}, \ref{tab:scrbrdgrr3}, \ref{tab:scrbrdfrxor}, \ref{tab:scrbrdfrgrr2}, \ref{tab:scrbrdfrflxor}, \ref{tab:scrbrdhg}). The time for encryptions and decryptions for both \texttt{DKC} schemes and symmetric schemes assumed to be the same and denoted as \texttt{edt} (for encryption/decryption time). ct stands for ciphertexts.

\begin{figure}
\includegraphics[height=5cm, angle=0]{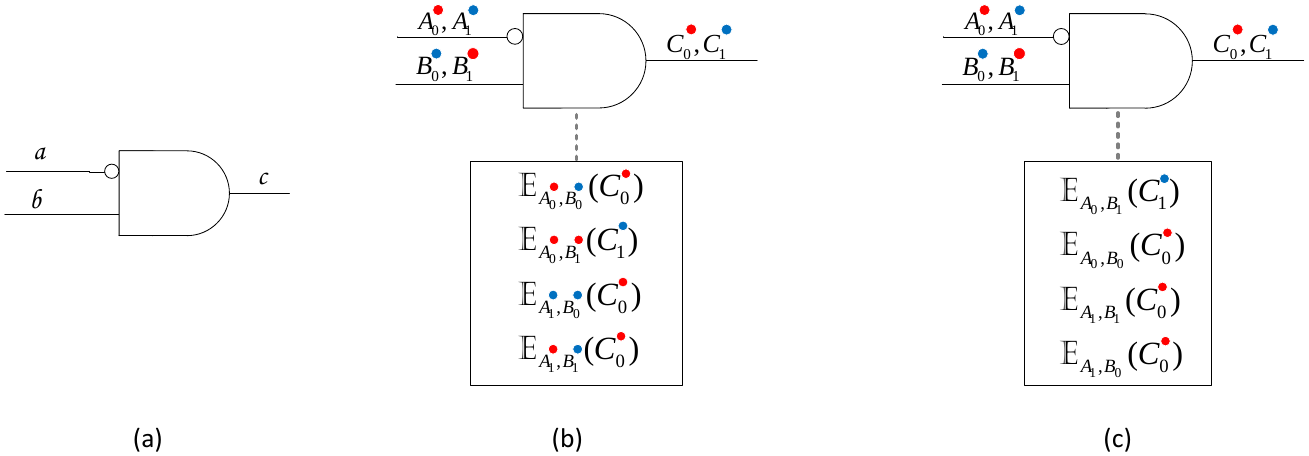}
\caption{ (a) The gate to be evaluated. (b) Label assignment. (c) Rearrangement of ciphertexts canonically with respect to input labels.}
\label{fig:pointandpermute}
\end{figure}

\section{Point and Permute} \label{sec:PP}{}

The evaluator needs to know which one of the ciphertexts for a gate must be decrypted during the evaluation process. However, he cannot be allowed to deduce the truth value of any of inputs or outputs. The oldest and yet secure method achieving is \textit{point and permute} (P\&P), and suggested by Beaver \textit{et al.} in \cite{BMR90}.

\texttt{Garbling:}
\begin{enumerate}
\setcounter{enumi}{0}
\item{Alice and Bob want to compute the output of the gate in Figure \ref{fig:pointandpermute} (a) where $a$ and $b$ is the input $c$ is the output.}
\item{Alice chooses masking values of wires such that each masking value has one of the two possible labels (the one for $a$ is either $A_0$ or $A_1$, the one for $b$ is either $B_0$ or $B_1$, and the one for $c$ is either $C_0$ or $C_1$), and for a given wire both masking values have different labels (see Figure \ref{fig:pointandpermute} (b)). The label needs to be something that can be directly detectable from the masking value (\textit{e.g.}, its last bit). For example, if the masking value on the wire $a$ corresponding to the truth value \texttt{FALSE} ($A_0$) has 0 on the last bit, then the masking value for the truth value \texttt{TRUE} ($A_1$) must have 1 on the last bit.  The truth value cannot be detected from the label of the masking value. Alice encrypts the possible output masking values of the gate with the corresponding input masking values ($E_{A_0,B_0}(C_0)$, $E_{A_0,B_1}(C_1)$, $E_{A_1,B_0}(C_0)$, and $E_{A_1,B_1}(C_0)$).} 
\item{Alice rearrange the ciphertexts with respect to the input labels, as in Figure \ref{fig:pointandpermute} (c). During the evaluation, Bob will know which ciphertext he must decrypt from the labels of the inputs. This way, ciphertexts are ordered unrelated to the truth values of wires and any information leakage is prevented.}
\end{enumerate}

The number of ciphertexts per gate that needs to be transmitted is 4 in this method. 4 encryption and 1 decryption are the computational cost for each gate (see Table \ref{tab:scrbrdpp}). 

\begin{table}[b]
\centering
\caption{Optimization Scoreboard (P\&P)}
\label{tab:scrbrdpp}
\begin{tabular}{l|l|l|l}
\small{\textbf{Method}} & \small{\textbf{Odd / Even Gate Size}} & \small{\textbf{Enc. Time per}} & \small{\textbf{Dec. Time per}} \\ 
\small{ } & \small{ } & \small{\textbf{Odd / Even Gate}} & \small{\textbf{Odd / Even Gate}} \\ \hline
\small{P\&P}   & \small{4 ct / 4 ct}                      & \small{4 \texttt{edt} / 4 \texttt{edt}}             & \small{1 \texttt{edt} / 1 \texttt{edt}}           \\
\multicolumn{4}{c}{ }\\
\multicolumn{4}{c}{{\scriptsize ct: ciphertexts; \texttt{edt}: total encryption and/or decryption time}}\\
\end{tabular}
\end{table}

\begin{figure}
\includegraphics[height=6.5cm, angle=0]{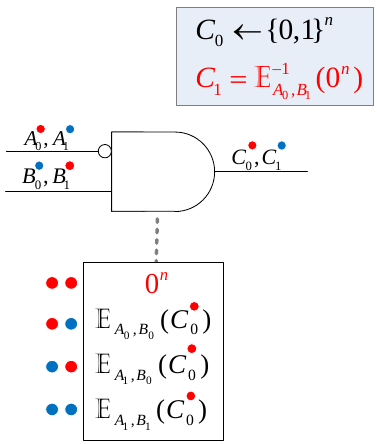}
\caption{Garbled row reduction 3 ciphertexts.}
\label{fig:GRR3}
\end{figure}

\section{Garbled Row Reduction 3  Ciphertexts} \label{sec:GRR3}{}

Instead of choosing the masking values of the output of a gate randomly as in P\&P (\ref{sec:PP}), in \cite{NPS99} Naor \textit{et al.} suggested a smarter way, called \textit{garbled row reduction 3 ciphertexts} (GRR3).

\texttt{Garbling:}
\begin{enumerate}
\setcounter{enumi}{0}
\item{Alice and Bob want to compute the output of the gate in Figure \ref{fig:GRR3}.}
\item{Alice choose the masking value of the first output in label order such that all bits of the resulting ciphertext is 0 (\textit{i.e.}, by decrypting  all 0, $C_1\gets E_{A_0,B_1}^{-1}(0^n)$). The masking value reached will still be pseudo-random.}
\item{Since there is no need to send the first ciphertext, sending 3 ciphertexts per gate suffices.}
\end{enumerate}

Although GRR3 results in smaller-sized garbled circuits than the ones resulted from P\&P (\ref{sec:PP}), it has little affect on the computation cost since the gain coming from one less encryptions goes to the decryption of the first ciphertext (see Table \ref{tab:scrbrdgrr3}).

\begin{table}[b]
\centering
\caption{Optimization Scoreboard (GRR3)}
\label{tab:scrbrdgrr3}
\begin{tabular}{l|l|l|l}
\small{\textbf{Method}} & \small{\textbf{Odd / Even Gate Size}} & \small{\textbf{Enc. Time per}} & \small{\textbf{Dec. Time per}} \\ 
\small{ } & \small{ } & \small{\textbf{Odd / Even Gate}} & \small{\textbf{Odd / Even Gate}} \\ \hline
\small{P\&P (\ref{sec:PP})}   & \small{4 ct / 4 ct}                      & \small{4 \texttt{edt} / 4 \texttt{edt}}             & \small{1 \texttt{edt} / 1 \texttt{edt}}           \\ \hline
\small{GRR3}   & \small{3 ct / 3 ct}                      & \small{4 \texttt{edt} / 4 \texttt{edt}}             & \small{1 \texttt{edt} / 1 \texttt{edt}}  \\
\multicolumn{4}{c}{ }\\
\multicolumn{4}{c}{{\scriptsize ct: ciphertexts; \texttt{edt}: total encryption and/or decryption time}}\\
\end{tabular}
\end{table}

\section{Free \texttt{XOR}} \label{sec:freeXOR}{}

One of the greatest jumps in the garbled circuit technology has been the free \texttt{XOR} technique, which is proposed by Kolesnikov and Schneider in \cite{KS08}. It basically eliminates the need for any ciphertext transmission and any calculation for \texttt{XOR} gates. The function can be compiled such that the number other gates are minimized. Usually they are just \texttt{AND} gates, since $(\texttt{XOR},\texttt{AND})$ is Turing complete.\footnote{The number of \texttt{AND} gates in the Boolean functions is called \textit{multiplicative complexity}. Reducing it at the expense of increasing \texttt{XOR}s is already an active research topic \cite{KK10}.}

\begin{figure}
\includegraphics[height=5.5cm, angle=0]{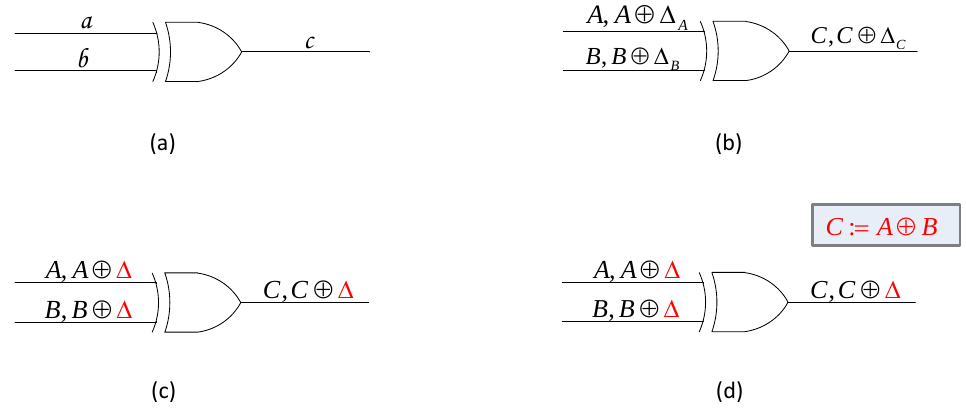}
\caption{(a) \texttt{XOR} gate with masked values on its wires. (b) \texttt{XOR} gate whose masked values interpreted with offsets. (c) \texttt{XOR} with the same offset in the masked values on wires. (d) \texttt{XOR} gate arranged for free \texttt{XOR} technique}
\label{fig:freexor}
\end{figure}

\texttt{Garbling:}
\begin{enumerate}
\setcounter{enumi}{0}
\item{Alice and Bob want to compute the output of the \texttt{XOR} gate in Figure \ref{fig:freexor} (a).}
\item{The masking value for \texttt{TRUE} in a wire $a$ can be written as the one for \texttt{FALSE} in that wire $A$ \texttt{XOR}ed with some offset $\triangle_A$, which is a random value having the same number of bits as $A$ and $B$, as in Figure \ref{fig:freexor} (b). The masking value for \texttt{FALSE} becomes $A$, and the masking value for \texttt{TRUE} becomes $A\oplus\triangle_A$. Alice also writes the masking values of $b$ and $c$ the same way.} 
\item{Alice sets the offsets of all wires be the same secret value $\triangle$ as in Figure \ref{fig:freexor} (c). Even if there are more than one gate in a circuit, all wires must be given the same offset so that the free \texttt{XOR} method can be applied. Offset must be kept as a secret by the garbler.}
\item{Alice choose the masking value for \texttt{FALSE} in the output, \texttt{XOR} of those for \texttt{FALSE} in the inputs as in Figure \ref{fig:freexor} (d). This makes transmitting any ciphertext for an \texttt{XOR} gate unnecessary.}
\end{enumerate}
 
 \texttt{Evaluating:}
\begin{enumerate}
\setcounter{enumi}{4}
\item{Bob just \texttt{XOR}s the masking value of the inputs to calculate the masking value of the output. }
\end{enumerate}

\begin{figure}
\includegraphics[height=4.5cm, angle=0]{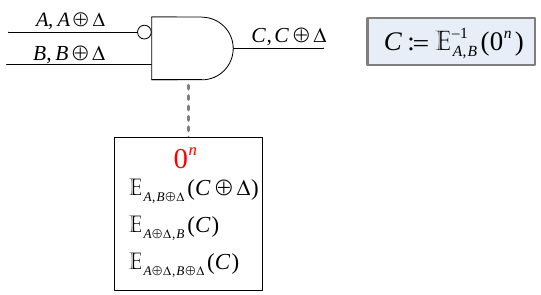}
\caption{Encryptions of a gate other than \texttt{XOR}  in the free \texttt{XOR} technique.}
\label{fig:freexorand}
\end{figure}

\texttt{AND} gates can be encrypted as in GRR3 (\ref{sec:GRR3}), and 3 cipher texts needs to be transmitted (see Figure \ref{fig:freexorand}). Labels still exist, and ciphertexts must be ordered accordingly. The offset must be chosen such that for a given wire both masking values have different labels (\textit{e.g.}, its \texttt{lsb} must be 1 if the label is the last bit). Since the same offset is used in both inputs and the payload, there is a need for a circularity assumption for the encryption scheme used \cite{CKK12}.

Free \texttt{XOR} technique, makes \texttt{XOR}s completely free for transmission and computation in both the garbler's side and the evaluator's side. This has a huge impact, not just for freeing \texttt{XOR}s but also permitting the minimization of the other gates at the expense of increasing \texttt{XOR}s (see Table \ref{tab:scrbrdfrxor}).

\begin{table}[b]
\centering
\caption{Optimization Scoreboard (Free \texttt{XOR})}
\label{tab:scrbrdfrxor}
\begin{tabular}{l|l|l|l}
\small{\textbf{Method}} & \small{\textbf{Odd / Even Gate Size}} & \small{\textbf{Enc. Time per}} & \small{\textbf{Dec. Time per}} \\ 
\small{ } & \small{ } & \small{\textbf{Odd / Even Gate}} & \small{\textbf{Odd / Even Gate}} \\ \hline
\small{P\&P (\ref{sec:PP})}   & \small{4 ct / 4 ct}                      & \small{4 \texttt{edt} / 4 \texttt{edt}}             & \small{1 \texttt{edt} / 1 \texttt{edt}}           \\ \hline
\small{GRR3 (\ref{sec:GRR3}{})}   & \small{3 ct / 3 ct}                      & \small{4 \texttt{edt} / 4 \texttt{edt}}             & \small{1 \texttt{edt} / 1 \texttt{edt}}  \\ \hline
\small{Free \texttt{XOR}}   & \small{3 ct / free}                      & \small{4 \texttt{edt} / free}             & \small{1 \texttt{edt} / free}  \\
\multicolumn{4}{c}{ }\\
\multicolumn{4}{c}{{\scriptsize ct: ciphertexts; \texttt{edt}: total encryption and/or decryption time}}\\
\end{tabular}
\end{table}

\section{Garbled Row Reduction 2 Ciphertexts} \label{sec:GRR2}{}

Pinkas \textit{et al.} proposed a method called \textit{garbled row reduction 2 ciphertexts} (GRR2) in order to reduce the number of transferred ciphertexts in \citep{PSSW09}. GRR2 is based on Shamir's secret sharing (\ref{sec:secretshare}). It is especially good for reducing the size in case of abundant \texttt{AND} gates \citep{PSSW09}.

\begin{figure}
\centering
\begin{subfigure}[b]{0.35\textwidth}
  \centering
	\includegraphics[width=\textwidth]{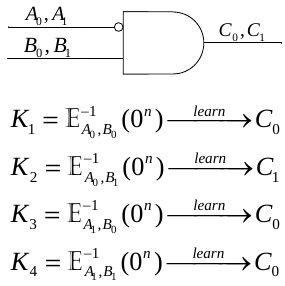}
\caption{}
\label{fig:sub1}
\end{subfigure}%
\quad
\begin{subfigure}[b]{0.57\textwidth}
  \centering
  \includegraphics[width=\textwidth]{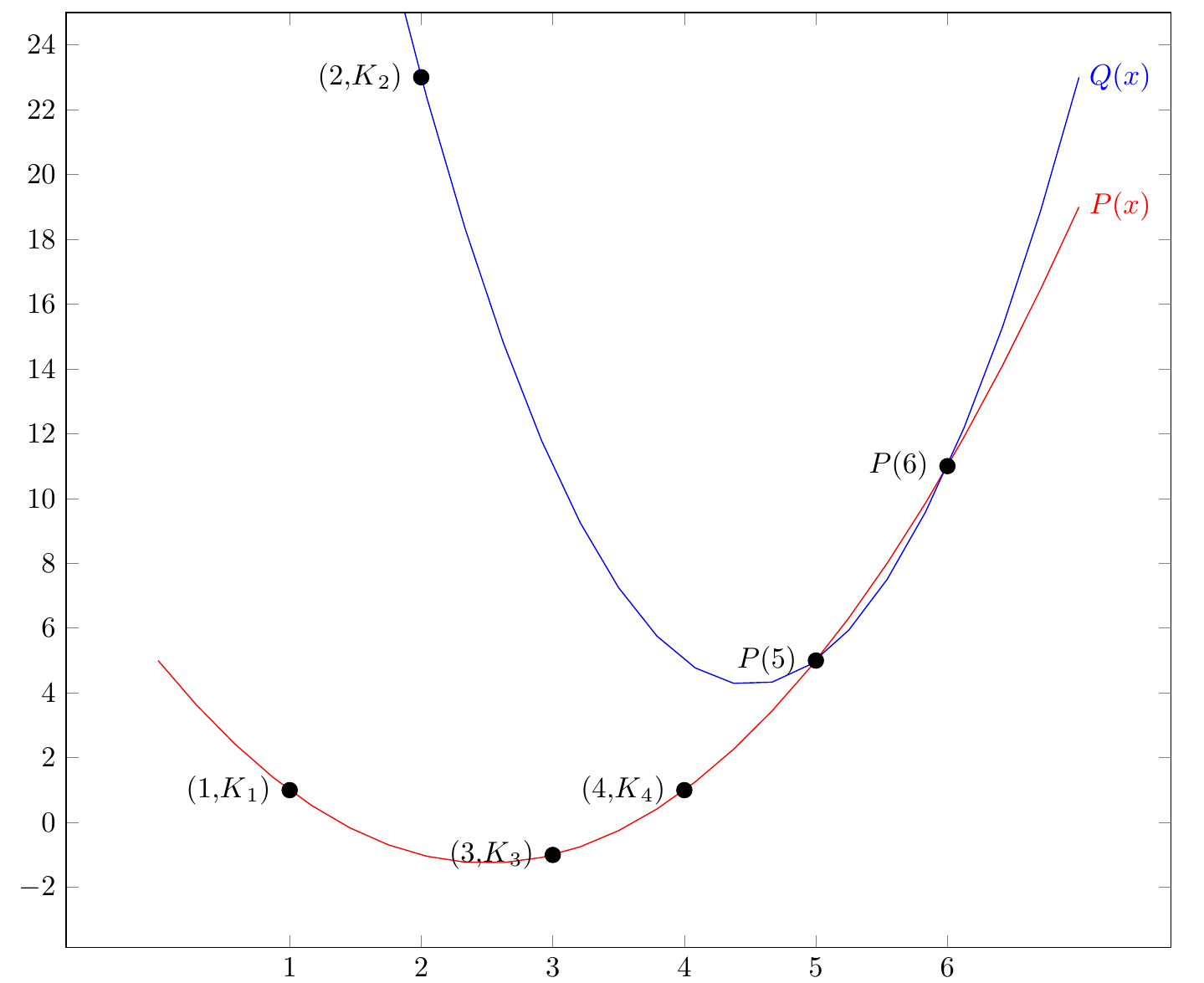}
  \caption{}
  \label{fig:sub2}
\end{subfigure}
\caption{(a) The odd gate to be garbled. (b) Plots of two polynomials obtained from $K_{1}$, $K_{2}$, $K_{3}$, and $K_{4}$.}
\label{fig:grr2}
\end{figure}

\texttt{Garbling:}
\begin{enumerate}
\setcounter{enumi}{0}
\item{Alice and Bob want to compute the output of the odd gate in Figure \ref{fig:grr2} (a).}
\item{Alice calculates $K_{1}$, $K_{2}$, $K_{3}$, and $K_{4}$ by decrypting all 0 for all possible input combinations (\textit{e.g.}, $K_{1}\gets E_{A_0,B_0}^{-1}(0^n)$, $K_{2}\gets E_{A_0,B_1}^{-1}(0^n)$, $K_{3}\gets E_{A_1,B_0}^{-1}(0^n)$, $K_{4}\gets E_{A_1,B_1}^{-1}(0^n)$).}
\item{Using the rows which give the same output (in this case the rows 1, 3, 4) Alice plots a $2^{nd}$ degree polynomial $P(x)$ (\textit{e.g.}, the red parabolas in Figure \ref{fig:grr2} (b)).}
\item{Alice also plots another $2^{nd}$ degree polynomial $Q(x)$ from the excluded row (here the row 2), $P(5)$, and $P(6)$ (\textit{e.g.}, the blue parabolas in Figure \ref{fig:grr2} (b)).}
\end{enumerate} 

\texttt{Evaluating:}
\begin{enumerate}
\setcounter{enumi}{4}
\item{Alice sends only the intersection points $P(5)$ and $P(6)$. Bob will get another point by decrypting all 0 with the masking values that he gets in the input. He will be able to reaching only one of the polynomials, not knowing which one. The output masking value will be the evaluation of this polynomial at $x=0$ (\textit{e.g.}, $C_{0}=P(0)$ and $C_{1}=Q(0)$).}
\end{enumerate}

The position in this  scheme leaks information. Moreover, since the wire masking values are not chosen but calculated pseudo-random values, it is impossible to directly use the P\&P (\ref{sec:PP}) technique. Instead, Pinkas \textit{et al.} proposed adding a one bit \textit{external value} $c_{i}$ for each wire. External values, like labes, are different for the \texttt{TRUE} and \texttt{FALSE} truth values unrelated to the truth value. Just like labels, external values are used for ordering. To calculate the external value of the output of a gate, 4 additional $M_{r}$ bits are sent. The evaluator, then, just needs to \texttt{XOR} the first bits of both input masking values and the related $M_{r}$ bit to find out the output external value. Since he does not know the masking values for other truth values of the input wires, he cannot find out the external values for the other output.

\texttt{Garbling:}
\begin{enumerate}
\setcounter{enumi}{0}
\item{For an even gate, Alice similarly calculates $K_{1}$, $K_{2}$, $K_{3}$, and $K_{4}$ as in the odd gate case, in order of the external values.}
\item{Somewhat differently from the previous procedure, she plots the two $1^{st}$ degree polynomials each passing through the two points which correspond to the same output value. For instance, if both $K_{1}$ and $K_{3}$ are for the rows corresponding to  \texttt{TRUE}, she plots $P(x)$ passing through $(1,K_{1})$ and $(3,K_{3})$ and $Q(x)$ passing through $(2,K_{2})$ and $(4,K_{4})$. She sends $P(5)$ and $Q(5)$, along with the 4 additional $M_{r}$ bits. She makes sure that ordering $P(5)$ and $Q(5)$ is according to the external value of the output of the gate just like using them the same as label bits, so that the evaluator know which one to use.}
\end{enumerate}

\texttt{Evaluating:}
\begin{enumerate}
\setcounter{enumi}{2}
\item{The evaluator decrypts all 0 with the masking values of the inputs. With two points in hand he plots the $1^{st}$ degree polinomial evaluate it at $x=0$ and reaches the output masking value.}
\end{enumerate}

Referring to Shamir's secret sharing (\ref{sec:secretshare}), two $t$-length values and 4 $M_r$ bits ($2t+4$) are needed to be sent per gate. For the sake of simplicity, we can take it as 2 ciphertexts per gate (see Table \ref{tab:scrbrdfrgrr2}).

\begin{table}[]
\centering
\caption{Optimization Scoreboard (GRR2)}
\label{tab:scrbrdfrgrr2}
\begin{tabular}{l|l|l|l}
\small{\textbf{Method}} & \small{\textbf{Odd / Even Gate Size}} & \small{\textbf{Enc. Time per}} & \small{\textbf{Dec. Time per}} \\ 
\small{ } & \small{ } & \small{\textbf{Odd / Even Gate}} & \small{\textbf{Odd / Even Gate}} \\ \hline
\small{P\&P (\ref{sec:PP})}   & \small{4 ct / 4 ct}                      & \small{4 \texttt{edt} / 4 \texttt{edt}}             & \small{1 \texttt{edt} / 1 \texttt{edt}}           \\ \hline
\small{GRR3 (\ref{sec:GRR3}{})}   & \small{3 ct / 3 ct}                      & \small{4 \texttt{edt} / 4 \texttt{edt}}             & \small{1 \texttt{edt} / 1 \texttt{edt}}  \\ \hline
\small{Free \texttt{XOR} (\ref{sec:freeXOR})}   & \small{3 ct / free}                      & \small{4 \texttt{edt} / free}             & \small{1 \texttt{edt} / free}  \\ \hline
\small{GRR2}   & \small{2 ct / 2 ct}                      & \small{4 \texttt{edt} / 4 \texttt{edt}}             & \small{1 \texttt{edt} / 1 \texttt{edt}}  \\
\multicolumn{4}{c}{ }\\
\multicolumn{4}{c}{{\scriptsize ct: ciphertexts; \texttt{edt}: total encryption and/or decryption time}}\\
\end{tabular}
\end{table}

Although GRR2 is good for reducing the sizes of odd gates, it has a major drawback: incompatibility with free \texttt{XOR} (\ref{sec:freeXOR}). This is because the output masking values of the gates garbled with the GRR2 technique are pseudo-random numbers which cannot be set to the same offset.

%

%
%

\section{Fle\texttt{XOR}}\label{sec:fleXOR}{}

The incompatibility of free \texttt{XOR} (\ref{sec:freeXOR}) and GRR2 (\ref{sec:GRR2}) causes an inconvenient situation where both may be better for different circuits depending on the proportion of \texttt{XOR} and \texttt{AND} gates. To solve this issue, Kolesnikov \textit{et al.} proposed the fle\texttt{XOR} technique in \cite{KMR14}. Fle\texttt{XOR} may reduce the number of ciphertexts for an \texttt{XOR} gate even if it has different offsets on its wires. With this technique, \texttt{XOR} gates requires 1 or less ciphertext most of the time. It may cost 2 ciphertexts, only if the output masked value of the XOR gate has different offset from its inputs. Actually, most of the time, the output masked value may be chosen such that it has the same offset at least one of the inputs 

\begin{figure}
\includegraphics[height=6.5cm, angle=0]{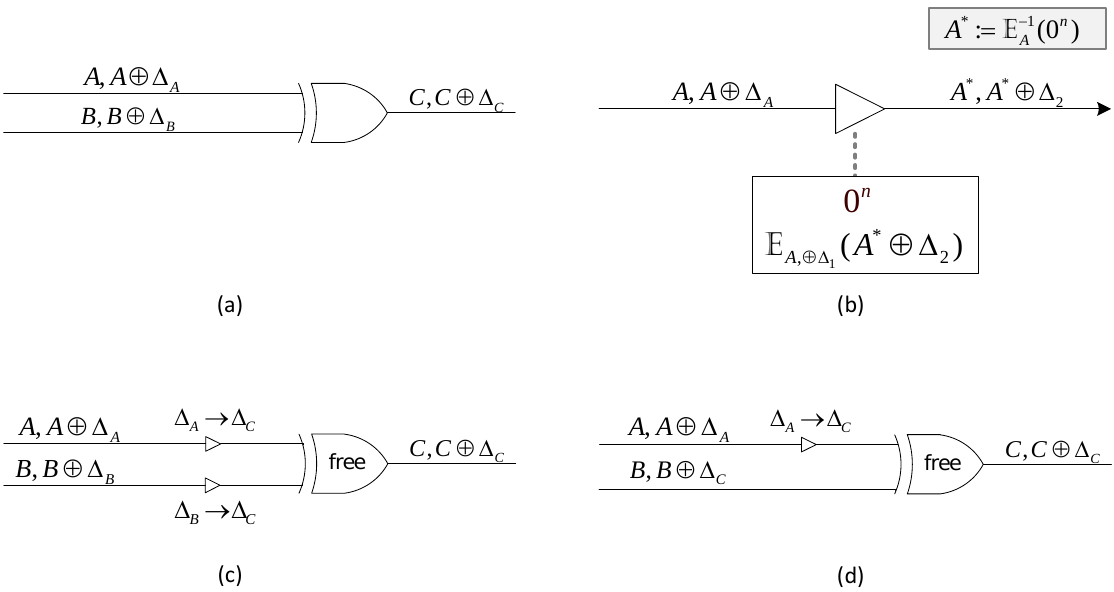}
\caption{ (a) An \texttt{XOR} gate with different offsets in its inputs and output. (b) A buffer gate to carry the offset of a wire. (c) An \texttt{XOR} gate offsets of whose inputs  are carried to the offset of its output by two imaginary buffer gates. (d) An \texttt{XOR} gate the offset of whose an input is carried to the offset of its output by an imaginary buffer gate.}
\label{fig:flexor}
\end{figure}

\texttt{Garbling:}
\begin{enumerate}
\setcounter{enumi}{0}
\item{Alice and Bob want to compute the output of the \texttt{XOR} gate in Figure \ref{fig:flexor} (a).}
\item{The idea is that if it was possible to carry the input wires to the same offset level with the output wire, which is $\triangle_{C}$, the \texttt{XOR} gate would be free. Figure \ref{fig:flexor} (b) depicts an imaginary buffer gate which can be used to carry the offset of a wire. Alice encrypts the output masked values with their corresponding inputs as $E_{A}(A^*)$ and $E_{A\oplus\triangle_{1}}(A^*\oplus\triangle_{2})$. She order them by P\&P (\ref{sec:PP}), and since $A^*$ can be any random value, she can let the first one in order all 0. Therefore, sending just one cipher text for a buffer gate suffices.}
\item{Alice needs at most two imaginary buffer gates for an \texttt{XOR} gate to carry the inputs to the same offset level as the outputs (see Figure \ref{fig:flexor} (c)).}
\item{Most of the time, one imaginary buffer per \texttt{XOR} gate will be enough since Alice can let the offset of the output the same as one of the inputs (see Figure \ref{fig:flexor} (d)). And if the inputs and the output have the same offset, the \texttt{XOR} gate will be free.}
\end{enumerate}

fle\texttt{XOR} technique can be combined with GRR2 (\ref{sec:GRR2}) in order to reduce the number of ciphertexts for \texttt{AND} gates (see Table \ref{tab:scrbrdfrflxor}). The combined scheme proposed by Kolesnikov \textit{et al.} can be seen in Algorithm \ref{alg:combflexor}.

\begin{table}[t]
\centering
\caption{Optimization Scoreboard (Fle\texttt{XOR})}
\label{tab:scrbrdfrflxor}
\begin{tabular}{l|l|l|l}
\small{\textbf{Method}} & \small{\textbf{Odd / Even Gate Size}} & \small{\textbf{Enc. Time per}} & \small{\textbf{Dec. Time per}} \\ 
\small{ } & \small{ } & \small{\textbf{Odd / Even Gate}} & \small{\textbf{Odd / Even Gate}} \\ \hline
\small{P\&P (\ref{sec:PP})}   & \small{4 ct / 4 ct}                      & \small{4 \texttt{edt} / 4 \texttt{edt}}             & \small{1 \texttt{edt} / 1 \texttt{edt}}           \\ \hline
\small{GRR3 (\ref{sec:GRR3}{})}   & \small{3 ct / 3 ct}                      & \small{4 \texttt{edt} / 4 \texttt{edt}}             & \small{1 \texttt{edt} / 1 \texttt{edt}}  \\ \hline
\small{Free \texttt{XOR} (\ref{sec:freeXOR})}   & \small{3 ct / free}                      & \small{4 \texttt{edt} / free}             & \small{1 \texttt{edt} / free}  \\ \hline
\small{GRR2 (\ref{sec:GRR2}{})}   & \small{2 ct / 2 ct}                      & \small{4 \texttt{edt} / 4 \texttt{edt}}             & \small{1 \texttt{edt} / 1 \texttt{edt}}  \\ \hline
\small{Fle\texttt{XOR}}   & \small{2 ct / \{0,1,2\} ct}                      & \small{4 \texttt{edt} / \{0,2,4\} \texttt{edt}}             & \small{1 \texttt{edt} / \{0,1,2\} \texttt{edt}} \\
\multicolumn{4}{c}{ }\\
\multicolumn{4}{c}{{\scriptsize ct: ciphertexts; \texttt{edt}: total encryption and/or decryption time}}\\
\end{tabular}
\end{table}

\begin{algorithm}
  	\scriptsize
	\caption{The combined fle\texttt{XOR} and GRR2 scheme proposed by Kolesnikov \textit{et al.} in \cite{KMR14}.}
	\label{alg:combflexor}
	\begin{algorithmic}[1] 
	\Procedure{\texttt{Gb}}{$1^k,f$} \Comment \texttt{Garbling phase}
		\State $(n,m,q,A',B',G)\gets f$
		\For{$i\in\{1,\dots,n\}$}  $t\gets \{0,1\}$, $X_i^0\gets \{0,1\}^{k-1}t$, $X_i^1\gets \{0,1\}^{k-1}\overline{t}$
		\EndFor
		\For{$g\in\{n+1,\ldots,n+q\}$ in a \textbf{safety-respecting} order} $a \gets A'(g)$, $b \gets B'(g)$
			\If {$g\in \texttt{XORGates}(f)$} 
				\If{$X_a^0\oplus X_a^1=X_b^0\oplus X_b^1$} $X_g^0\gets X_a^0\oplus X_b^0$, $X_g^1\gets X_a^0\oplus X_b^1$, $P[g]\gets \bot$
				\Else 
					\If{$\texttt{C}_{X_a^0}=0$} $X_a^{\bar{0}}\gets H(X_a^0,g\parallel 00)$, $X_a^{\bar{1}}\gets X_a^{\bar{0}}\oplus X_b^0\oplus X_b^1$
						\State $X_g^0\gets X_a^{\bar{0}}\oplus X_b^0$, $X_g^1\gets X_a^{\bar{0}}\oplus X_b^1$, $P[g]\gets H(X_a^1,g\parallel 00) \oplus X_a^ {\bar{1}}$
					\Else \texttt{ }      $X_a^{\bar{1}}\gets H(X_a^1,g\parallel 00)$, $X_a^{\bar{0}}\gets X_a^{\bar{1}}\oplus X_b^0\oplus X_b^1$ 
						\State $X_g^0\gets X_a^{\bar{0}}\oplus X_b^0$, $X_g^1\gets X_a^{\bar{0}}\oplus X_b^1$, $P[g]\gets H(X_a^0,g\parallel 00) \oplus X_a^ {\bar{0}}$
					\EndIf
				\EndIf
				\State $\texttt{C}_{X_g^0}\gets \texttt{C}_{X_a^0}\oplus \texttt{C}_{X_b^0}$, $\text{C}_{X_g^1}\gets \bar{\texttt{C}_{X_g^0}}$
			\Else 
				\State $\text{C}_{X_g^0}\gets\{0,1\}$ $\texttt{C}_{X_g^1}\gets \bar{\texttt{C}_{X_g^0}}$
				\For{$(i,j)\in \{0,1\}^2$} $V_{ij}\parallel m_{ij} \gets H(X_{ai},X_{bj},g\parallel i\parallel j)$, $c_{ij}\gets \texttt{C}_{X_g^{w_{ai}\wedge w_{bj}}}\oplus  m_{ij}$
				\EndFor
				\State $Q \gets interp \{(2i+j,V_{ab})\texttt{ } | \texttt{ }w_{ai}\wedge w_{bj}=0\}$
				\State $R \gets interp \{(2i+j,V_{ab}\text{ } |\texttt{ }w_{ai}\wedge w_{bj}=1),(4,Q(4)),(5,Q(5))\}$
				\State $X_g^0\gets Q(-1)$, $X_g^1\gets R(-1)$, $P[g]\gets (Q(4),Q(5),c_{00},c_{01},c_{10},c_{11})$
			\EndIf
           	\EndFor
		\State $F\gets(n,m,q,A',B',P)$, $e\gets(X_1^0,X_1^1,\ldots,X_n^0,X_n^1)$, $d\gets(X_{n+q-m+1}^0,X_{n+q-m+1}^1,\ldots,X_{n+q}^0,X_{n+q}^1)$
            	\State \textbf{return} $(F,e,d)$ 
        \EndProcedure
        \Statex
        \Procedure{\texttt{En}}{$e,x$} \Comment \texttt{Encoding phase}
		\State $(X_1^0,X_1^1,\ldots,X_n^0,X_n^1)\gets e$, $x_1\ldots x_n\gets x$, $X\gets (X_1^{x_1},\ldots,X_n^{x_n})$, \textbf{return} $X$ 
        \EndProcedure
        \Statex
        \Procedure{\texttt{Ev}}{$F,X$} \Comment \texttt{Evaluating phase}
		\State $(n,m,q,A',B',P)\gets F$, $(X_1,\ldots,X_n)\gets X$
		\For{$g\gets n+1 \textbf{ to } n+q$} $a \gets A'(g)$, $b \gets B'(g)$
                		\If {$g\in \texttt{XORGates}(f)$}
				\If{$P[g]\gets \bot$} $X_g\gets X_a\oplus X_b$
				\Else 
					\If{$\texttt{C}_{X_a}=0$} $X_a^{\bar{}}\gets H(X_a,g\parallel 00)$
					\Else \texttt{ }       $X_a^{\bar{}}\gets P[g]\oplus H(X_a,g\parallel 00)$
					\EndIf
				\EndIf
				\State $\texttt{C}_{X_g}\gets \texttt{C}_{X_a}\oplus \texttt{C}_{X_b}$
			\Else 
				\State$V*\parallel m* \gets H(X_{a},X_{b},g\parallel {C}_{X_a}\parallel {C}_{X_b})$
				\State $R* \gets interp \{(2{C}_{X_a}+{C}_{X_b},V*),(4,Q(4)),(5,Q(5))\}$
				\State $X_g\gets R*(-1)$, $\texttt{C}_{X_g}\gets c_{\texttt{C}_{X_a}\texttt{C}_{X_b}}\oplus  m*$
			\EndIf
           	\EndFor
		\State \textbf{return} $(X_{n+q-m+1},\ldots,X_{n+q})$ 
        \EndProcedure
        \Statex
        \Procedure{\texttt{De}}{$d,Y$} \Comment \texttt{Decoding phase}
		\State $(Y_1,\ldots,Y_m)\gets Y$, $(Y_1^0,Y_1^1,\ldots,Y_m^0,Y_m^1)\gets d$
		\For{$i\in\{1,\ldots,m\}$}
                		\If{$Y_i=Y_i^0$} $y_i\gets 0$
			\ElsIf{$Y_i=Y_i^1$} $y_i\gets 1$
			\Else \textbf{ return} $\bot$
			\EndIf
           	\EndFor
		\State \textbf{return} $y\gets y_1\ldots y_m$
        \EndProcedure
    \end{algorithmic}
\end{algorithm}

The notation used in Algorithm \ref{alg:combflexor} is similar to the one in Algorithm \ref{alg:yaosgcpproc}. $\texttt{XORGates}(f)$ denotes the set of $\texttt{XOR}$ gates in $f$. $\texttt{C}_{X_i}$ denotes the external value of the wire whose masking value is $X_i$. $V_{ij}$ denotes the value used in the interpolation related to the order $ij$. $m_{ij}$ denotes the one bit value used to mask the external value. $X_{ai}$ denotes the masking value on the wire $a$, $i$ being the external value. $w_{ai}$ denotes the truth value on the wire $a$, $i$ being the external value. $c_{ij}$ denotes the bits sent for the calculation of the external value of the output of a gate, $ij$ being the order coming from the input external values \cite{KMR14}.

\section{Half Gates} \label{sec:halfgates}{}

The \textit{half gates} method, which is proposed by Zahur \textit{et al.} in \cite{ZRE15}, proves that sending 2 ciphertexts can be enough for an \texttt{AND} gate while \texttt{XOR} gates are still free. The same offset is kept throughout the whole circuit wires, like the free \texttt{XOR} (\ref{sec:freeXOR}). It is based upon the idea that if one of the sides knows the truth value on an input wire of an \texttt{AND} gate, it is enough to send just one ciphertext. The method divides the \texttt{AND} gate into two \texttt{AND} gates where one of the parties knows the truth value on an input wire. The name of the method comes from this division.

$A$, $B$, $C$, $C_1$, and $C_2$ are the masking values for the wires $a$, $b$, $c$ (output of the \texttt{AND} gate), $c_1$ (output of the garbler half gate), and $c_2$ (output of the evaluator half gate), respectively. $\triangle$ denotes the common offset as in free \texttt{XOR} (\ref{sec:freeXOR}).

\texttt{Garbling:}
\begin{enumerate}
\setcounter{enumi}{0}
\item{Alice and Bob want to compute the output of an \texttt{AND} gate whose inputs are $a$ and $b$.}
\item{An \texttt{AND} gate can be written as an \texttt{XOR} of two \texttt{AND} gates as in Equation~\eqref{eq:andgatedivided} where $r$ is a randomly chosen bit only known to Alice. Alice chooses it to be the label bit of the $B$, which is the masked value for \texttt{FALSE} on the wire $b$. $r$ is still unknown to Bob.}
\end{enumerate}

\begin{equation} 
\label{eq:andgatedivided}
 a\wedge b=(a\wedge r)\oplus [a\wedge(b\oplus r)] 
\end{equation}

\texttt{Garbler Half Gate:}
\begin{enumerate}
\setcounter{enumi}{2}
\item{$a\wedge r$ is the garbler half gate, whereas $a\wedge(b\oplus r)$ is the evaluator half gate. For the output of the garbler half gate $c_{1}\gets a\wedge r$, Alice needs to send $E_{B}(C_{1})$ and $E_{B\oplus\triangle} (C_{1}\oplus r\triangle)$. Since she knows the value of $r$, there is just 2 input combinations. She orders the ciphertexts with respect to the label bit of $b$. Row reduction (\ref{sec:GRR3}) is also possible by letting the $1^{st}$ ciphertext in all 0. She calculates the $2^{nd}$ ciphertext from the value she reaches by decrypting the first one. Thus, sending just 1 ciphertext is enough for the garbler half gate.}
\item{During the evaluation of the garbler half gate, Bob decrypts the related cipher text depending on the label bit of the masking value on the wire $b$. Since the order is by labels he can not learn the truth value of $b$.}
\end{enumerate}

\texttt{Evaluator Half Gate:}
\begin{enumerate}
\setcounter{enumi}{4}
\item{For the evaluator half gate, Alice needs to let Bob learn $q=b\oplus r$ without learning $b$ or $r$. Actually, it is whatever Bob gets as the label bit of the masked value on wire $b$. This was the main reason why $r$ was chosen as the label bit of $B$ in the beginning.}
\item{To garble the evaluator half gate $c_{2}\gets a\wedge q$, there are two ways Alice may go depending on the value of $r$. If $r$ is \texttt{FALSE}, Alice sends two ciphertexts $E_{B}(C_{2})$ and $E_{B\oplus\triangle} (C_{2}\oplus A)$ in this order strictly. Otherwise, Alice sends two ciphertexts $E_{B\oplus\triangle} (C_{2})$ and $E_{B}(C_{2}\oplus A)$ in this order strictly. Moreover, the $1^{st}$ ciphertext can be let all 0 and the $2^{nd}$ one can be calculated from it. Therefore, sending only one ciphertext for the evaluator half gate also suffices.}
\item{If Bob gets \texttt{FALSE} as $q$, he decrypts the first ciphertext using the masking value on the wire $b$, arriving at the masking value of the output of the evaluator half gate. Otherwise, he decrypts the second ciphertext  using the value on the wire $b$, and \texttt{XOR}s the result with the masking value on the wire $a$, arriving at the masking value of the output of the half gate.}
\end{enumerate}

The evaluator does not learn the truth values of $a$, $b$, $r$, or $c_{2}$ (if $q= 1$, of course, otherwise he learns  $c_{2}$).  In the end, the results of the half gates must be \texttt{XOR}ed, in order to obtain the final output of the \texttt{AND} gate.

With the half gates technique, an \texttt{AND} gate costs 2 cipher texts and \texttt{XOR}s are free, which makes the half gates technique the optimum from size point of view among the methods developed so far (see Table \ref{tab:scrbrdhg}). Zahur \textit{et al.} have also proven that decreasing the size of an \texttt{AND} gate further is impossible.

\begin{table}[b]
\centering
\caption{Optimization Scoreboard (Half Gates)}
\label{tab:scrbrdhg}
\begin{tabular}{l|l|l|l}
\small{\textbf{Method}} & \small{\textbf{Odd / Even Gate Size}} & \small{\textbf{Enc. Time per}} & \small{\textbf{Dec. Time per}} \\ 
\small{ } & \small{ } & \small{\textbf{Odd / Even Gate}} & \small{\textbf{Odd / Even Gate}} \\ \hline
\small{P\&P (\ref{sec:PP})}   & \small{4 ct / 4 ct}                      & \small{4 \texttt{edt} / 4 \texttt{edt}}             & \small{1 \texttt{edt} / 1 \texttt{edt}}           \\ \hline
\small{GRR3 (\ref{sec:GRR3}{})}   & \small{3 ct / 3 ct}                      & \small{4 \texttt{edt} / 4 \texttt{edt}}             & \small{1 \texttt{edt} / 1 \texttt{edt}}  \\ \hline
\small{Free \texttt{XOR} (\ref{sec:freeXOR})}   & \small{3 ct / free}                      & \small{4 \texttt{edt} / free}             & \small{1 \texttt{edt} / free}  \\ \hline
\small{GRR2 (\ref{sec:GRR2}{})}   & \small{2 ct / 2 ct}                      & \small{4 \texttt{edt} / 4 \texttt{edt}}             & \small{1 \texttt{edt} / 1 \texttt{edt}}  \\ \hline
\small{Fle\texttt{XOR} (\ref{sec:fleXOR}{})}   & \small{2 ct / \{0,1,2\} ct}                      & \small{4 \texttt{edt} / \{0,2,4\} \texttt{edt}}             & \small{1 \texttt{edt} / \{0,1,2\} \texttt{edt}}  \\ \hline
\small{Half Gates}   & \small{2 ct / free}                      & \small{4 \texttt{edt} / free}             & \small{2 \texttt{edt} / free}  \\
\multicolumn{4}{c}{ }\\
\multicolumn{4}{c}{{\scriptsize ct: ciphertexts; \texttt{edt}: total encryption and/or decryption time}}\\ 
\end{tabular}
\end{table}

\textbf{The Complete Scheme.} For a boolean circuit $f$, a numeric index is assigned to each wire in the circuit. The sets of input wires, output wires, output wires of \texttt{XOR} gates in $f$$ are denoted as \texttt{Inputs}(f)$, $\texttt{Outputs}(f)$, and $\texttt{XORGates}(f)$, respectively. These functions can also be applied to garbled version $F$ of $f$ as $\texttt{Inputs}(F)$, $\texttt{Outputs}(F)$, and $\texttt{XORGates}(F)$. $v_{i}$ denotes the one bit truth value on the $i^{th}$ wire in a circuit. If the output wire of a gate has index $i$, that gate is named as $i^{th}$ gate. The wire masking values for \texttt{FALSE} and \texttt{TRUE} on the $i^{th}$ wire is denoted as $W_{i}^0, W_{i}^1 \in \{0, 1\}^k$, respectively. The security parameter of the scheme is denoted as $k$. For each wire masking value $W$, the label bit is its least significant bit $\texttt{lsb}W$. For the $i^{th}$ wire, define $p_{i}= \texttt{lsb}W_{i}^0$. Being named as the permute bit of the wire, that value is a secret kept by the generator. Intuitively, if label bit a masking value on a wire is $s_{i}$, that masking value is $W _{i}^{s_{i}\oplus p_{i}} $, and corresponds to the truth value $s_{i}\oplus p_{i}$. $W_{i}$ implies that the evaluator does not know $v_{i}$. The free \texttt{XOR} offset is denoted as $R \in \{0, 1\}^k$. We have $\texttt{lsb}R = 1$ so that $\texttt{lsb}W_{i}^0\neq\texttt{lsb}W_{i}^1$, and the complementary masking values on wires have different label bits. Sometimes $\wedge$ is omitted and two symbols is  juxtaposed  to imply \texttt{AND} ($ab = a\wedge b$). $H:\{0,1\}^k\times \mathbb{Z} \rightarrow \{0,1\}^k$ denotes a hash-function that is usable in garbled circuits.

\begin{equation} 
\label{eq:oddgategeneralization}
 (v\textsubscript{a},v\textsubscript{b}) \rightarrow(a\textsubscript{a} \oplus v\textsubscript{a})\wedge(a\textsubscript{b} \oplus v\textsubscript{b})\oplus a\textsubscript{c}
\end{equation}

The technique can be further generalized such that it can be applied any odd gate (\texttt{OR}, \texttt{NOR}, \texttt{NAND}, etc.), since all of them can be written as in Equation~\eqref{eq:oddgategeneralization} where $a\textsubscript{a}$, $a\textsubscript{b}$, $a\textsubscript{c}$ are constants. For example, an \texttt{AND} gate results from setting all to \texttt{FALSE},  an \texttt{OR} gate results from setting all to \texttt{TRUE}. The construction of half gate is shown step-by-step in Table \ref{tab:halfgates}. Note that the $a$ values does not affect what the evaluator does.

\vspace{2mm}
\begin {table}[H]
\caption {The construction of half gates for computing Equation~\eqref{eq:oddgategeneralization} \cite{ZRE15}.} \label{tab:halfgates} 
\scriptsize
\begin{center}
\begin{tabular}{ >{\arraybackslash}m{2.7in} | >{\arraybackslash}m{2.7in}}
\toprule[1.5pt] 
{\bf Generator half gate: $p_b$ known to generator} & {\bf Evaluator half gate: $v_b\oplus p_b$ known to evaluator} \\ 
\underline{Computes:}\newline $f_G(v_a,p_b)\gets(v_a\oplus a_a)(p_b\oplus a_b) \oplus a_c$  &  \underline{Computes:}\newline $f_E(v_a,v_b\oplus p_b)\gets(v_a\oplus a_a)(v_b\oplus p_b)$   \\
\underline{Before GRR and Permutation:}\newline $H(W_a^0)\oplus f_G(0,p_b)R\oplus W_{Gc}^0$\newline $H(W_a^1)\oplus f_G(1,p_b)R\oplus W_{Gc}^0$      &  \underline{Before GRR:}\newline $H(W_b^{p_b})\oplus W_{Ec}^0$\newline $H(W_b^{p_b\oplus 1})\oplus W_{Ec}^0\oplus W_a^{a_a}$   \\
\underline{After GRR and permutation:}\newline $T_{Gc}\gets H(W_a^0)\oplus H(W_a^1) \oplus (p_b\oplus a_b)R$\newline $W_{Gc}^0\gets H(W_a^{p_a})\oplus  f_G(p_a,p_b)R$&  \underline{After GRR (permutation not needed):}\newline $T_{Ec}\gets H(W_b^0)\oplus H(W_b^1) \oplus W_a^{a_a}$\newline $W_{Ec}^0\gets H(W_b^{p_b})$   \\
\underline{Generator sends $T_{Gc}$}&\underline{Generator sends $T_{Ec}$ }  \\

\bottomrule[1.25pt]
\end {tabular}
\end{center}
\end {table}

The complete garbling procedure for an entire circuit proposed by Zahur \textit{et al.} is shown in Algorithm \ref{alg:comphalfgates} \cite{ZRE15}. All gates are assumed to be either an \texttt{AND} or an \texttt{XOR} gate. Since \textsc{\texttt{De}} never returns $\bot$, this scheme does not satisfy the authenticity criterion. In order to make it authentic, Zahur \textit{et al.} propose the following changes:

\begin{itemize}
\item{The \textbf{for}-loop on Line \ref{lst:labelarrange} of Algorithm \ref{alg:comphalfgates} must be changed as:}
\end{itemize}

\begin{algorithm}
  	\scriptsize
	\label{alg:dehalfgates}
	\begin{algorithmic}
		\For{$i\in \texttt{Outputs}(f)$}  
			\State $j\gets \texttt{NextIndex}()$
			\State $d_i\gets (H(W_i^0,j),H(W_i^1,j))$
		\EndFor
	\end{algorithmic}
\end{algorithm}

\begin{itemize}
\item{The \textbf{for}-loop on Line \ref{lst:decode} of Algorithm \ref{alg:comphalfgates} must be changed as:}
\end{itemize}

\begin{algorithm}[H]
  	\scriptsize
	\label{alg:dehalf-gates}
	\begin{algorithmic}
		\For{$d_i\in d$} $j\gets \texttt{NextIndex}()$, parse $(h_0,h_1)\gets d_i$
			\If{$H(Y_i,j)=h_0$} $y_i\gets 0$
			\ElsIf{$H(Y_i,j)=h_1$} $y_i\gets 1$
			\Else \textbf{ return} $\bot$
			\EndIf
		\EndFor
	\end{algorithmic}
\end{algorithm}


\begin{algorithm}
  	\scriptsize
	\caption{The complete half gates garbling scheme proposed by Zahur \textit{et al.} in \cite{ZRE15}.}
	\label{alg:comphalfgates}
	\begin{algorithmic}[1] 
	\Procedure{\texttt{Gb}}{$1^k,f$} \Comment \texttt{Garbling phase}
		\State $R\twoheadleftarrow \{0,1\}^{k-1}1$
		\For{$i\in \texttt{Inputs}(f)$}  
			\State $W_i^0\twoheadleftarrow \{0,1\}^k$, $W_i^1\gets W_i^0\oplus R$, $e_i\gets W_i^0$
		\EndFor
		\For{$i\notin \texttt{Inputs}(f)$} \{\textit{in topo. order}\}
			\State $\{a,b\} \gets \texttt{GateInputs}(f,i)$
			\If{$i\in \texttt{XORGates}(f)$} $W_i^0\gets W_a^0\oplus W_b^0$
			\Else \texttt{ }$(W_i^0,T_{Gi},T_{Ei})\gets \textsc{\texttt{GbAnd}}(W_a^0,W_b^0)$, $F_i\gets T_{Gi},T_{Ei}$
			\EndIf
			\State $W_i^1\gets W_i^0\oplus R$
		\EndFor
		\For{$i\in \texttt{Outputs}(f)$}  \label{lst:labelarrange}
			\State $d_i\gets \texttt{lsb} (W_i^0)$
		\EndFor
            	\State \textbf{return} $(F,e,d)$ 
        \EndProcedure
        \Statex
        \algrenewcommand\algorithmicprocedure{\textbf{private procedure}}
	\Procedure{\texttt{GbAnd}}{$W_a^0,W_b^0$} \Comment  \texttt{Garbling AND gates}
		\State $p_a\gets \texttt{lsb} (W_a^0)$, $p_b\gets \texttt{lsb} (W_b^0)$
		\State $j\gets \texttt{NextIndex}()$, $j'\gets \texttt{NextIndex}()$
		\State \{\textit{First half gate}\}
		\State $T_G\gets H(W_a^0,j)\oplus H(W_a^1,j)\oplus p_bR$
		\State $W_G^0\gets H(W_a^0,j)\oplus p_aT_G$
		\State \{\textit{Second half gate}\}
		\State $T_E\gets H(W_b^0,j')\oplus H(W_b^1,j')\oplus W_a^0$
		\State $W_E^0\gets H(W_b^0,j')\oplus p_b(T_E\oplus W_a^0)$
		\State \{\textit{Combine two halves}\}
		\State $W_0\gets W_G^0\oplus W_E^0$
		\State \textbf{return} $(W^0,T_G,T_E)$ 
        \EndProcedure
        \Statex
        \algrenewcommand\algorithmicprocedure{\textbf{procedure}}
        \Procedure{\texttt{En}}{$e,x$} \Comment \texttt{Encoding phase}
		\For{$e_i\in e$}  $X_i \gets e_i\oplus x_iR$
		\EndFor
		\State \textbf{return} $X$ 
        \EndProcedure
        \Statex
        \Procedure{\texttt{Ev}}{$F,X$} \Comment  \texttt{Evaluating phase}
		\For{$i\in \texttt{Inputs}(F)$}  
			\State $W_i\gets X_i$
		\EndFor
		\For{$i\notin \texttt{Inputs}(F)$} \{\textit{in topo. order}\}
			\State $\{a,b\} \gets \text{GateInputs}(F,i)$
			\If{$i\in \texttt{XORGates}(F)$} $W_i\gets W_a\oplus W_b$
			\Else \text{ } $s_a\gets \texttt{lsb} (W_a)$, $s_b\gets \texttt{lsb} (W_b)$, $j\gets \texttt{NextIndex}()$, $j'\gets \texttt{NextIndex}()$
			\State $T_{Gi},T_{Ei}\gets F_i$, $W_{Gi}\gets H(W_a,j)\oplus s_aT_{Gi}$, $W_{Ei}\gets H(W_b,j')\oplus s_b(T_{Ei}\oplus W_a)$
			\State $W_i\gets W_{Gi}\oplus W_{Ei}$
			\EndIf
			\State $W_i^1\gets W_i^0\oplus R$
		\EndFor
		\For{$i\in \texttt{Outputs}(F)$}  $Y_i\gets W_i$
		\EndFor
		\State \textbf{return} $Y$ 
        \EndProcedure
        \Statex
        \Procedure{\texttt{De}}{$d,Y$} \Comment \texttt{Decoding phase}
		\For{$d_i\in d$}  $y_i \gets d_i\oplus \text{lsb}Y_i$ \label{lst:decode}
		\EndFor 
		\State \textbf{return} $y$
        \EndProcedure
    \end{algorithmic}
\end{algorithm}

 
\section{Our Compatibility Analysis of Garbled Circuit Optimizations}

We conclude this chapter with a useful table which reflects the compatibility of garbled circuit optimizations with each other (see Table \ref{tab:gcoptcompatibility}). \footnotesize{\checkmark} \normalsize and $\mathsf{X}$ stand for compatible and non-compatible, respectively. For the use of external value (Ext. Val.), see Section \ref{sec:GRR2}{}.

\vspace{7mm}
\begin{table}[H]
\centering
\caption{Compatibility of Garbled Circuit Optimization Techniques.}
\label{tab:gcoptcompatibility}
\begin{tabular}{c|c|c|c|c|c|c}
\cellcolor{black!50}                    & \footnotesize{\textbf{P\&P}}          & \footnotesize{\textbf{GRR3}}  & \footnotesize{\textbf{Free \texttt{XOR}}} & \footnotesize{\textbf{GRR2}}          & \footnotesize{\textbf{Fle\texttt{XOR}}}        & \footnotesize{\textbf{Half Gates}} \\ \hline
\footnotesize{\textbf{P\&P}}       &\cellcolor{black!50}                        & \footnotesize{\checkmark}     & \footnotesize{\checkmark}        & \footnotesize{\checkmark (Ext. Val.)} & \footnotesize{\checkmark (Ext. Val.)} & \footnotesize{\checkmark}          \\ \hline
\footnotesize{\textbf{GRR3}}       & \footnotesize{\checkmark}             & \cellcolor{black!50}               & \footnotesize{\checkmark}        & $\mathsf{X}$         & \footnotesize{\checkmark}             & \footnotesize{\checkmark}          \\ \hline
\footnotesize{\textbf{Free \texttt{XOR}}}  & \footnotesize{\checkmark}            & \footnotesize{\checkmark}     &  \cellcolor{black!50}                 & $\mathsf{X}$         & $\mathsf{X}$         & \footnotesize{\checkmark}          \\ \hline
\footnotesize{\textbf{GRR2}}      & \footnotesize{\checkmark (Ext. Val.)} & {$\mathsf{X}$} & $\mathsf{X}$   &         \cellcolor{black!50}               & \footnotesize{\checkmark}             & $\mathsf{X}$      \\ \hline
\footnotesize{\textbf{Fle\texttt{XOR}}}     & \footnotesize{\checkmark (Ext. Val.)} & \footnotesize{\checkmark}     & $\mathsf{X}$   & \footnotesize{\checkmark}             &       \cellcolor{black!50}                 & $\mathsf{X}$      \\ \hline
\footnotesize{\textbf{Half Gates}} & \footnotesize{\checkmark}             & \footnotesize{\checkmark}     & \footnotesize{\checkmark}        & $\mathsf{X}$         &$\mathsf{X}$         &      \cellcolor{black!50}  \\
\multicolumn{7}{c}{ }\\
\multicolumn{7}{c}{{\scriptsize Ext. Val.: External Value (\ref{sec:GRR2})}}\\
\end{tabular}
\end{table}

 

\chapter{Practical Implementations of Yao's Protocol} \label{chap:implementations}
\lhead{Chapter \ref{chap:implementations}. \emph{Practical Implementations of Yao's Protocol}}

Various implementations have been developed so far based on Yao's protocol. Many of them utilize Yao's protocol for MPC applications, although some targets \textit{Private Function Evaluation} (PFE). A comprehensive catalogue of them would have been far from the reach of just a master's thesis work. So, we will explain only some of them which are supposed to be helpful for people to see Yao's protocol in practise. They also reflect the importance of Yao's protocol and the areas it can be applied in the future. First, we will start with introducing some of the generic MPC solutions that use Yao's protocol. We compare those generic implementations in terms of their use of garbled circuit optimizations.  At the end, we will present some real-world applications.

\section{Generic Usage of Yao's Protocol in Practice} \label{sec:GenericUsageofYao'sProtocolinPractice}{}

\subsection{Pipelined Implementation (\texttt{FastGC})}\label{sub:PipelinedImplementationFastGC}{}

The memory required to store the entire garbled circuit is generally a limitation. Huang \textit{et al.} proposed pipelining optimization in their framework in \cite{HEKM11} to reduce the required memory. The garbled circuit generation and evaluation procedures can be done simultaneously, eliminating the need for keeping the entire garbled circuit in memory and the need for preparation of the entire garbled circuit before its transmission to the evaluator, which results in a decrease in total Yao's protocol time. \texttt{FastGC} framework automates pipelined implementation, so that the only need remaining is the construction of the desired circuit  \cite{HEKM11}.

At the beginning of the computation the circuit structure is instantiated by both the garbler and the evaluator. While the protocol is being executed, the generator garbles each gate in topological\footnote{Safety-respecting if the garbling method is fle\texttt{XOR}.} order, and transmits it over the network as soon as it is produced. When a garbled gate is received by the evaluator, it is associated with the corresponding gate of the circuit and evaluated. A gate is eliminated as soon as it has been evaluated, so that the memory use would be minimal. This technique is called \textit{pipelined implementation}. Note that it also reduces total Yao's protocol time of  at the expense that both parties needs to be online at the same time.

\subsection{Garbled RAM} \label{sub:GarbledRAM}{}

The notion of \textit{garbled RAM} was introduced by Lu and Ostrovsky in \cite{LO13}. Gentry \textit{et al.} have later improved it using \textit{identity-based encryption} (IBE)\footnote{Identity based encryption (IBE) is a form of public key encryption (\ref{sub:publickeycryptography}) where a user's public key is his identity. In generic public key cryptosystems, private keys are chosen randomly and public keys are produced from them. However, in IBE the private keys are generated from users' public keys \cite{BF01}. \label{ftnt:ibe}} in \cite{GHRW14} for provable security. It differs from Yao's garbled circuits in that it permits direct garbling of a RAM program, \textit{without converting it into a boolean circuit}. A RAM program whose run-time is $T$ can be converted into a Turing Machine whose run-time is $O(T^3)$ resulting in a boolean circuit of size $O(T^3\texttt{log}T)$, whereas the size and computation time of a garbled RAM program is only proportional to its running time on a RAM \cite{GHRW14}. The inefficiency is even more prominent in the setting of \textit{big data} \cite{GHRW14}. In this case, efficient programs, such as binary search, run in sub-linear time with the size of the data, however their boolean circuit representations run in linear time with the size of the data.

Just like  garbled circuits, garbled RAM includes a garbler who garbles the program, and sends it to the evaluator. Evaluator evaluates the garbled program using the garbled inputs and, unlike the case of Yao's protocol, outputs the actual output of the RAM program. Like the garbled circuits, garbled RAM targets security againist semi-honest adversaries ((\ref{sub:semi-honest}). Gentry \textit{et al.}'s scheme of  garbled RAM is explained in detail below  \cite{GHRW14}.

The notation $P^D (x)$ denotes a RAM program $P$ which accesses a memory containing data $D$ and takes an input $x$. Imagining $D$ as a \textit{huge database} controlled by the evaluator and $P$ as a \textit{database query} that has read or write access to the database and whose parameter is a value $x$ (like $P$ searches $x$ in $D$) would help for understanding the notions. 

A garbled RAM scheme can be used to garble $P$, $D$, $x$ into $\bar{P}$, $\bar{D}$, $\bar{x}$, such that $\bar{P}$, $\bar{D}$, $\bar{x}$ reveals only $P^D(x)$. Furthermore, the sizes of $\bar{P}$, $\bar{D}$, $\bar{x}$ are only proportional to their corresponding plain texts. Similar to Yao's the garbled circuits, garbling $x$ consists of providing a subset of masking values.

A RAM program $P$ can be represented as a colleciton of \texttt{CPU}-Step Circuits which execute a single \texttt{CPU} step. Equation~\eqref{eq:cpustepj} shows the execution of \texttt{CPU} step $j$. The input to the circuit $C_\texttt{CPU}^P$ is the current \texttt{CPU} $\texttt{state}_j$ and a bit $b_j^\texttt{read}$ which resides in the memory location assigned in the previous cycle. Its outputs are an updated $\texttt{state}_{j+1}$, the next reading location $i_(j+1)^\texttt{read}$, a location $i_{j+1}^\texttt{write}$ for writing to (maybe $\bot$), a bit $b_{j+1}^\texttt{write}$ to write into that location. The start of the computation $P^D(x)$ is in the initial state $\texttt{state}_1=x$  and $b_0^\texttt{read}=0$, and it proceeds step-by-step. In each step $j$, first $b_j^\texttt{read}$ is set to $D[i_j^\texttt{read}]$, and if $i_j^\texttt{write}\neq\bot$, $D[i_j^\texttt{write}]$ is set to $b_j^\texttt{write}$. The output of the last \texttt{CPU} step is the output of the computation $y=P^D(x)$ as \texttt{state}. $P$ has \textit{read-only} memory access if it never overwrites any values in memory $D$ (\textit{i.e.}, $i_j^\texttt{write}$ is always $\bot$).

\begin{equation} 
\label{eq:cpustepj}
C_\texttt{CPU}^P (\texttt{state}_j,b_j^\texttt{read},i_j^\texttt{write},b_j^\texttt{write}) = (\texttt{state}_{j+1},i_{j+1}^\texttt{read},i_{j+1}^\texttt{write},b_{j+1}^\texttt{write})
\end{equation}

Gentry \textit{et al.} propose their scheme with security against unprotected memory access (UMA) in which the initial contents of the memory $D$ and the complete memory access pattern of \texttt{MemAccess} (including the contents) may be learned by the intruder, \cite{GHRW14}. They also propose that encrypting the memory contents and applying oblivious RAM is enough for transforming any garbled RAM scheme with UMA security into one providing full security.

\textbf{Read-only Solution.} \label{subsub:ReadonlySolution}{}The garbled memory is made of $\bar{D}[i]$'s, each containing an IBE\textsuperscript{\ref{ftnt:ibe}} secret key $sk_{(i,b)}$ for the public key $(i,b)$ where $i$ is the location and $b$ is the data bit $D[i]$. Another future of $\bar{D}[i]$ is that it can remain and be used by the future programs. The garbled input $\bar{x}_j$ to the \texttt{CPU} step $j$ is the masking value for the $\texttt{state}_j$, and $\bar{x}_0$ is the masking input $\bar{x}$. The \texttt{CPU} step in Equation~\eqref{eq:cpustepj} simply becomes the one in Equation~\eqref{eq:cpustepjro}.

\begin{equation} 
\label{eq:cpustepjro}
C_\texttt{CPU}^P (\texttt{state}_j,b_j^\texttt{read}) = (\texttt{state}_{j+1},i_{j+1}^\texttt{read})
\end{equation}

\eqref{eq:garbledcpucircuitj} shows the garbled circuit $\bar{C}_\texttt{CPU,j}^P$ of the step $j$. The problem with garbling the \texttt{CPU} step $j$ is that the location of $b_j^\texttt{read}$ is not pre-known since it is the output of the previous cycle. Let $\bar{b}_{0j}^\texttt{read}$ denote the masking value of $b_j^\texttt{read}$ for \texttt{FALSE} and $\bar{b}_{1j}^\texttt{read}$ denote the masking value of $b_j^\texttt{read}$ for \texttt{TRUE}. Each garbled step $j$ outputs a translation mapping $\texttt{translate}_{j+1}= (ct_{0(j+1)}, ct_{1(j+1)})$ where $ct_{b(j+1)}=E((i_{j+1}^\texttt{read},b),r_{bj},\bar{b}_{b(j+1)}^\texttt{read})$ calculated\footnote{$r_{bj}$ is the randomization value to provide semantic security.} by using IBE so that the evaluator can only learn the masking value of $D[i_{j+1}^\texttt{read}]$ using the key $\bar{D}[i_{j+1}^\texttt{read}]$. $\bar{b}_{0(j+1)}^\texttt{read}$, $\bar{b}_{1(j+1)}^\texttt{read}$, $r_{0j}$ and $r_{1j}$ are hardcoded in the step circuit $j$ and cannot be learned directly by the evaluator due to the garbling process. $i_{j+1}^\texttt{read}$ is not private since the target is UMA security, and so it does not require a masking value.

\begin{equation} 
\label{eq:garbledcpucircuitj}
(\bar{x}_{j+1},i_{j+1}^\texttt{read},\texttt{translate}_{j+1}) \gets \bar{C}_\texttt{CPU,j}^P(\bar{x}_j,\bar{b}_{bj}^\texttt{read})  
\end{equation}

Each garbled cycle $j$ starts with the decryption of $ct_{bj}$ (the evaluator may know which one to decrypt due to UMA security) to get $\bar{b}_{bj}^\texttt{read}$, except for the first cycle where $\bar{b}_{bj}^\texttt{read}=\bot$. The last cycle directly outputs $y = P^D(x)$.

\textbf{Writing to the Memory.} \label{subsub:WritingtotheMemory}{}Similar to the  read-only case, the garbled memory is made of $\bar{D}[i]$'s, each containing an \textit{timed IBE} secret key $sk_{(u,i,b)}$ for the public key $(u,i,b)$ where $u$ is the cycle that $i$ is written last time. The full step given in Equation~\eqref{eq:cpustepj} needs to be evaluated. Equation~\eqref{eq:garbledcpuwritecircuitj}  shows the garbled circuit $\bar{C}_\texttt{CPU,j}^P$ of the step $j$. Unlike the read-only case, each step $j$ writes $sk_{(j,i,b)}$ to the garbled memory address $i_j^\texttt{write}$ (if they are not $\bot$), and outputs $sk_{(j+1,i,b)}$ and $i_{j+1}^\texttt{write}$ for writing in the next cycle. Each garbled step $j$ outputs a translation mapping $\texttt{translate}_{j+1}= (ct_{0(j+1)}, ct_{1(j+1)})$ where $ct_{b(j+1)}=E((u_{j+1},i_{j+1}^\texttt{read},b),r_{bj},\bar{b}_{b(j+1)}^\texttt{read})$ calculated by using timed IBE. Here, the assumption is that there exists a polynomial size circuit \texttt{WriteTime} such that $u_{j+1} = \texttt{WriteTime}(j,\bar{x}_j,i_{j+1}^\texttt{read})$, and step $j$ can call it. Just like the read-only case, the evaluator can only learn the masking value of $D[i_{j+1}^\texttt{read}]$ using the key $\bar{D}[i_{j+1}^\texttt{read}]$. $\bar{b}_{0(j+1)}^\texttt{read}$, $\bar{b}_{1(j+1)}^\texttt{read}$, $r_{0j}$ and $r_{1j}$ are hardcoded in the step circuit $j$ and cannot be learned directly by the evaluator due to the garbling process. $i_{j+1}^\texttt{read}$ and $i_{j+1}^\texttt{write}$ are not private since the target is UMA security.

\begin{equation} 
\label{eq:garbledcpuwritecircuitj}
(\bar{x}_{j+1},i_{j+1}^\texttt{read},\texttt{translate}_{j+1},i_{j+1}^\texttt{write},sk_{(j+1,i,b)}) \gets \bar{C}_\texttt{CPU,j}^P(\bar{x}_j,\bar{b}_{bj}^\texttt{read},i_j^\texttt{write},sk_{(j,i,b)})  
\end{equation}

\textbf{Full Security.} \label{subsub:FullSecurity}{}Gentry \textit{et al.} propose that any garbled RAM scheme that only provides UMA security and only supports program executions with \texttt{WriteTime} calls can be transformed into a fully secure garbled RAM scheme for arbitrary programs \cite{GHRW14}. This transformation uses \textit{oblivious RAM} (ORAM)\footnote{Oblivious RAM (ORAM), first proposed by Goldreich and Ostrovsky \textit{et al.} \cite{GO96}, permit a user to hide its access pattern to a remote storage. Although  the physical storage locations accessed can be observed by an adversary,  it is ensured  by ORAM that anything about the real access pattern may not be learned  \cite{SDS+12}.} to first compile the original program $P$ into a new program $P^*$ that stores/accesses its memory using ORAM. This ensures that the memory contents and access pattern of the compiled program do not reveal anything about those of the original program. Some ORAM schemes already ensure that the compiled program provides \texttt{WriteTime} calls.

\subsection{\texttt{JustGarble}} \label{sub:JustGarble}{}

In \cite{BHKR13}, Bellare \textit{et al.} proposed \texttt{JustGarble} framework, which targets optimized garbling of any circuit. It is entirely open-source and can be freely downloaded from \url{http://cseweb.ucsd.edu/groups/justgarble}. It implements Ga (P\&P (\ref{sec:PP})), GaX (Free \texttt{XOR} (\ref{sec:freeXOR}) without GRR3 (\ref{sec:GRR3})), and GaXR (Free \texttt{XOR} with GRR3), using constant key 128-bit \texttt{AES} as the \texttt{DKC} (\ref{part:dkcschemes}) as in Equation~\eqref{eq:encschemeaesconstkey}. It works both ways: garble a boolean circuit, and evaluate a garbled circuit.

\texttt{JustGarble} uses a circuit representation called Simple Circuit Description (SCD). It is based on the circuit formulation from \cite{BHR12}. An SCD file consists of values $n$, $m$, $q$, and arrays $A$, $B$, and $G$. If $G$ is not present the file is a topological circuit representation. In \texttt{JustGarble}, there are modules for building circuits, garbling boolean circuits, and evaluating garbled circuits. The Build module is useful for constructing circuits, permitting working at the individual gate level or higher. SCD files are written with constructed circuits. The Garble module is utilized for realizing the \textsc{\texttt{Gb}} algorithm of the three  garbling schemes given. Garble takes a circuit $f = (n, m, q, A, B, G)$ described in an SCD as input and outputs the garbled tables $P$ that compose of the related garbled circuit $F = (n, m, q, A, B, P )$. The inputs to the Evaluate module are a topological circuit$ \bar{f} = (n,m,q,A,B)$, the garbled tables $P$ needed  for evaluating, and a garbled input $X$. The garbled output $Y$ is produced. JustGarble also composes of procedures to realize \texttt{De}, mapping the garbled output $Y$ to the plain output $y$ \cite{BHKR13}.

The \texttt{JustGarble} implementation of GaXR (Free \texttt{XOR} (\ref{sec:freeXOR}) with GRR3 (\ref{sec:GRR3})) for 36.5K gate optimized \texttt{AES} boolean circuit whose 82\% are \texttt{XOR} gates has resulted in  5.40 bytes per gate (bpg) as the size, 35.0 cycles per gate (cpg) as the evaluation time, and 63.3 cpg as the garbling time. The JustGarble implementation of GaX for the same circuit, however, has yielded 23.2 cpg as the evaluation time, 55.6 cpg as the garbling time, and 11.5 as the size \cite{BHKR13}. (With a 3.201 GHz processor, evaluating the garbled circuit is 7.25 nsec/gate and garbling it is 17.4 nsec/gate.)

\subsection{\texttt{ABY}} \label{sub:ABY}{}

\texttt{ABY} is a framework for 2PC, proposed by Demmler \textit{et al.} in \cite{DSZ15}. Most of the time, a mixture of MPC primitives (GMW protocol (\ref{sec:gmwprotocol}), Yao, HE (\ref{sec:homoenc})$\ldots$) may yield more efficient implementations than what would have been if just one of them is used.  Based on this idea, \texttt{ABY} uses \textbf{A}rithmetic sharing, \textbf{B}oolean sharing, and \textbf{Y}ao sharing (\ref{sec:secretshare}). The framework aims security in the semi-honest model (\ref{sub:semi-honest}). \texttt{ABY} works like a virtual machine, and high-level languages can be compiled to it. Variables may be either in Cleartext (i.e. one of the parties knows its value, e.g. inputs and outputs) or secret shared among the two parties. \texttt{ABY} also allows efficient conversion between the different types of sharings. The user of the framework may decide which sharings to be used depending on the application.


\subsection{\texttt{Obliv-C}}

\texttt{Obliv-C} is built by Zahur and Evans as an extension of C programming language with secure computation infrastructure \cite{ZE15}. It supports various C features like \texttt{pointers}, \texttt{typedef}, \texttt{struct}, \textit{etc.}, and provides new data types and constructions so that programs would run on private inputs. It is especially designed for scalable MPC protocols, and to enhance research on new MPC techniques by easing implementation such that just writing a new library is enough instead of building a new compiler for each technique.  The source code for \texttt{Obliv-C} can be found at \url{https://oblivc.org}.

\subsection{\texttt{ObliVM}}

Liu \textit{et al.} proposed \texttt{ObliVM} as a programming framework for MPC \cite{DBLP:conf/sp/2015}. It offers a domain-specific language (\texttt{ObliVM-lang}) useful for compilation of programs into suitable representations required for MPC protocols. It also provides high-level programming constructions for MPC infrastructure which can be adapted by non-specialist programmers on security as well. The source code for \texttt{ObliVM} can be found at \url{http://oblivm.com}.

\subsection{\texttt{Frigate}} \label{sub:Frigate}{}

\texttt{Frigate} is designed by Mood \textit{et al.} as a compiler and a circuit interpreter for MPC  \cite{MGC+16}. It can implement any function that can be written as a boolean circuit and run any MPC primitive that operates on boolean circuits. \texttt{Frigate} permits the use of C-like language with  constructs and operators specifically designed for representing  Boolean circuit efficiently. To improve the efficiency,  the compiler is designed to favor \texttt{XOR} gates, utilizing structures like  Boyar \textit{et al.}'s full adder with four \texttt{XOR} and one \texttt{AND} \cite{BPP00}. \texttt{Frigate} is also \textit{significantly fast} in terms of compilation, interpretation and execution times. The source code for \texttt{Frigate} can be found at \url{https://bitbucket.org/bmood/frigaterelease}.

\subsection{Comparison Based on Garbling Optimizations Used}

Now, we compare the generic frameworks for Yao's protocol based on their use of garbled circuit optimizations (see Table \ref{tab:comparefwgcopt}). \footnotesize{\checkmark} \normalsize and $\mathsf{X}$ stand for compatible and non-compatible, respectively. 

\vspace{7mm}
\begin{table}[H]
\centering
\caption{Comparison of Generic Frameworks Techniques Based on Their Use of Garbled Circuit Optimizations.}
\label{tab:comparefwgcopt}
\begin{tabular}{c|c|c|c|c|c|c}
                  & \footnotesize{\textbf{P\&P}}          & \footnotesize{\textbf{GRR3}}  & \footnotesize{\textbf{Free \texttt{XOR}}} & \footnotesize{\textbf{GRR2}}          & \footnotesize{\textbf{Fle\texttt{XOR}}}        & \footnotesize{\textbf{Half Gates}} \\ \hline
\footnotesize{\textbf{\texttt{JustGarble} (2013) \cite{BHKR13}}}       &  \footnotesize{\checkmark}  & \footnotesize{\checkmark}     & \footnotesize{\checkmark}        & $\mathsf{X}$ & $\mathsf{X}$ & $\mathsf{X}$          \\ \hline
\footnotesize{\textbf{\texttt{ABY} (2015) \cite{DSZ15}}}       & \footnotesize{\checkmark}             &         \footnotesize{\checkmark}         & \footnotesize{\checkmark}        & $\mathsf{X}$         & $\mathsf{X}$             & $\mathsf{X}$         \\ \hline
\footnotesize{\textbf{\texttt{Obliv-C} (2015) \cite{ZE15}}}  & \footnotesize{\checkmark}            & \footnotesize{\checkmark}     &     \footnotesize{\checkmark}            & \footnotesize{\checkmark}         & \footnotesize{\checkmark}         & \footnotesize{\checkmark}          \\ \hline
\footnotesize{\textbf{\texttt{ObliVM} (2015) \cite{DBLP:conf/sp/2015}}}      & \footnotesize{\checkmark} & \footnotesize{\checkmark} & \footnotesize{\checkmark}   &     $\mathsf{X}$                   & $\mathsf{X}$               & $\mathsf{X}$      \\ \hline
\footnotesize{\textbf{\texttt{Frigate} (2016) \cite{MGC+16}}}     & \footnotesize{\checkmark} & \footnotesize{\checkmark}     & \footnotesize{\checkmark}   & \footnotesize{\checkmark}             &    \footnotesize{\checkmark}                & \footnotesize{\checkmark}      \\ 
\end{tabular}
\end{table}

\texttt{Obliv-C} and \texttt{Frigate} make the use of any garbled circuit optimization possible since they permit the alterations of garbling schemes although those garbling circuit techniques are not built-in. On the other hand, \texttt{JustGarble}, \texttt{ABY}, and  \texttt{ObliVM} do not allow changing the built-in garbling constructions, therefore, is limited for the use of state-of-the-art garbled circuit optimization techniques. All of the frameworks allow compilations optimized for reducing the number of odd gates. We can deduce that \texttt{Frigate} is the optimum for working with garbled circuits since it offer maximum optimization options while being the most efficient one.

\section{Real-World Applications} \label{sec:SomePracticalApplicationsinRealWorld}

We give two real-world examples indicating the importance of Yao's protocol in practise.

\subsection{Secure Computation of Satellite Collusion Probabilities} \label{sub:SecureComputationofSatelliteCollusionProbabilities}{}

Satellite operators are very eager to protecting their satellites since they are extremely costly. One of the issues that operators are interested in is preventing collisions with other satellites. However, the operators also want to keep the trajectories of their satellites private, which makes coordination between different operators difficult. Hemenway \textit{et al.} proposed an 2PC framework that combines GMW protocol (\ref{sec:gmwprotocol})  and Yao's protocol for high-precision computation of satellite collusion probabilities in \cite{HLOW16}.  The framework does not target just the semi-honest model (\ref{sub:semi-honest}) since in the case of satellite operators, it does not provide sufficient security. Instead, first, they prove the security of the protocol in semi-honest (\ref{sub:semi-honest}) setting. Then, they strengthen their construction by using standard arithmetic \texttt{MAC}s against malicious adversaries (\ref{sub:malicious}).

For the sake of simplicity, the model of each satellite is a spherical object on a linear path in any short time window. Each satellite may deviate from its position, \textbf{p}, and the distribution of these deviations are assumed to be covariance matrix$^2$ \textbf{C}. The private input of a satellite $a$ includes four parts: its position \textbf{p}$_a$ in $\mathbb{R}^3$, its velocity \textbf{v}$_a$ in $\mathbb{R}^3$, the covariance matrix \textbf{C}$_a$ in $\mathbb{R}^{3\times3}$, and its radius $R_a$ in $\mathbb{R}$. The algorithm which needs to be calculated securely for satellites $a$ and $b$ is the conjunction analysis calculation, which returns the collision probability $p$ (see Algorithm \ref{alg:conjunctionanalysis}). 

\begin{algorithm}
  	\scriptsize
	\caption{The conjunction analysis calculation proposed in \cite{HLOW16}.}
	\label{alg:conjunctionanalysis}
	\begin{algorithmic}[1] 
		\State \textbf{Inputs: } $\{\textbf{v}_i,\textbf{C}_i,\textbf{p}_i,R_i\}_{i\in a,b}$
		\State $\textbf{v}_r\gets \textbf{v}_b - \textbf{v}_a$, $\textbf{i}\gets \dfrac{\textbf{v}_r}{|\textbf{v}_r|}$, $\textbf{j}\gets \dfrac{\textbf{v}_b\times \textbf{v}_a}{|\textbf{v}_b\times \textbf{v}_a|}$, $\textbf{k}\gets\textbf{i}\times\textbf{j}$, $\textbf{Q}\gets \begin{bmatrix} \textbf{j} & \textbf{k} \end{bmatrix}$, $\textbf{C} \gets \textbf{Q}^T (\textbf{C}_a+\textbf{C}_b)\textbf{Q}$ 
		\State $(\textbf{u},\textbf{v})\gets \texttt{Eigenvectors}(\textbf{C})$, $(\sigma_x^2,\sigma_y^2)\gets \texttt{Eigenvalues}(\textbf{C})$, $\sigma_x\gets\sqrt{\sigma_x^2}$, $\sigma_y\gets\sqrt{\sigma_y^2}$
		\State $\textbf{u}\gets\dfrac{\textbf{u}}{|\textbf{u}|}$, $\textbf{v}\gets\dfrac{\textbf{v}}{|\textbf{v}|}$, $\textbf{U}\gets\begin{bmatrix} \textbf{u} & \textbf{v} \end{bmatrix}$, $\begin{bmatrix} x_m \\ y_m \end{bmatrix}\gets \textbf{U}^T\textbf{Q}^T(\textbf{p}_b-\textbf{p}_a)$
		\State $p \gets \dfrac{1}{2\pi\sigma_x\sigma_y} \displaystyle\int_{-R}^R\int_{-\sqrt{R^2-x^2}}^{\sqrt{R^2-x^2}}f(x,y)dydx$ where $f(x,y)= \texttt{exp}\left[\dfrac{-1}{2}\left[\left(\dfrac{x-x_m}{\sigma_x}\right)^2+\left(\dfrac{y-y_m}{\sigma_y}\right)^2\right]\right]$ \label{lst:integral}
		\State \textbf{return } $p$
	\end{algorithmic}
\end{algorithm}

Hemenway \textit{et al.} propose GMW protocol (\ref{sec:gmwprotocol}) for computing integer addition and multiplications \cite{HLOW16}. To compute comparison and shift operations, they are represented as Boolean circuits and then evaluated using Yao's protocol. For compatibility with GMW protocol the garbled circuit must take secret inputs of both parties and the output of the gate must be computed as an arithmetic secret sharing (\ref{sec:secretshare}) among both sides.

\begin{itemize}
\item{A shift operation is computed as follows: $x_0$ and $x_1$ are Alice and Bob's arithmetic shares, respectively. $(x_0 + x_1)$ needs to be shifted by an amount $N$ which is known publicly. This can be accomplished by using Yao's protocol to compute Algorithm \ref{alg:shiftop}, where Alice is the garbler, and Bob is the evaluator. Bob uses OT (\ref{sec:oblivioustrans}) to get the masking values for his inputs.}
\end{itemize}

\begin{algorithm}
	\caption{The shift operation computation  proposed in \cite{HLOW16}.}
	\label{alg:shiftop}
	\begin{algorithmic}[1] 
		\State \textbf{Hardwired:} $M = 2m$ and $c$, which are a modulus and a shift constant, respectively.
		\State \textbf{Inputs:} $x_0$ and $x_1$, held by Alice and Bob, respectively. In addition a random $R$ is provided by Alice.
		\State $x \gets x_0 + x_1\text{ } (\texttt{mod }M )$ using standard $m$-bit addition circuit.
		\State $y \gets x >> c$ by dropping $c$ rightmost wires.
		\State \textbf{Return:} $z_1 \gets y + R\text{ } (\texttt{mod }M )$ to Bob. She sets $z_0 = -R \text{ }(\texttt{mod }M )$ for herself.
	\end{algorithmic}
\end{algorithm}

\begin{itemize}
\item{A shift comparison is  computed as follows: $x_0$ and $x_1$ are Alice and Bob's arithmetic shares, respectively.  They would like to detect whether $(x_0 + x_1)$ is positive or not. This can be done by using Yao's protocol to compute Algorithm  \ref{alg:comparisonop}, where Alice is the garbler, and Bob is the evaluator. Bob uses OT (\ref{sec:oblivioustrans}) to get the masking values for his inputs.}
\end{itemize}

\begin{algorithm}
	\caption{The comparison operation computation proposed in \cite{HLOW16}.}
	\label{alg:comparisonop}
	\begin{algorithmic}[1] 
		\State \textbf{Hardwired:} $M = 2m$, which is a modulus.
		\State \textbf{Inputs:} $x_0$ and $x_1$, held by Alice and Bob, respectively. In addition a random $R$ is provided by Alice.
		\State $x \gets x_0 + x_1\text{ } (\texttt{mod }M )$ using standard $m$-bit addition circuit.
		\State $b\gets sgn(x)$
		\State \textbf{Return:} $z_1 \gets b + R\text{ } (\texttt{mod }M )$ to Bob. She sets $z_0 = -R \text{ }(\texttt{mod }M )$ for herself.
	\end{algorithmic}
\end{algorithm}

Now, we return the computation of Algorithm \ref{alg:conjunctionanalysis}. In the rest of this section, we provide the methods proposed by Hemenway \textit{et al.} for the implementation of functions in Algorithm \ref{alg:conjunctionanalysis} \cite{HLOW16}.

\textbf{Circuit Representation for Division:} Integer division is implemented by repeated subtractions.

\textbf{Circuit Representation for} \texttt{exp}(): The function \texttt{exp}() must be implemented by representing it as a degree-24 Taylor series. Then the Taylor coefficients can be hard-coded constants in the circuit \cite{HLOW16}.

\textbf{Circuit Representation for} $\sqrt{\cdot}$: Iterative Babylonian Algorithm can be used to approximate a square root. The Babylonian Algorithm computes Equation~\eqref{eq:babylonianalg} on an input $S$, and an initial estimate $x_0$ \cite{HLOW16}.

\begin{equation} 
\label{eq:babylonianalg}
 x_{n+1}=\dfrac{1}{2}\left( x_n+ \dfrac{S}{x_n}\right)
\end{equation}

The double integral on Line \ref{lst:integral} of Algorithm \ref{alg:conjunctionanalysis} can be written as in Equation~\eqref{eq:integral} where $g(x)$ is a sum of \texttt{erf}s. Simpson's Rule approximation to this integral (\textit{i.e.}, using arcs of parabola) is suggested by Alfano in \cite{Alf05}.

\begin{equation} 
\label{eq:integral}
 p=\dfrac{3}{\sqrt{8\pi}\sigma_x}\displaystyle\int_{-R}^R g(x)dx
\end{equation}

\textbf{Circuit Representation for} \texttt{erf}(): approximate $1 - \texttt{erf}(x)$ using the degree 96 rational function in Equation~\eqref{eq:erfapprox} where $a_1 = .3275911$, $a_2 = .254829592$, $a_3 = .0092705272$, $a_4 = .0001520143$, $a_5 = .0002765672$, and $a_6 = .0000430638$.

\begin{equation} 
\label{eq:erfapprox}
 1 - \texttt{erf}(x) \approx \dfrac{1}{\left(1+a_1x+a_2x^2 +a_3x^3 +a_4x^4 +a_5x^5 +a_6x^6\right)^{16}}
\end{equation}

Hemenway \textit{et al.} demonstrate that their framework is highly efficient. The collision probability calculation scheme proposed requires numerical estimation of a complicated integral. The work of Hemenway \textit{et al.} proves that evaluating very complex functions is now possible by using MPC technology \cite{HLOW16}.

\subsection{Privacy-Preserving Data Mining} \label{sub:ppdatamining}{}

\textit{Privacy-preserving data mining} deals with the problem of how to run data mining algorithms on private data \cite{LP08}. Mainly, privacy-preserving data mining is applied to two classic settings \cite{LP08}: 
\begin{enumerate}
\item{Instead of a single party having the whole data set, two or more parties hold different parts of it. Running a data mining algorithm on the union of the parties' databases is aimed while each party's input is being kept private \cite{LP08}.}

\item{Some part of statistical data that needs to be released may be confidential. Hence, it can be first altered so that }
\begin{enumerate}
\item{no one's privacy is compromised by it,}
\item{Data mining algorithms can be run on the modified data set to obtain meaningful results \cite{LP08}.}
\end{enumerate}
\end{enumerate}

Although both privacy problems are important, we will only deal with the first one where MPC techniques suit better. An example of the first type problem occurs in the field of medical research \cite{LP08}. A group of hospitals would like to mine their patient data jointly for the medical research purposes but they also need to keep their patients' personal data private. Another example would be a cooperation scenario of intelligence agencies. These agencies cannot grant each other free access to their confidential databases because of the high security standards they must obey \cite{LP08}.

The relationship of privacy preserving techniques and MPC is so wide that we cannot cover it here comprehensively. Instead here we will examine and explain common notions of classification problem and ID3 algorithm and their relationships with MPC.

\textbf{Classification problem.} The input of a \textit{classification problem} is a database structured such that each of its rows is a \textit{transaction} and each of its columns is an \textit{attribute} which may have different values (\textit{e.g.}, each row may be a patient, and each column may be a different type of symptoms that is found in the patient) \cite{LP08}. One of those attributes in the database is the main one, named as the class attribute (\textit{e.g.}, it may represent whether the patient has lung cancer or not) \cite{LP08}. We aim to use the database for prediction of the class of a new transaction by examining only its non-class attributes  \cite{LP08}.

Another example would be credit risk analysis of a bank that wishes to identify which customers are likely to be profitable before giving them a loan \cite{LP00}. Then the class attribute is defined as \texttt{Profitable-customer} (its values may be \texttt{YES} or \texttt{NO}) by the bank. The database attributes used for prediction include: \texttt{Home-Owner}, \texttt{Income}, \texttt{Years-of-Credit}, and \texttt{Other-Delinquent-Accounts}. In order to ensure proper decision making, various rules are defined by the bank. For example \cite{LP00}:

\textbf{If} ($\texttt{Other-Delinquent-Accounts} = 0$) and ($\texttt{Income} > 30k$ or $\texttt{Years-of-Credit} > 3$) 
\textbf{then} \texttt{Profitable-customer} = \texttt{YES} [\texttt{accept credit-card application}]

The collection of those rules that cover all possible transactions can be used for classification of a customer as profitable or not. The classification may include a probability of error \cite{LP00}.

\textbf{Decision tree.} Being a rooted tree, a \textit{decision tree} has internal nodes, each corresponding to an attribute, and the edges leaving each node, corresponding to the possible values of the attribute \cite{LP08}. The tree also has leaves, each containing the expected class value for a transaction that has the attribute values in the path from the root to that leaf. By using a decision tree, the class of a new transaction can be predicted by following the nodes from the root until the leaf. \cite{LP08}.

\begin{figure}
 
\centering
\label{fig:credit}
 
\begin{tikzpicture}
  [
    grow                    = right,
    sibling distance        = 6em,
    level distance          = 10.6em,
    edge from parent/.style = {draw, -latex},
    every node/.style       = {font=\footnotesize},
    sloped
  ]
 
  \node [env] {\texttt{Other-}\\\texttt{Delinquent-}\\\texttt{Accounts}}
    child { node [env] {\texttt{Profitable-}\\\texttt{customer}\\ $=$ \texttt{NO}}
      edge from parent node [below] {$>0$?} }
    child { node [env] {\texttt{Income}}
      child { node [env] {\texttt{Years-of-}\\\texttt{Credit}}
        child { node [env] {\texttt{Profitable-}\\\texttt{customer}\\ $=$ \texttt{NO}}
          edge from parent node [below] {$<1$ year?} }
        child { node [env] {\texttt{Profitable-}\\\texttt{customer}\\ $=$ \texttt{NO}}
          edge from parent node [above] {1-3}
                           node [below] {years?} }
        child { node [env] {\texttt{Profitable-}\\\texttt{customer}\\ $=$ \texttt{YES}}
                edge from parent node [above] {$>3$ years?} }
        edge from parent node [below] {$<30k?$} }
      child { node [env] {\texttt{Years-of-}\\\texttt{Credit}$\ldots$}
              edge from parent node [above] {$>30k?$}
                }
       edge from parent node [above] {0?} };
\end{tikzpicture}

\caption{A decision tree for credit eligibility.}

\end{figure}
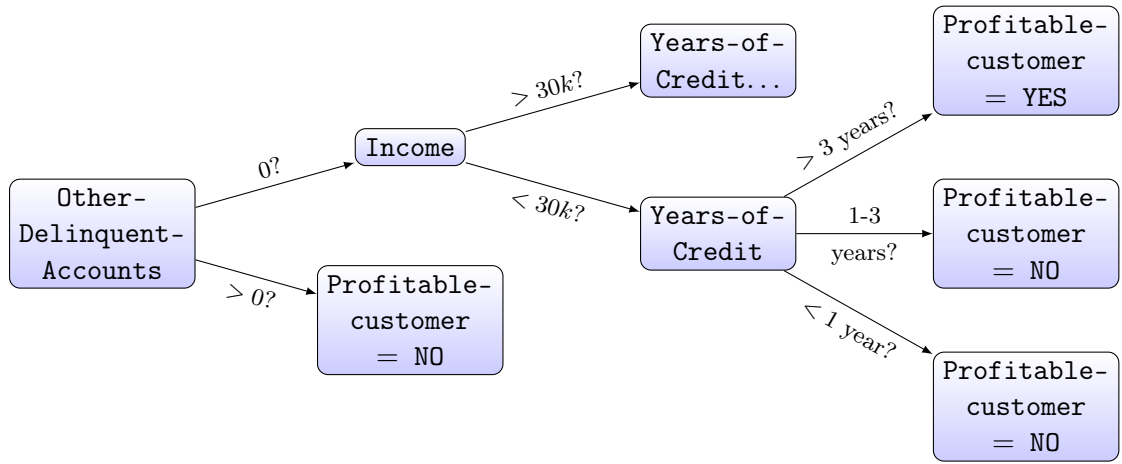

Figure 5.1 shows an example decision tree for identifying profitable customers as in the previous scenario. However, it reflects only a small portion of the tree. The whole tree would have many more nodes, edges, and leaves.

\textbf{ID3 Algorithm.} One of the well-known ways for designing decision trees is the use of \textit{ID3 algorithm} \cite{LP08}. The construction of the tree starts from the root node, goes top-down  recursively. At each node the attribute is chosen based on its ability of  classifying the transactions on its own. If an attribute is chosen for a node, the remaining transactions are partitioned by it, resulting in a smaller database which composes of the related transactions \cite{LP08}.

The main principle of ID3 is choosing the attribute which is best at predicting the class of the transaction. This is done by searching the attribute that decreases the information of the class to the maximum degree \cite{LP08}. Namely, by choosing the attribute maximizing the \textit{information gain}, which is the difference between the entropies of the class attribute for all transactions and for ones having the same value for a give attribute \cite{LP08}. The resulting decision tree is a smaller one consistent with the database due to the greedy algorithm used in searching \cite{LP08}.

\textbf{Privacy preserving distributed computation of ID3.}  We include a setting involving two parties, each having a database with different transactions to which the same set of attributes applies \cite{LP08}. The parties aim at computing a decision tree of the union of their databases by using the ID3 algorithm \cite{LP08}. Lindell and Pinkas describe an efficient privacy preserving protocol to solve this problem in \cite{LP00}.

According to Lindell and Pinkas, direct application of Yao's protocol faces some major problems, mainly the large sizes of input databases require too many OTs (\ref{sec:oblivioustrans}), resulting in huge communication and computation costs \cite{LP08}. Moreover, the boolean circuit conversion of ID3 results in a very large circuit, because of myriad repetitions of information gain calculation which is the basic step of the algorithm \cite{LP08}.

Lindell and Pinkas observe that MPC of each node can be done separately \cite{LP00,LP08}. Starting with the root node, for each node a secure computation is invoked. Its output is revealed to both parties and the computation goes with the next node in the path. This does not compromise the protocol security since the assigned attribute to each node is also a part of the final output. Just like the non-privacy preserving implementation of ID3, both parties separately partition the rest of their transactions after an attribute is assigned to a node. This way, Lindell and Pinkas reduce the whole protocol to proper attribute assignments for node, namely the ones resulting in the highest information gains \cite{LP08}. They also show how to apply Yao's protocol to proper attribute assignment \cite{LP08}.


\chapter{Private Function Evaluation} 

\label{Chapter6} 

\lhead{Chapter 6. \emph{Private Function Evaluation}} 


Consider the case that one invents an algorithm which can be used for efficiently diagnosing various diseases based on some information about a person's general health \cite{Pil15}. It is obvious that this algorithm would be precious, and healthcare institutions would volunteer to pay millions in order to use it. However, the inventor of the algorithm would prefer keeping it as a secret since he is regularly payed for it a lot of money. The problem is that medical institutions generally prefer keeping their patients' data private, preventing them from just giving it to the algorithm owner. Here the following question might be asked: How can those parties compute an algorithm which is known by only one of the parties while its input is known by only the other one? This problem is known as \textit{private function evaluation} (PFE) \cite{Pil15}. The problem may also be widened to involve the case that the algorithm owner may also have his private inputs.

PFE is a special case of MPC in which $n$ participants needs to compute a private function $f$ using their private inputs ($x_1,\ldots, x_n)$), resulting in $f(x_1,\ldots, x_n)$. One of the parties $P_1$ holds a boolean circuit $\mathcal{C}_f$ of the function $f$, while each party $P_i$  holds a private input $x_i$, and the parties aim to learn only the output of the circuit $\mathcal{C}_f(x_1,\ldots, x_n)$ while $f$ or all other partys' inputs remain unknown to each of them except for $P_1$ who already knows $f$ \cite{MS13}. The difference of this scheme from the standard MPC setting is that here the function $f$ and its boolean circuit representation $\mathcal{C}$ are not known publicly. There are many situations where such a PFE scheme would be useful, \textit{e.g.} the ones where the function itself contains private information, or reveals security weaknesses, or the ones where service providers may prefer hiding their function or its specific implementation as their Intellectual Property. Design of efficient special or generic PFE protocols is considered in a variety of papers in literature \cite{MS13}. 

Most generic PFE solutions target the MPC of a \textit{universal circuit} $U_g$ taking the circuit $\mathcal{C}$ with a number of gates less than $g$ and the inputs $x_1,\ldots, x_n$ of parties  as input, and outputing $f(x_1,\ldots, x_n)$. The works based on this approach mainly aim to reduce the size of universal circuits, and to optimize their implementations with the help of various MPC techniques, such as Yao's protocol. However, they have a main source of inefficiency  the massive sizes of known universal circuits. The complexity in their designs and implementations also increases the need for searching better alternatives.

In this section, we will explain the concepts and constructions for PFE proposed by Mohassel and Sadeghian in \cite{MS13}, especially for two-party case where Yao's protocol is involved.\footnote{For a clear explanation of multi-party case where GMW protocol is privately evaluated, we must refer the reader to \cite{Pil15}.} The target security is in the semi-honest (\ref{sub:semi-honest}) setting. Their work remains the most efficient PFE scheme to this date.

\section{Mohassel and Sadeghian's Generic PFE Scheme \cite{MS13}}\label{sec:genericpfe}

Mohassel and Sadeghian present a generic PFE framework in \cite{MS13}. In addition to the private inputs of parties which is hidden by any proper MPC scheme, hiding the topology of a boolean circuit $\mathcal{C}$ and the functionality of its gates suffices for hiding a circuit completely \cite{MS13}.

There are three types of information that Mohassel and Sadeghian's PFE scheme does not intend to hide about a circuit \cite{MS13}: 

\begin{enumerate}
\item{The number of its inputs,}
\item{The number of its outputs,}
\item{The number of its gates.}
\end{enumerate}

\begin{figure}
\includegraphics[height=5cm, angle=0]{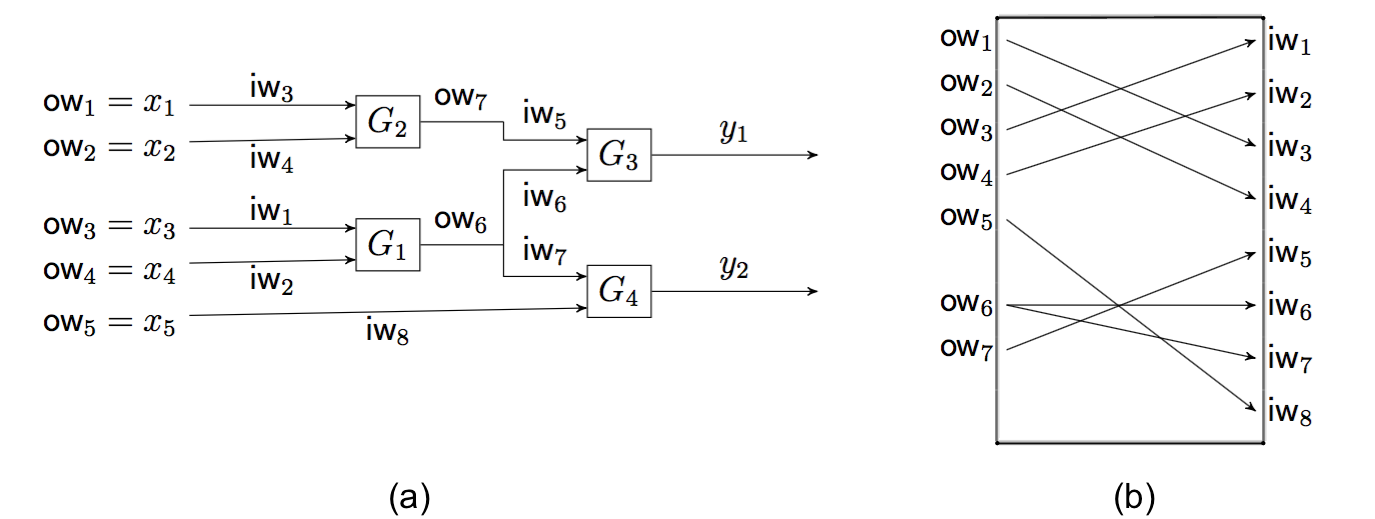}
\caption{(a) An example circuit \cite{Pil15}. (b) The mapping of the circuit \cite{Pil15}.}
\label{fig:mapping}
\end{figure}

Mohassel and Sadeghian suggest two different functionalities that make up the complete task of PFE  \cite{MS13}:

\begin{enumerate}
\item{\textbf{Circuit Topology Hiding} (\texttt{CTH}) \textbf{Functionality.} The full description of the topology of a circuit $\mathcal{C}$ can be accomplished with the use of a mapping $\pi_{\mathcal{C}} : \texttt{OW}\rightarrow \texttt{IW}$. Let $g$, $n$ and $m$ denote the size, the number of inputs and the number of outputs of $\mathcal{C}$, respectively. \texttt{OW} (outgoing wires) is the union of the input wires of the circuit and the output wires of its non-output gates: $ \{ \texttt{ow}_1 = x_1,\ldots,\texttt{ow}_n = x_n,\texttt{ow}_{n+1}=\texttt{Output}(G_1),\ldots,\texttt{ow}_{n+g-m}=\texttt{Output}(G_{g-m})\}$. \texttt{IW} (incoming wires) is the set of input wires to all the gates in the circuit: $\{\texttt{iw}_1,\ldots,\texttt{iw}_{2g}\}$. $\pi_{\mathcal{C}}$ maps $i$ to $j$ (\textit{i.e.}, $\pi_{\mathcal{C}}(i) \rightarrow j)$, if and only if $\texttt{ow}_i \in \texttt{OW}$ and $\texttt{iw}_j \in \texttt{IW}$ correspond to the same wire in the circuit $\mathcal{C}$. Because an outgoing wire can correspond to more than one incoming wire, $\pi_{\mathcal{C}}$ is rarely a function. However, its inverse $\pi_{\mathcal{C}}^{-1}$ is a function since a wire can be either an output of only one gate or an input. Figure \ref{fig:mapping} shows an example circuit (a) and its mapping $\pi_{\mathcal{C}}$ (b). The main target of the \texttt{CTH} \textit{functionality} is the oblivious application of this mapping $\pi_{\mathcal{C}} : \texttt{OW}\rightarrow \texttt{IW}$.}
\end{enumerate}

It is useful to include a computation of the number of possible mapping since it is directly related to the security of the PFE scheme. Although one may expect the number of possible mappings to be $M^N$ due to the ability of any $\texttt{ow}_i$ to go to any $\texttt{iw}_j$, the exact value is smaller since an $\texttt{ow}_i$ must go at least one $\texttt{iw}_j$. Since $\pi^{-1}$ is an onto function, computing the number of possible onto functions suffices. Applying the inclusion-exclusion principle, we get Equation~\eqref{eq:numberofmappings} which shows the number ($\rho$) of possible mappings for a circuit where \texttt{OW} has $M$ elements and \texttt{IW} has $N$ elements \cite{Maz10}.

\begin{equation} 
\label{eq:numberofmappings}
\rho=\sum_{i=0}^{M} (-1)^i {M\choose i} (M-i)^N
\end{equation}

%

\begin{enumerate}
\setcounter{enumi}{1}
\item{\textbf{Private Gate Evaluation} (\texttt{PGE}) \textbf{Functionality.} The \texttt{PGE} \textit{functionality} deals with hiding the functionality of each gate in a circuit. It can be seen as a black-box gate mechanism where only one of the parties ($P_1$) knows its functionality. The input of the mechanism is the shares of all parties for both inputs of the hidden gate, and it returns to the parties their shares for the output of the gate.}
\end{enumerate}

\section{\texttt{CTH} Functionality Realization}\label{sec:cthrealization}

Before describing Mohassel and Sadeghian's construction in more detail, the concept of an \textit{extended permutation} needs to be explained \cite{MS13}. A mapping $\pi : \{1\ldots M\} \rightarrow \{1\ldots N\}$ can be regarded as a permutation if it is one-to-one and onto (\textit{i.e.} a bijection).  This notion can be generalized to an extended permutation as follows: Given the positive integers $M$ and $N$, a mapping $\pi : \{1\ldots M\} \rightarrow \{1\ldots N\}$ is called as an extended permutation (\texttt{EP}) if and only if there exists exactly one $x \in \{1\ldots M\}$ for every $y \in \{1\ldots N\}$ such that $\pi(x) = y$. $x$ is often denoted by $\pi^{-1}(y)$. Unlike the mapping of a standard permutation, the mapping of an \texttt{EP} may also replicate or discard elements in the domain, allowing the domain to be larger or smaller than the range.

$n+q-m$ \textit{oblivious mapping} (\texttt{OMAP}) queries and $2q$ \texttt{Reveal} queries are needed to be implemented in order to realize the \texttt{CTH} functionality (an \texttt{OMAP} query for each $\texttt{ow}_i$, and a \texttt{Reveal} query for each $\texttt{iw}_i$). These \texttt{OMAP}/\texttt{Reveal} queries can be combined to construct a problem known as \textit{oblivious evaluation of the extended permutation} (\texttt{OEP}) to which Mohassel and Sadeghian's \texttt{CTH} scheme mainly address. 

\textbf{\texttt{OEP} Definition.} Two-party \texttt{OEP} Problem 2-$\texttt{OEP}(\vec{\pi},\vec{x},\vec{t})$ is defined as follows: The first party $P_1$ holds an \texttt{EP} $\pi : \{1\ldots M\} \rightarrow \{1\ldots N\}$, and a blinding vector for outputs $\vec{t} = (t_1,\ldots,t_N)$; whereas the other party $P_2$ holds a vector of inputs $ \vec{x} = (x_1,\ldots,x_M)$. Both the $x_i$s and $t_i$s are $\ell$-bit strings. The protocol ends in $P_2$ learning $(x_{\pi^{-1}(1)}\oplus t_1,\ldots,x_{\pi^{-1}(N)}\oplus t_N)$, while $P_1$ learning nothing.

Mohassel and Sadeghian construct a solution for \texttt{OEP} from switching networks which they observe as more efficient than the previous constructions.

\textbf{Switching Networks.} A \textit{switching network} \texttt{SN} composes of \textit{2-switches} which are interconnected. Its inputs are $N$ $\ell$-bit strings and a set of selection bits of each switches, while its outputs are $N$ $\ell$-bit strings. Each switch takes two $\ell$-bit strings and two selection bits as input, outputting two $\ell$-bit strings. Each of the outputs may get the value of any of the input strings depending on the selection bits. This means for input values $(x_0,x_1)$ and output values $(y_0,y_1)$, there are four different switch output possibilities. The two selection bits $s_0$ and $s_1$ are used for determining the switch output. In particular, the switch will output $y_0 = x_{s_0}$, and $y_1 = x_{s_1}$.

The mapping $\pi : \{1\ldots N\} \rightarrow \{1\ldots N\}$ of an \texttt{SN} is defined as $\pi(i)=j$ if and only if after the \texttt{SN} is evaluated, the string on the output wire $j$ becomes that on the input wire $i$. There is no need for the mapping $\pi$ to be a function because the value of any input wire can be mapped to any number of output wires. However, its inverse $\pi^{-1}$ must always be a function.

A \textit{permutation network} \texttt{PN} is a switching network whose mapping is a permutation of its inputs. In contrast to switching networks, permutation networks compose of \textit{1-switches}. Unlike 2-switches, they have only one selection bit $s$. For an input $(x_0,x_1)$, a 1-switch outputs one of the two possible outputs: $(x_0,x_1)$ if $s=0$, and $(x_1,x_0)$ if $s=1$. 1-switch may also be called a \textit{permutation cell}.

Waksman proposed an efficient construction for a permutation network in \cite{Wak68}. Mainly, his work suggests that a permutation network with $N=2^k$ can be constructed with  $N \texttt{log}_2 N - N + 1 $ switches, that the switch depth of the constructed \texttt{PN} will be $2 \texttt{log}_2N - 1$, and that its computational complexity will be $O(N  \texttt{log}_2 N )$.

\textbf{Extended Permutation from Switching Networks.} Mohassel and Sadeghian propose the general method for construction of an extended permutation from switching and permutation networks \cite{MS13}.  However,  extended permutations differ from switching networks in that the number of their inputs $M$ and that of their outputs $N$ need not be equal ($M \leq N$) \cite{MS13}. $N - M$ additional dummy inputs are added to the real inputs of an \texttt{EP}  $\pi : \{1\ldots M\} \rightarrow \{1\ldots N\}$ in order to simulate it as an \texttt{SN}.

\begin{figure}[]
\includegraphics[height=6cm, angle=0]{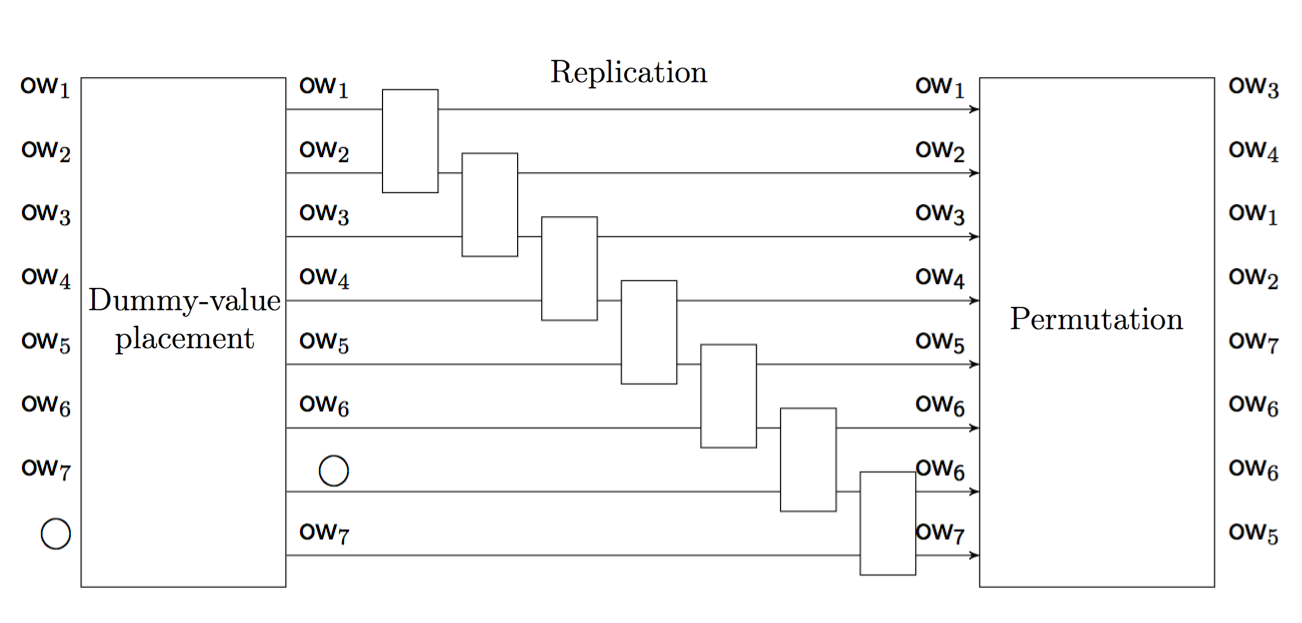}
\caption{The switching network for \texttt{EP} of the circuit in Figure \ref{fig:mapping} \cite{Pil15}.}
\label{fig:ep}
\end{figure}

Mohassel and Sadeghian divide a switching network into three components \cite{MS13}:

\begin{enumerate}
\item{\textbf{Dummy-value placement component.} This component takes  $N$ input strings composing of real and dummy ones. For each real input that $\pi$ maps to $k$ different outputs, the dummy-value placement component's output is the real string followed by $k-1$ dummy strings. An efficient implementation of this process can be via a Waksman permutation network \cite{Wak68}.}
\item{\textbf{Replication component.} This component takes the output of the dummy-value placement component as input. If a value is real, it goes unchanged. If it is a dummy value, it is replaced by the real value which precedes it. This can be computed by a series of $N - 1$ 2-switches whose selection bits $(s_0,s_1)$ are either (0,0) or (0,1). If the selection bits are (0,0), that means $x_1$ is dummy, and $x_0$ goes both of the outputs. If they are (0,1), that means both inputs are real, and both are kept on the outputs in the same order. At the end of this step, all the dummy inputs are replaced by the necessary copies of the real inputs.}
\item{\textbf{Permutation component.} This component takes the output wires of the replication component as input and outputs a permutation of them so that each string is placed on its final location according to the prescription of mapping $\pi$. An efficient implementation of this process can also be via a Waksman permutation network \cite{Wak68}.} 
\end{enumerate}

Adding up the three components, the number of switches needed for implementation of  \texttt{EP} is $2(N \texttt{log}_2 N - N + 1) + N - 1 = 2N \texttt{log}_2 N - N + 1$. The topology of the whole switching network is the same for all $N$ input \texttt{EP}s and the output depends on the selection bits.

\textbf{Oblivious Evaluation of Switching Networks (OSN).} Now, we can return to our \texttt{OEP} problem. If the \texttt{EP} construct from switched and permutation networks can be evaluated oblivously, we have a solution. Mohassel and Sadeghian propose a method for oblivious evaluation of their building blocks, \textit{i.e.}, 1-switches and 2-switches \cite{MS13}.

Recall that $P_1$ holds the selection bits of the switching network, and an output blinding vector $\vec{t}$ while $P_2$ holds the input vector $\vec{x}$. $P_2$ must learn the switching network's blinded output which is the \texttt{EP} of her input vector blinded with the vector $\vec{t}$; while $P_1$ learns $\bot$.

\textbf{Secure evaluation of a single 2-switch.} The express the general idea of the secure computation of whole network, Mohassel and Sadeghian  describe the secure evaluation of its building block,  a single 2-switch $u$ \cite{MS13}. Let the input wires of the 2-switch be $w_i$ and $w_j$, and its output wires be $w_k$ and $w_l$. $P_2$ assigns four uniformly random values $r_i$, $r_j$, $r_k$, $r_l$ to the four wires of the switch. $P_1$ has the blinded values $x_i \oplus r_i$ and $x_j \oplus r_j$ as his shares for the two input wires. The aim is letting $P_1$ obtain his output shares which is the blinded values on the output wires (see Table \ref{tab2swtichout}). In fact, there are four possible output pairs ($x_i \oplus r_k$, $x_i \oplus r_l$), ($x_i \oplus r_k$, $x_j\oplus r_l$),  ($x_j \oplus r_k$, $x_i \oplus r_l$), or ($x_j\oplus r_k$, $x_j \oplus r_l$) which $P_1$ may obtain based on the values of his selection bits $s_{0u}$ and $s_{1u}$.

\begin{table}[]
\centering
\caption{$P_1$ must learn one of these ($y_0$,$y_1$) according to his selection bits.}
\label{tab2swtichout}
\begin{tabular}{c|c|c}
{($s_{0u}$,$s_{1u}$)} & {$y_0$}   & {$y_1$}  \\ \hline
(0,0)                  & $x_i \oplus r_k$ & $x_i\oplus r_l$ \\
(0,1)                  & $x_i \oplus r_k$ & $x_j\oplus r_l$ \\
(1,0)                  & $x_j \oplus r_k$ & $x_i\oplus r_l$ \\
(1,1)                  & $x_j \oplus r_k$ & $x_j\oplus r_l$
\end{tabular}
\end{table}

\begin{table}[b]
\centering
\caption{$P_1$ gets one of these ($T_0$,$T_1$) by engaging in 1-out-of-4 OT (\ref{sec:oblivioustrans}) with $P_2$.}
\label{tab2swtichmasks}
\begin{tabular}{c|c|c}
{($s_{0u}$,$s_{1u}$)} & {$T_0$}   & {$T_1$}  \\ \hline
(0,0)                  & $r_i \oplus r_k$ & $r_i\oplus r_l$ \\
(0,1)                  & $r_i \oplus r_k$ & $r_j\oplus r_l$ \\
(1,0)                  & $r_j \oplus r_k$ & $r_i\oplus r_l$ \\
(1,1)                  & $r_j \oplus r_k$ & $r_j\oplus r_l$
\end{tabular}
\end{table}

$P_2$ prepares a table with four rows: ($r_i\oplus r_k$, $r_j\oplus r_l$), ($r_i\oplus r_k$, $r_i\oplus r_l$), ($r_j\oplus r_k$, $r_i\oplus r_l$), and ($r_j \oplus r_k$, $r_j \oplus r_l$) as shown in Table \ref{tab2swtichmasks}. Then, $P_1$ and $P_2$ engage in a 1-out-of-4 OT (\ref{sec:oblivioustrans}) in which $P_2$ inputs the four rows that he just prepared, and $P_1$ inputs his selection bits for the switch $u$. Suppose that $P_1$'s selection bits are (0,0). This means $P_1$ retrieves the first row, \textit{i.e.}, ($r_i \oplus r_k$, $r_j \oplus r_l$). He then \texttt{XOR}s $x_i \oplus r_i$ and $r_i \oplus r_k$, as well as $x_j \oplus r_j$ and $r_i \oplus r_l$, reaching his output shares $x_i \oplus r_k$ and $x_i \oplus r_l$.

\textbf{Constant round protocol.} Using the OT-based protocol proposed for 2-switches, the entire switching network can be securely computed in constant round since the protocol permits parallel OT (\ref{sec:oblivioustrans}) runs \cite{MS13}. In an \textit{offline} stage, a set of random strings for each wire is generated, and a table for each switch is prepared by $P_2$. Then $P_1$ and $P_2$ run the parallel OTs (\ref{sec:oblivioustrans}) as described above, leading to that a single row of each table is learned by $P_1$  according to his selection bits.

In the \textit{online} stage, $P_2$ blinds his input vector with the blinding strings on the inputs of the input switches before sending them to $P_1$. $P_1$ is now able to compute the entire switching network. He just need to perform sequential \texttt{XOR}s (in topological order) to reach the blinded values on the output wires. He then applies his own blinding vector $\vec{t}$ and sends the result to $P_2$. $P_2$  removes her blinding, and obtains the output of the \texttt{OEP} \cite{MS13}.

\textbf{Efficiency of the Mohassel and Sadeghian's \texttt{OEP}.} As we mentioned before, to implement an extended permutation $\pi: {1\ldots M} \rightarrow {1\ldots N}$, $2N \texttt{log} N - N + 1$ switches are needed. In fact, 1-out-of-2 OTs (\ref{sec:oblivioustrans}) are enough to implement \texttt{PN}s which consist of  1-switches. Moreover, 2-switches in replication component can also be implemented with 1-out-of-2 OTs (\ref{sec:oblivioustrans}) since their outputs have 2 possibilities unlike the generic 2-switches \cite{MS13}. To sum up, this protocol costs $2N \texttt{log} N - N + 1$ 1-out-of-2 OTs (\ref{sec:oblivioustrans}). Mohassel and Sadeghian suggests the use of OT extension \cite{IKNP03}, which reduce total number of public key operations for their \texttt{OEP} to a constant value depending on the security parameter of protocol, \textit{i.e.} $O(k)$  \cite{MS13}. In this case, the number of symmetric key operations will be twice the number of OTs, which is $4N \texttt{log} N - 2N + 2$  \cite{MS13}.

\begin{figure}
\includegraphics[height=3cm, angle=0]{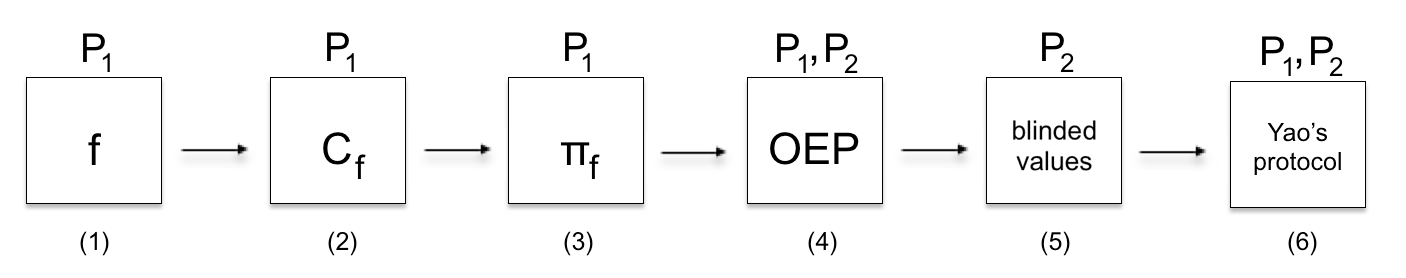}
\caption{Basic procedures of topology hiding: (1)The function $f$ known by $P_1$. (2) Circuit representation of $f$. (3) Circuit mapping of $f$. (4) \texttt{OEP} for $P_2$ learning blinded values. (5) The blinded values learnt by $P_2$. (6) Yao's protocol with the blinded values.}
\label{fig:oep}
\end{figure}

Figure \ref{fig:oep} summarizes the basic procedures of topology hiding via \texttt{OEP}. $P_1$ owns a function $f$ (1). He converts $f$ to a circuit representation $C_f$ (2). Then he extracts the circuit mapping $\pi_f$ (3). $P_1$ and $P_2$ engage in an $\texttt{OEP}$ (4) of $\pi_f$ where $P_2$ learns the blinded values of her input masking values (5) which she will later use in Yao's protocol (6).

Mohassel and Sadeghian show applications of their framework to arithmetic circuits, GMW protocol (\ref{sec:gmwprotocol}), and Yao's protocol. Since the main topic of this thesis Yao's protocol and two-party cases, we will continue with its application to Yao's protocol \cite{MS13}.

\section{Two-Party PFE of Yao's Protocol}\label{twopartypfe}

Alice and Bob would like to compute a function $f(x_0,x_1)$, where $x_0$ is Alice's input, $x_1$ is Bob's input.   Bob acts like $P_1$, and Alice acts like $P_2$ in Mohassel and Sadeghian's generic PFE scenario \cite{MS13}. So, only Bob knows $f$, and the topology of the circuit $\mathcal{C}$ of $f$. Since (\texttt{NAND}) is Turing complete, all gates in the circuit are let to be a \texttt{NAND} gate, so that the need for \texttt{PGE} functionality can be eliminated. Alice may learn the number of gates, but she should know the circuit topology. Now, one may ask the following question: How can someone garble a circuit which she does not know? Well, cryptography can achieve many incredible things.

The protocol goes as follows \cite{MS13}:

\texttt{Offline Preparation:}
\begin{enumerate}
\setcounter{enumi}{0}
\item{Bob sorts the gates topologically and computes the extended permutation $\pi_\mathcal{C}$ corresponding to circuit $\mathcal{C}$.}
\item{Alice randomly generates a masking value pair ($W_{i}^0$,$W_i^1$) for each $\texttt{ow}_i \in \texttt{OW}$. This yields a total of $M = n + g - o$ pairs. Each masking value is $k$ bits long, where $k$ is the security parameter. The \texttt{lsb} of 2 masking values belonging to the same pair must be different so that they have different labels.}
\item{Alice generates a bit vector $\vec{v} = (v_1,\ldots,v_M)$ where $v_i = \texttt{lsb} W_i^0$. She arranges each masking value pair with respect to their labels. So, they become ($W_i^{v_i}$,$W_i^{\bar{v_i}}$). This arrangement will be important during the garbled circuit evaluation. Moreover, she assigns those pairs to 2 vectors $\vec{p}=(p_1,\ldots,p_M)$ and $\vec{q}=(q_1,\ldots,q_M)$ where $p_i=W_i^{v_i}$ and $q_i=W_i^{\bar{v_i}}$.}
\item{Bob generates a random bit vector $\vec{v}' = (v_1',\ldots,v_N')$  where $v_j'$ is a random bit. This yields a total of  $N = 2 g$ bits. He also generates random blinding pairs ($t_j^0$,$t_j^1$) for each $\texttt{iw}_j \in \texttt{IW}$ such that $\texttt{lsb}(t_j^0)=\texttt{lsb}(t_j^1)$. He assigns those pairs to 2 blinding vectors $\vec{t^0}=(t_1^0,\ldots,t_N^0)$ and $\vec{t^1}=(t_1^1,\ldots,t_N^1$).}
\end{enumerate}

\texttt{Oblivious Evaluation of Switching Networks:}
\begin{enumerate}
\setcounter{enumi}{4}
\item{Alice and Bob engage in an \texttt{OEP} protocol where his input is the extended permutation $\pi_\mathcal{C}$ and $\vec{v}'$, while her input is $\vec{v}$. As a result, Alice learns $v''= (v_1'',\ldots,v_N'')$ where $v_j'' = v_{\pi^{-1}(j)} \oplus v_j'$.}
\item{Alice and Bob engage in a slightly modified \texttt{OEP} protocol where his input is the extended permutation $\pi_\mathcal{C}$ and $\vec{t^0}$, while her input is $\vec{p}$. The output is a vector $p'=(p_1',\ldots,p_N')$ where $p_j' = p_{\pi^{-1}(j)} \oplus t_j^0$. The modification is that the output is not learned by Alice but fed to a new permutation network $\hat{\texttt{SN}}$.}
\item{Alice and Bob engage in a slightly modified \texttt{OEP} protocol where his input is the extended permutation $\pi_\mathcal{C}$ and $\vec{t^1}$, while her input is $\vec{q}$. The output is a vector $q'=(q_1',\ldots,q_N')$ where $q_j' = q_{\pi^{-1}(j)} \oplus t_j^1$. The modification is that the output is not learned by Alice but fed to $\hat{\texttt{SN}}$ as well.}
\item{$\hat{\texttt{SN}}$ is a switching network including $N$ 1-swiches whose switch depth is 1. Each 1-switch $u_j$ takes ($p_j'$,$q_j'$) as input, $v_j'$ as the selection bit and outputs either ($p_j'$,$q_j'$) if $v_j'=0$ or ($q_j'$,$p_j'$) if $v_j'=1$.}
\item{After oblivious evaluation of $\hat{\texttt{SN}}$, Alice learns the output which is a set of $N$ pairs whose $j^{th}$ element is either ($p_j'$,$q_j'$) if $v_j'=0$ or ($q_j'$,$p_j'$) if $v_j'=1$.}
\end{enumerate}

\texttt{Garbling:}
\begin{enumerate}
\setcounter{enumi}{9}
\item{Alice needs to arrange the blinded pairs into their original position since the truth values must be known for garbling. This can be done by using $v''$. If $v_j''=0$, the pair remains unchanged, otherwise it is swapped. $j^{th}$ element of the output will be ($\hat{W}_{\pi^{-1}(j)} ^0$,$\hat{W}_{\pi^{-1}(j)}^1$) where $\hat{W}_i^b$ means a blinded value for $W_i^b$.}
\item{For all gates, Bob tells Alice which two of the incoming wires and which one of the outgoing wires belong to the same gate. He also tell her the outgoing wires corresponding to his and her input bits.}
\item{Alice garbles each gate by encrypting the masking values on the outgoing wires using the blinded values on the incoming wires as the keys. She sends Bob the garbled gates and the masking values for her inputs in \texttt{OW}. Bob gets his input masking values from her using 1-out-of-2 OT (\ref{sec:oblivioustrans}).}
\end{enumerate}

\texttt{Evaluating:}
\begin{enumerate}
\setcounter{enumi}{12}
\item{Using the circuit mapping $\pi_\mathcal{C}$, his blinding vectors $\vec{t^0}=(t_1^0,\ldots,t_N^0)$ and $\vec{t^1}=(t_1^1,\ldots,t_N^1$), and the garbled gates told by Alice, Bob evaluates the whole garbled circuit in topological order. When an outgoing wire $i$ is mapped to an incoming wire $j$, the masking value $W_i$ on that outgoing wire is \texttt{XOR}ed with $t_j^{\texttt{lsb}W_i}$ on the incoming wire $j$. These \texttt{XOR}ed (blinded) values are used as the decryption keys in the corresponding garbled gates to reveal the next masking value on the outgoing wire of the gate.}
\item{In the end, Bob reaches the output masking values. He tells Alice those output masking values. She decodes them and reaches $f(x_0,x_1)$. Alice tells Bob the output.}
\end{enumerate}

\textbf{Complexity}. The steps 5, 6 and 7 can be combined for only one \texttt{OEP}. Hence, this protocol requires $2N \texttt{log}_2 N - N + 1$ OTs for \texttt{OEP} and $N$ OTs for $\hat{\texttt{SN}}$, \textit{i.e.} $2N \texttt{log}_2 N  + 1$ OTs in total.  OTs for Bob's input masking values increases the total OT requirement of complete two-party PFE protocol but they do not change its round complexity since they can be implemented in parallel with the OTs for \texttt{OEP}.

\chapter{Conclusion and Discussions}\label{chap:conclusion}
\lhead{Chapter \ref{chap:conclusion}. \emph{Conclusion and Discussions}}

In this thesis, we were interested in surveying all known Yao's protocol optimizations and showing practical applications of Yao's protocol.

We have presented P\&P (\ref{sec:PP}), GRR3 (\ref{sec:GRR3}), free \texttt{XOR} (\ref{sec:freeXOR}), GRR2 (\ref{sec:GRR2}), fle\texttt{XOR} (\ref{sec:fleXOR}), and half gates (\ref{sec:halfgates}) techniques in the descending order for size of garbled gates. We have compared those optimizations in terms of communication and computation complexities, and showed their compatibilities with each other.

What else can be done for optimization? Well, in science, especially in cryptography there is no end. Although Zahur \textit{et al.} have proved that the half gates method gives the most size-optimum technique for an odd gate and yet compatible with free \texttt{XOR} \cite{ZRE15}, there are still two more optimization parameters that can be improved. There may be faster and/or more secure garbling techniques in the future. To improve on the size parameter there is a need for a revolutionary change in the traditional approach. This improvement may be a method which garbles a group of gates together instead of garbling each gate separately, resulting in a lower size.

We have also presented some generic  applications as well as some real-world application examples. The generic applications include pipelining method (\ref{sub:PipelinedImplementationFastGC}) which is useful for reducing total protocol time when the both parties of a garbling scheme is online at the same time. We also included garbling RAM (\ref{sub:GarbledRAM}) which is a quite useful technique especially for applications within the realm of big data. Some generic MPC tools \texttt{JustGarble} (\ref{sub:JustGarble}), \texttt{ABY} (\ref{sub:ABY}), \texttt{Obliv-C}, \texttt{ObliVM}, and \texttt{Frigate} (\ref{sub:Frigate}) are also introduced briefly. We compared them in terms of the use of garbling optimization techniques. At the end of the chapter, we have given some real-world applications, including MPC of satellite collusion probabilities (\ref{sub:SecureComputationofSatelliteCollusionProbabilities}) and privacy preserving data mining.

We have explained private function evaluation, and Mohassel \textit{et al.}'s PFE scheme. It is the most efficient PFE scheme known. Although their PFE scheme is limited for use right now, we know that cryptography is one of the fastest fields in computing science. It is hard to say whether it will be in use soon but someday generic PFE schemes will be in every day use for many applications, where one of the parties is also willing to hide her function since the path to developing such a technique is already open.


\addtocontents{toc}{\vspace{0em}} 

\appendix 



\addtocontents{toc}{\vspace{0em}} 

\backmatter


\label{Bibliography}

\lhead{\emph{Bibliography}} 

\bibliographystyle{unsrtnat}

\bibliography{Bibliography} 

\begin{thebibliography}{60}
\providecommand{\natexlab}[1]{#1}
\providecommand{\url}[1]{\texttt{#1}}
\expandafter\ifx\csname urlstyle\endcsname\relax
  \providecommand{\doi}[1]{doi: #1}\else
  \providecommand{\doi}{doi: \begingroup \urlstyle{rm}\Url}\fi

\bibitem[Mohassel and Sadeghian(2013)]{MS13}
P.~Mohassel and S.~Sadeghian.
\newblock How to hide circuits in mpc an efficient framework for private
  function evaluation.
\newblock In \emph{Advances in Cryptology -- EUROCRYPT 2013: 32nd Annual
  International Conference on the Theory and Applications of Cryptographic
  Techniques, Athens, Greece, May 26-30, 2013. Proceedings}, pages 557--574,
  Berlin, Heidelberg, 2013. Springer Berlin Heidelberg.
\newblock ISBN 978-3-642-38348-9.
\newblock \doi{10.1007/978-3-642-38348-9_33}.
\newblock Available at \url{http://dx.doi.org/10.1007/978-3-642-38348-9_33}.

\bibitem[Schneider(2011)]{Sch11}
T.~Schneider.
\newblock \emph{Engineering Secure Two-Party Computation Protocols -- Advances
  in Design, Optimization, and Applications of Efficient Secure Function
  Evaluation}.
\newblock PhD thesis, Ruhr-University Bochum, Germany, Information Sciences,
  2011.
\newblock Available at \url{http://thomaschneider.de/papers/S11Thesis.pdf}.

\bibitem[Bellare et~al.(2012)Bellare, Hoang, and Rogaway]{BHR12}
M.~Bellare, V.~Hoang, and P.~Rogaway.
\newblock Foundations of garbled circuits.
\newblock In \emph{Proceedings of the 2012 ACM Conference on Computer and
  Communications Security}, CCS '12, pages 784--796, New York, NY, USA, 2012.
  ACM.
\newblock ISBN 978-1-4503-1651-4.
\newblock \doi{10.1145/2382196.2382279}.
\newblock Available at \url{http://doi.acm.org/10.1145/2382196.2382279}.

\bibitem[Pullonen(2015)]{Pil15}
P.~Pullonen.
\newblock Private function evaluation for mpc.
\newblock 2015.
\newblock Available at
  \url{https://courses.cs.ut.ee/MTAT.07.022/2015_spring/uploads/Main/pille-report-s15.pdf}.

\bibitem[Zahur et~al.(2015)Zahur, Rosulek, and Evans]{ZRE15}
S.~Zahur, M.~Rosulek, and D.~Evans.
\newblock Two halves make a whole - reducing data transfer in garbled circuits
  using half gates.
\newblock In \emph{Advances in Cryptology - {EUROCRYPT} 2015 - 34th Annual
  International Conference on the Theory and Applications of Cryptographic
  Techniques, Sofia, Bulgaria, April 26-30, 2015, Proceedings, Part {II}},
  pages 220--250, 2015.
\newblock \doi{10.1007/978-3-662-46803-6_8}.
\newblock Available at \url{http://dx.doi.org/10.1007/978-3-662-46803-6_8}.

\bibitem[Yao(1982)]{Yao82}
A.~Yao.
\newblock Protocols for secure computations.
\newblock In \emph{Proceedings of the 23rd Annual Symposium on Foundations of
  Computer Science}, SFCS '82, pages 160--164, Washington, DC, USA, 1982. IEEE
  Computer Society.
\newblock \doi{10.1109/SFCS.1982.88}.
\newblock Available at \url{http://dx.doi.org/10.1109/SFCS.1982.88}.

\bibitem[Yao(1986)]{Yao86}
A.~Yao.
\newblock How to generate and exchange secrets.
\newblock In \emph{Proceedings of the 27th Annual Symposium on Foundations of
  Computer Science}, SFCS '86, pages 162--167, Washington, DC, USA, 1986. IEEE
  Computer Society.
\newblock ISBN 0-8186-0740-8.
\newblock \doi{10.1109/SFCS.1986.25}.
\newblock Available at \url{http://dx.doi.org/10.1109/SFCS.1986.25}.

\bibitem[Goldreich et~al.(1987)Goldreich, Micali, and Wigderson]{GMW87}
O.~Goldreich, S.~Micali, and A.~Wigderson.
\newblock How to play any mental game.
\newblock In \emph{Proceedings of the Nineteenth Annual ACM Symposium on Theory
  of Computing}, STOC '87, pages 218--229, New York, NY, USA, 1987. ACM.
\newblock ISBN 0-89791-221-7.
\newblock \doi{10.1145/28395.28420}.
\newblock Available at \url{http://doi.acm.org/10.1145/28395.28420}.

\bibitem[Bogdanov et~al.(2012)Bogdanov, Talviste, and Willemson]{BTW12}
D.~Bogdanov, R.~Talviste, and J.~Willemson.
\newblock Deploying secure multi-party computation for financial data analysis.
\newblock In \emph{Financial Cryptography and Data Security: 16th International
  Conference, FC 2012, Kralendijk, Bonaire, Februray 27-March 2, 2012, Revised
  Selected Papers}, pages 57--64, Berlin, Heidelberg, 2012. Springer Berlin
  Heidelberg.
\newblock ISBN 978-3-642-32946-3.
\newblock \doi{10.1007/978-3-642-32946-3_5}.
\newblock Available at \url{http://dx.doi.org/10.1007/978-3-642-32946-3_5}.

\bibitem[Lindell and Pinkas(2009)]{LP08}
Y.~Lindell and B.~Pinkas.
\newblock Secure multiparty computation for privacy-preserving data mining.
\newblock \emph{The Journal of Privacy and Confidentiality}, 1\penalty0
  (1):\penalty0 59--98, 2009.
\newblock Available at \url{http://repository.cmu.edu/jpc/vol1/iss1/5}.

\bibitem[Hemenway et~al.(2016)Hemenway, Lu, Ostrovsky, and Welser~IV]{HLOW16}
B.~Hemenway, S.~Lu, R.~Ostrovsky, and W.~Welser~IV.
\newblock High-precision secure computation of satellite collision
  probabilities.
\newblock In \emph{Security and Cryptography for Networks: 10th International
  Conference, SCN 2016, Amalfi, Italy, August 31 -- September 2, 2016,
  Proceedings}, pages 169--187, Cham, 2016. Springer International Publishing.
\newblock ISBN 978-3-319-44618-9.
\newblock \doi{10.1007/978-3-319-44618-9_9}.
\newblock Available at \url{http://dx.doi.org/10.1007/978-3-319-44618-9_9}.

\bibitem[Cramer et~al.(1997)Cramer, Gennaro, and Schoenmakers]{CGS97}
R.~Cramer, R.~Gennaro, and B.~Schoenmakers.
\newblock A secure and optimally efficient multi-authority election scheme.
\newblock In \emph{Advances in Cryptology --- EUROCRYPT '97: International
  Conference on the Theory and Application of Cryptographic Techniques
  Konstanz, Germany, May 11--15, 1997 Proceedings}, pages 103--118, Berlin,
  Heidelberg, 1997. Springer Berlin Heidelberg.
\newblock ISBN 978-3-540-69053-5.
\newblock \doi{10.1007/3-540-69053-0_9}.
\newblock Available at \url{http://dx.doi.org/10.1007/3-540-69053-0_9}.

\bibitem[Bogetoft et~al.(2009)Bogetoft, Christensen, Damg{\aa}rd, Geisler,
  Jakobsen, Kr{\o}igaard, Nielsen, Nielsen, Nielsen, Pagter, Schwartzbach, and
  Toft]{BCD+09}
P.~Bogetoft, D.~Christensen, I.~Damg{\aa}rd, M.~Geisler, T.~Jakobsen,
  M.~Kr{\o}igaard, J.~Nielsen, J.~Nielsen, K.~Nielsen, J.~Pagter,
  M.~Schwartzbach, and T.~Toft.
\newblock Financial cryptography and data security.
\newblock chapter Secure Multiparty Computation Goes Live, pages 325--343.
  Springer-Verlag, Berlin, Heidelberg, 2009.
\newblock ISBN 978-3-642-03548-7.
\newblock \doi{10.1007/978-3-642-03549-4_20}.
\newblock Available at \url{http://dx.doi.org/10.1007/978-3-642-03549-4_20}.

\bibitem[Naor et~al.(1999)Naor, Pinkas, and Sumner]{NPS99}
M.~Naor, B.~Pinkas, and R.~Sumner.
\newblock Privacy preserving auctions and mechanism design.
\newblock In \emph{Proceedings of the 1st ACM Conference on Electronic
  Commerce}, EC '99, pages 129--139, New York, NY, USA, 1999. ACM.
\newblock ISBN 1-58113-176-3.
\newblock \doi{10.1145/336992.337028}.
\newblock Available at \url{http://doi.acm.org/10.1145/336992.337028}.

\bibitem[Kulkarni and Namboodiri(2013)]{KN13}
R.~Kulkarni and A.~Namboodiri.
\newblock Secure hamming distance based biometric authentication.
\newblock In \emph{2013 International Conference on Biometrics (ICB)}, pages
  1--6, June 2013.
\newblock \doi{10.1109/ICB.2013.6613008}.
\newblock Available at \url{http://dx.doi.org/10.1109/ICB.2013.6613008}.

\bibitem[Bringer et~al.(2013)Bringer, Chabanne, and Patey]{BCP13}
J.~Bringer, H.~Chabanne, and A.~Patey.
\newblock Shade: Secure hamming distance computation from oblivious transfer.
\newblock In \emph{Financial Cryptography and Data Security: FC 2013 Workshops,
  USEC and WAHC 2013, Okinawa, Japan, April 1, 2013, Revised Selected Papers},
  pages 164--176, Berlin, Heidelberg, 2013. Springer Berlin Heidelberg.
\newblock ISBN 978-3-642-41320-9.
\newblock \doi{10.1007/978-3-642-41320-9_11}.
\newblock Available at \url{http://dx.doi.org/10.1007/978-3-642-41320-9_11}.

\bibitem[Kiraz et~al.(2015)Kiraz, Gen\c{c}, and Karda\c{s}]{KGK15}
M.~Kiraz, Z.~Gen\c{c}, and S.~Karda\c{s}.
\newblock Security and efficiency analysis of the hamming distance computation
  protocol based on oblivious transfer.
\newblock \emph{Security and Communication Networks}, 8\penalty0 (18):\penalty0
  4123--4135, 2015.
\newblock \doi{10.1002/sec.1329}.
\newblock Available at \url{http://dx.doi.org/10.1002/sec.1329}.

\bibitem[Launchbury et~al.(2014)Launchbury, Archer, DuBuisson, and
  Mertens]{LADM14}
J.~Launchbury, D.~Archer, T.~DuBuisson, and E.~Mertens.
\newblock Application-scale secure multiparty computation.
\newblock In \emph{Programming Languages and Systems: 23rd European Symposium
  on Programming, ESOP 2014, Held as Part of the European Joint Conferences on
  Theory and Practice of Software, ETAPS 2014, Grenoble, France, April 5-13,
  2014, Proceedings}, pages 8--26, Berlin, Heidelberg, 2014. Springer Berlin
  Heidelberg.
\newblock ISBN 978-3-642-54833-8.
\newblock \doi{10.1007/978-3-642-54833-8_2}.
\newblock Available at \url{http://dx.doi.org/10.1007/978-3-642-54833-8_2}.

\bibitem[Kiraz and Schoenmakers(2008)]{KiS08}
M.~Kiraz and B.~Schoenmakers.
\newblock An efficient protocol for fair secure two-party computation.
\newblock In \emph{Topics in Cryptology -- CT-RSA 2008: The Cryptographers'
  Track at the RSA Conference 2008, San Francisco, CA, USA, April 8-11, 2008.
  Proceedings}, pages 88--105, Berlin, Heidelberg, 2008. Springer Berlin
  Heidelberg.
\newblock ISBN 978-3-540-79263-5.
\newblock \doi{10.1007/978-3-540-79263-5_6}.
\newblock Available at \url{http://dx.doi.org/10.1007/978-3-540-79263-5_6}.

\bibitem[Kiraz(2008)]{Kir08}
M.~Kiraz.
\newblock \emph{Secure and Fair Two-Party Computation}.
\newblock PhD thesis, Technische Universiteit Eindhoven, 2008.
\newblock Available at \url{http://alexandria.tue.nl/extra2/200811317.pdf}.

\bibitem[Huang et~al.(2011)Huang, Evans, Katz, and Malka]{HEKM11}
Y.~Huang, D.~Evans, J.~Katz, and L.~Malka.
\newblock Faster secure two-party computation using garbled circuits.
\newblock In \emph{Proceedings of the 20th USENIX Conference on Security},
  SEC'11, pages 35--35, Berkeley, CA, USA, 2011. USENIX Association.
\newblock Available at \url{http://dl.acm.org/citation.cfm?id=2028067.2028102}.

\bibitem[Kiraz and Schoenmakers(2006)]{KiS06}
M.~Kiraz and B.~Schoenmakers.
\newblock A protocol issue for the malicious case of yao's garbled circuit
  construction.
\newblock In \emph{In Proceedings of 27th Symposium on Information Theory in
  the Benelux}, 2006.
\newblock Available at
  \url{http://citeseerx.ist.psu.edu/viewdoc/download?doi=10.1.1.140.2627&rep=rep1&type=pdf}.

\bibitem[Kumar(2003)]{Ana14}
A.~Kumar.
\newblock \emph{Fundamentals of Digital Circuits}.
\newblock Prentice-Hall Of India Pvt. Limited, 2003.
\newblock ISBN 9788120317451.
\newblock Available at \url{https://books.google.com.tr/books?id=BOVkrtiLUcEC}.

\bibitem[Pinkas et~al.(2009)Pinkas, Schneider, Smart, and Williams]{PSSW09}
B.~Pinkas, T.~Schneider, N.~Smart, and S.~Williams.
\newblock Secure two-party computation is practical.
\newblock In \emph{Advances in Cryptology -- ASIACRYPT 2009: 15th International
  Conference on the Theory and Application of Cryptology and Information
  Security, Tokyo, Japan, December 6-10, 2009. Proceedings}, pages 250--267,
  Berlin, Heidelberg, 2009. Springer Berlin Heidelberg.
\newblock ISBN 978-3-642-10366-7.
\newblock \doi{10.1007/978-3-642-10366-7_15}.
\newblock Available at \url{http://dx.doi.org/10.1007/978-3-642-10366-7_15}.

\bibitem[Vollmer(1999)]{Vol99}
H.~Vollmer.
\newblock \emph{Introduction to Circuit Complexity: A Uniform Approach}.
\newblock Springer-Verlag New York, Inc., Secaucus, NJ, USA, 1999.
\newblock ISBN 3540643109.

\bibitem[Sipser(1996)]{Sip96}
M.~Sipser.
\newblock \emph{Introduction to the Theory of Computation}.
\newblock International Thomson Publishing, 1st edition, 1996.
\newblock ISBN 053494728X.

\bibitem[Kojevnikov and Kulikov(2010)]{KK10}
A.~Kojevnikov and A.~Kulikov.
\newblock Circuit complexity and multiplicative complexity of boolean
  functions.
\newblock In \emph{Programs, Proofs, Processes: 6th Conference on Computability
  in Europe, CiE 2010, Ponta Delgada, Azores, Portugal, June 30 -- July 4,
  2010. Proceedings}, pages 239--245, Berlin, Heidelberg, 2010. Springer Berlin
  Heidelberg.
\newblock ISBN 978-3-642-13962-8.
\newblock \doi{10.1007/978-3-642-13962-8_27}.
\newblock Available at \url{http://dx.doi.org/10.1007/978-3-642-13962-8_27}.

\bibitem[Boyar et~al.(2000)Boyar, Peralta, and Pochuev]{BPP00}
J.~Boyar, R.~Peralta, and D.~Pochuev.
\newblock On the multiplicative complexity of boolean functions over the basis
  ($\wedge,\oplus,1$).
\newblock \emph{Theoretical Computer Science}, 235\penalty0 (1):\penalty0 43 --
  57, 2000.
\newblock ISSN 0304-3975.
\newblock \doi{http://dx.doi.org/10.1016/S0304-3975(99)00182-6}.
\newblock Available at
  \url{http://www.sciencedirect.com/science/article/pii/S0304397599001826}.

\bibitem[Daemen and Rijmen(2002)]{AES-FIPS}
J.~Daemen and V.~Rijmen.
\newblock \emph{The Design of Rijndael: AES - The Advanced Encryption
  Standard}.
\newblock Springer Verlag, Berlin, Heidelberg, New York, 2002.
\newblock ISBN 3-540-42580-2.

\bibitem[Rivest et~al.(1978)Rivest, Shamir, and Adleman]{RSA78}
R.~Rivest, A.~Shamir, and L.~Adleman.
\newblock A method for obtaining digital signatures and public-key
  cryptosystems.
\newblock \emph{Commun. ACM}, 21\penalty0 (2):\penalty0 120--126, February
  1978.
\newblock ISSN 0001-0782.
\newblock \doi{10.1145/359340.359342}.
\newblock Available at \url{http://doi.acm.org/10.1145/359340.359342}.

\bibitem[El~Gamal(1985)]{ElGamal85}
T.~El~Gamal.
\newblock A public key cryptosystem and a signature scheme based on discrete
  logarithms.
\newblock In \emph{Proceedings of CRYPTO 84 on Advances in Cryptology}, pages
  10--18, New York, NY, USA, 1985. Springer-Verlag New York, Inc.
\newblock ISBN 0-387-15658-5.
\newblock Available at \url{http://dl.acm.org/citation.cfm?id=19478.19480}.

\bibitem[Rogaway and Shrimpton(2004)]{RS04}
P.~Rogaway and T.~Shrimpton.
\newblock Cryptographic hash-function basics: Definitions, implications, and
  separations for preimage resistance, second-preimage resistance, and
  collision resistance.
\newblock In \emph{Fast Software Encryption: 11th International Workshop, FSE
  2004, Delhi, India, February 5-7, 2004. Revised Papers}, pages 371--388,
  Berlin, Heidelberg, 2004. Springer Berlin Heidelberg.
\newblock ISBN 978-3-540-25937-4.
\newblock \doi{10.1007/978-3-540-25937-4_24}.
\newblock Available at \url{http://dx.doi.org/10.1007/978-3-540-25937-4_24}.

\bibitem[Koblitz and Menezes(2015)]{KM15}
N.~Koblitz and A.~Menezes.
\newblock The random oracle model: a twenty-year retrospective.
\newblock \emph{Designs, Codes and Cryptography}, 77\penalty0 (2):\penalty0
  587--610, 2015.
\newblock ISSN 1573-7586.
\newblock \doi{10.1007/s10623-015-0094-2}.
\newblock Available at \url{http://dx.doi.org/10.1007/s10623-015-0094-2}.

\bibitem[Handschuh(2011)]{Han11}
H.~Handschuh.
\newblock Sha-0, sha-1, sha-2 (secure hash algorithm).
\newblock In \emph{Encyclopedia of Cryptography and Security}, pages
  1190--1193, Boston, MA, 2011. Springer US.
\newblock ISBN 978-1-4419-5906-5.
\newblock \doi{10.1007/978-1-4419-5906-5_615}.
\newblock Available at \url{http://dx.doi.org/10.1007/978-1-4419-5906-5_615}.

\bibitem[Lindell et~al.(2008)Lindell, Pinkas, and Smart]{LPS08}
Y.~Lindell, B.~Pinkas, and N.~Smart.
\newblock Implementing two-party computation efficiently with security against
  malicious adversaries.
\newblock In \emph{Security and Cryptography for Networks: 6th International
  Conference, SCN 2008, Amalfi, Italy, September 10-12, 2008. Proceedings},
  pages 2--20, Berlin, Heidelberg, 2008. Springer Berlin Heidelberg.
\newblock ISBN 978-3-540-85855-3.
\newblock \doi{10.1007/978-3-540-85855-3_2}.
\newblock Available at \url{http://dx.doi.org/10.1007/978-3-540-85855-3_2}.

\bibitem[Kreuter et~al.(2012)Kreuter, a.~shelat, and Shen]{KsS12}
B.~Kreuter, a.~shelat, and C.~Shen.
\newblock Billion-gate secure computation with malicious adversaries.
\newblock In \emph{Presented as part of the 21st USENIX Security Symposium
  (USENIX Security 12)}, pages 285--300, Bellevue, WA, 2012. USENIX.
\newblock ISBN 978-931971-95-9.
\newblock Available at \url{http://eprint.iacr.org/2012/179}.

\bibitem[Bellare et~al.(2013)Bellare, Hoang, Keelveedhi, and Rogaway]{BHKR13}
M.~Bellare, V.~Hoang, S.~Keelveedhi, and P.~Rogaway.
\newblock Efficient garbling from a fixed-key blockcipher.
\newblock In \emph{Proceedings of the 2013 IEEE Symposium on Security and
  Privacy}, SP '13, pages 478--492, Washington, DC, USA, 2013. IEEE Computer
  Society.
\newblock ISBN 978-0-7695-4977-4.
\newblock \doi{10.1109/SP.2013.39}.
\newblock Available at \url{http://dx.doi.org/10.1109/SP.2013.39}.

\bibitem[Shamir(1979)]{Sha79}
A.~Shamir.
\newblock How to share a secret.
\newblock \emph{Commun. ACM}, 22\penalty0 (11):\penalty0 612--613, November
  1979.
\newblock ISSN 0001-0782.
\newblock \doi{10.1145/359168.359176}.
\newblock Available at \url{http://doi.acm.org/10.1145/359168.359176}.

\bibitem[Demmler et~al.(2015)Demmler, Schneider, and Zohner]{DSZ15}
D~Demmler, T~Schneider, and M~Zohner.
\newblock {ABY} -- a framework for efficient mixed-protocol secure two-party
  computation.
\newblock In \emph{22. Annual Network and Distributed System Security Symposium
  (NDSS'15)}. The Internet Society, February 8-11, 2015.
\newblock \doi{10.14722/ndss.2015.23113}.
\newblock Available at \url{http://encrypto.de/code/ABY}.

\bibitem[Naskar and Sengupta(2010)]{NS10}
R.~Naskar and I.~Sengupta.
\newblock Secret sharing and proactive renewal of shares in hierarchical
  groups.
\newblock \emph{CoRR}, abs/1006.1192, 2010.
\newblock Available at
  \url{http://dblp.uni-trier.de/db/journals/corr/corr1006.html#abs-1006-1192}.

\bibitem[Chou and Orlandi(2015)]{CO15}
T.~Chou and C.~Orlandi.
\newblock The simplest protocol for oblivious transfer.
\newblock In \emph{Progress in Cryptology -- LATINCRYPT 2015: 4th International
  Conference on Cryptology and Information Security in Latin America,
  Guadalajara, Mexico, August 23-26, 2015, Proceedings}, pages 40--58, Cham,
  2015. Springer International Publishing.
\newblock ISBN 978-3-319-22174-8.
\newblock \doi{10.1007/978-3-319-22174-8_3}.
\newblock Available at \url{http://dx.doi.org/10.1007/978-3-319-22174-8_3}.

\bibitem[Ishai et~al.(2003{\natexlab{a}})Ishai, Kilian, Nissim, and
  Petrank]{YKNP03}
Y.~Ishai, J.~Kilian, K.~Nissim, and E.~Petrank.
\newblock Extending oblivious transfers efficiently.
\newblock In \emph{Advances in Cryptology - CRYPTO 2003: 23rd Annual
  International Cryptology Conference, Santa Barbara, California, USA, August
  17-21, 2003. Proceedings}, pages 145--161, Berlin, Heidelberg,
  2003{\natexlab{a}}. Springer Berlin Heidelberg.
\newblock ISBN 978-3-540-45146-4.
\newblock \doi{10.1007/978-3-540-45146-4_9}.
\newblock Available at \url{http://dx.doi.org/10.1007/978-3-540-45146-4_9}.

\bibitem[Gentry(2009)]{Gen09}
C.~Gentry.
\newblock \emph{A Fully Homomorphic Encryption Scheme}.
\newblock PhD thesis, Stanford University, Stanford, CA, USA, 2009.
\newblock AAI3382729, Available at
  \url{https://crypto.stanford.edu/craig/craig-thesis.pdf}.

\bibitem[Kolesnikov et~al.(2014)Kolesnikov, Mohassel, and Rosulek]{KMR14}
V.~Kolesnikov, P.~Mohassel, and M.~Rosulek.
\newblock Flexor: Flexible garbling for xor gates that beats free-xor.
\newblock In \emph{Advances in Cryptology -- CRYPTO 2014: 34th Annual
  Cryptology Conference, Santa Barbara, CA, USA, August 17-21, 2014,
  Proceedings, Part II}, pages 440--457, Berlin, Heidelberg, 2014. Springer
  Berlin Heidelberg.
\newblock ISBN 978-3-662-44381-1.
\newblock \doi{10.1007/978-3-662-44381-1_25}.
\newblock Available at \url{http://dx.doi.org/10.1007/978-3-662-44381-1_25}.

\bibitem[Beaver et~al.(1990)Beaver, Micali, and Rogaway]{BMR90}
D.~Beaver, S.~Micali, and P.~Rogaway.
\newblock The round complexity of secure protocols.
\newblock In \emph{Proceedings of the Twenty-second Annual ACM Symposium on
  Theory of Computing}, STOC '90, pages 503--513, New York, NY, USA, 1990. ACM.
\newblock ISBN 0-89791-361-2.
\newblock \doi{10.1145/100216.100287}.
\newblock Available at \url{http://doi.acm.org/10.1145/100216.100287}.

\bibitem[Kolesnikov and Schneider(2008)]{KS08}
V.~Kolesnikov and T.~Schneider.
\newblock Improved garbled circuit: Free xor gates and applications.
\newblock In \emph{Proceedings of the 35th International Colloquium on
  Automata, Languages and Programming, Part II}, ICALP '08, pages 486--498,
  Berlin, Heidelberg, 2008. Springer-Verlag.
\newblock ISBN 978-3-540-70582-6.
\newblock \doi{10.1007/978-3-540-70583-3_40}.
\newblock Available at \url{http://dx.doi.org/10.1007/978-3-540-70583-3_40}.

\bibitem[Choi et~al.(2012)Choi, Katz, Kumaresan, and Zhou]{CKK12}
S.~Choi, J.~Katz, R.~Kumaresan, and H.~Zhou.
\newblock On the security of the ``free-xor'' technique.
\newblock In \emph{Theory of Cryptography}, volume 7194 of \emph{Lecture Notes
  in Computer Science}, pages 39--53. Springer, 2012.
\newblock \doi{10.1007/978-3-642-28914-9_3}.
\newblock Available at \url{http://dx.doi.org/10.1007/978-3-642-28914-9_3}.

\bibitem[Lu and Ostrovsky(2013)]{LO13}
S.~Lu and R.~Ostrovsky.
\newblock How to garble ram programs?
\newblock In \emph{Advances in Cryptology -- EUROCRYPT 2013: 32nd Annual
  International Conference on the Theory and Applications of Cryptographic
  Techniques, Athens, Greece, May 26-30, 2013. Proceedings}, pages 719--734,
  Berlin, Heidelberg, 2013. Springer Berlin Heidelberg.
\newblock ISBN 978-3-642-38348-9.
\newblock \doi{10.1007/978-3-642-38348-9_42}.
\newblock Available at \url{http://dx.doi.org/10.1007/978-3-642-38348-9_42}.

\bibitem[Boneh and Franklin(2001)]{BF01}
D.~Boneh and M.~Franklin.
\newblock Identity-based encryption from the weil pairing.
\newblock In \emph{Proceedings of the 21st Annual International Cryptology
  Conference on Advances in Cryptology}, CRYPTO '01, pages 213--229, London,
  UK, UK, 2001. Springer-Verlag.
\newblock ISBN 3-540-42456-3.
\newblock Available at \url{http://dl.acm.org/citation.cfm?id=646766.704155}.

\bibitem[Gentry et~al.(2014)Gentry, Halevi, Lu, Ostrovsky, Raykova, and
  Wichs]{GHRW14}
C.~Gentry, S.~Halevi, S.~Lu, R.~Ostrovsky, M.~Raykova, and D.~Wichs.
\newblock Garbled ram revisited.
\newblock In \emph{Advances in Cryptology -- EUROCRYPT 2014: 33rd Annual
  International Conference on the Theory and Applications of Cryptographic
  Techniques, Copenhagen, Denmark, May 11-15, 2014. Proceedings}, pages
  405--422, Berlin, Heidelberg, 2014. Springer Berlin Heidelberg.
\newblock ISBN 978-3-642-55220-5.
\newblock \doi{10.1007/978-3-642-55220-5_23}.
\newblock Available at \url{http://dx.doi.org/10.1007/978-3-642-55220-5_23}.

\bibitem[Goldreich and Ostrovsky(1996)]{GO96}
O.~Goldreich and R.~Ostrovsky.
\newblock Software protection and simulation on oblivious rams.
\newblock \emph{J. ACM}, 43\penalty0 (3):\penalty0 431--473, May 1996.
\newblock ISSN 0004-5411.
\newblock \doi{10.1145/233551.233553}.
\newblock Available at \url{http://doi.acm.org/10.1145/233551.233553}.

\bibitem[Stefanov et~al.(2013)Stefanov, van Dijk, Shi, Fletcher, Ren, Yu, and
  Devadas]{SDS+12}
E.~Stefanov, M.~van Dijk, E.~Shi, C.~Fletcher, L.~Ren, X.~Yu, and S.~Devadas.
\newblock Path oram: An extremely simple oblivious ram protocol.
\newblock In \emph{Proceedings of the 2013 ACM SIGSAC Conference on Computer
  \&\#38; Communications Security}, CCS '13, pages 299--310, New York, NY, USA,
  2013. ACM.
\newblock ISBN 978-1-4503-2477-9.
\newblock \doi{10.1145/2508859.2516660}.
\newblock Available at \url{http://doi.acm.org/10.1145/2508859.2516660}.

\bibitem[Zahur and Evans(2015)]{ZE15}
S.~Zahur and D.~Evans.
\newblock Obliv-c: {A} language for extensible data-oblivious computation.
\newblock \emph{{IACR} Cryptology ePrint Archive}, 2015:\penalty0 1153, 2015.
\newblock Available at \url{http://eprint.iacr.org/2015/1153}.

\bibitem[Liu et~al.(2015)Liu, Wang, Nayak, Huang, and Shi]{DBLP:conf/sp/2015}
C.~Liu, X.~Wang, K.~Nayak, Y.~Huang, and E.~Shi.
\newblock Oblivm: {A} programming framework for secure computation.
\newblock In \emph{2015 {IEEE} Symposium on Security and Privacy, {SP} 2015,
  San Jose, CA, USA, May 17-21, 2015}, pages 359--376, 2015.
\newblock \doi{10.1109/SP.2015.29}.
\newblock Available at \url{http://dx.doi.org/10.1109/SP.2015.29}.

\bibitem[Mood et~al.(2016)Mood, Gupta, Carter, Butler, and Traynor]{MGC+16}
B.~Mood, D.~Gupta, H.~Carter, K.~Butler, and P.~Traynor.
\newblock Frigate: {A} validated, extensible, and efficient compiler and
  interpreter for secure computation.
\newblock In \emph{{IEEE} European Symposium on Security and Privacy,
  EuroS{\&}P 2016, Saarbr{\"{u}}cken, Germany, March 21-24, 2016}, 2016.
\newblock Available at \url{http://dx.doi.org/10.1109/EuroSP.2016.20}.

\bibitem[Alfano(January-March 2005)]{Alf05}
S.~Alfano.
\newblock A numerical implementation of spherical object collision probability.
\newblock \emph{Journal of the Astronautical Sciences}, 53\penalty0
  (1):\penalty0 103--109, January-March 2005.
\newblock Available at
  \url{http://centerforspace.com/downloads/files/pubs/JAS.V53.N01.pdf}.

\bibitem[Lindell and Pinkas(2000)]{LP00}
Y.~Lindell and B.~Pinkas.
\newblock Privacy preserving data mining.
\newblock In \emph{Proceedings of the 20th Annual International Cryptology
  Conference on Advances in Cryptology}, CRYPTO '00, pages 36--54, London, UK,
  UK, 2000. Springer-Verlag.
\newblock ISBN 3-540-67907-3.
\newblock Available at \url{http://dl.acm.org/citation.cfm?id=646765.704129}.

\bibitem[Mazur(2010)]{Maz10}
D.~Mazur.
\newblock \emph{Combinatorics : a guided tour}.
\newblock MAA textbooks. Mathematical Association of America, Washington, DC,
  2010.
\newblock ISBN 978-0-88385-762-5.
\newblock Available at \url{http://opac.inria.fr/record=b1133224}.

\bibitem[Waksman(1968)]{Wak68}
A.~Waksman.
\newblock A permutation network.
\newblock \emph{J. ACM}, 15\penalty0 (1):\penalty0 159--163, January 1968.
\newblock ISSN 0004-5411.
\newblock \doi{10.1145/321439.321449}.
\newblock Available at \url{http://doi.acm.org/10.1145/321439.321449}.

\bibitem[Ishai et~al.(2003{\natexlab{b}})Ishai, Kilian, Nissim, and
  Petrank]{IKNP03}
Y.~Ishai, J.~Kilian, K.~Nissim, and E.~Petrank.
\newblock Extending oblivious transfers efficiently.
\newblock In \emph{Advances in Cryptology - CRYPTO 2003: 23rd Annual
  International Cryptology Conference, Santa Barbara, California, USA, August
  17-21, 2003. Proceedings}, pages 145--161, Berlin, Heidelberg,
  2003{\natexlab{b}}. Springer Berlin Heidelberg.
\newblock ISBN 978-3-540-45146-4.
\newblock \doi{10.1007/978-3-540-45146-4_9}.
\newblock Available at \url{http://dx.doi.org/10.1007/978-3-540-45146-4_9}.

\end{thebibliography}

\end{document}